%% file: main.tex
\RequirePackage{rotating}
\PassOptionsToPackage{usenames,dvipsnames}{xcolour}
\documentclass[twocolumn,twocolappendix]{openjournal}

\pdfoutput=1 %for arXiv submission
\usepackage{amsmath,amstext}
\usepackage[T1]{fontenc}
\usepackage{apjfonts}
\usepackage{ae,aecompl}
\usepackage[utf8]{inputenc}
\usepackage{hyperref}
\usepackage[figure,figure*]{hypcap}
\usepackage{natbib}
\usepackage{xcolor}

\usepackage{url}
\usepackage{mdwlist}
\usepackage{listings}
\usepackage{rotating}
\usepackage{multirow}
\usepackage{lineno}
\shorttitle{Catalog-to-cosmology framework for LSST}
\shortauthors{Prat, Zuntz et al.\ (LSST~DESC)}

\begin{document}
\title{The catalog-to-cosmology framework for weak lensing and galaxy clustering for LSST}
\input{authorlist}

\begin{abstract}
We present \textsc{TXPipe}, a modular, automated and reproducible pipeline for 
ingesting catalog data and performing all the calculations required to obtain quality-assured two-point measurements of lensing and clustering, and their covariances, with the metadata necessary for parameter estimation. The pipeline is developed within the Rubin Observatory Legacy Survey of Space and Time (LSST) Dark Energy Science Collaboration (DESC), and designed for cosmology analyses using LSST data. In this paper, we present the pipeline for the so-called ``3$\times$2pt'' analysis -- a combination of three two-point functions that measure the auto- and cross-correlation between galaxy density and shapes. We perform the analysis both in real and harmonic space using \textsc{TXPipe} and other LSST-DESC tools. We validate the pipeline using Gaussian simulations and show that it accurately measures data vectors and recovers the input cosmology to the accuracy level required for the first year of LSST data under this simplified scenario. We also apply the pipeline to a realistic mock galaxy sample extracted from the \textsc{CosmoDC2} simulation suite \citep{Korytov2019}. \textsc{TXPipe} establishes a baseline framework that can be built upon as the LSST survey proceeds. Furthermore, the pipeline is designed to be easily extended to science probes beyond the 3$\times$2pt analysis.  
\end{abstract}


%\tableofcontents%-----------------------------
%===========================
% BEGINNING OF THE MAIN TEXT
%===========================

\section{Introduction}

The large-scale structure (LSS) contains rich information on both the geometry of spacetime and the growth of cosmic structure. Among the most direct avenues for probing the LSS is examining the statistical properties of the large-scale distribution of galaxies \citep{Eisenstein2005,Springel2006}, which are biased tracers of the distribution of mass. In addition, one can use the phenomenon of {\it weak gravitational lensing} -- the small deflection of photon trajectories due to the perturbation of spacetime from mass -- to map the distribution of mass directly. The weak lensing-inferred mass distribution is typically measured using the distortion of observed galaxy shapes \citep[for a review of weak lensing, see e.g.][]{Bartelmann2001}. Recent analyses of galaxy surveys have further shown that it is even more effective to combine galaxy clustering and weak lensing in a multi-probe approach to jointly infer cosmology \cite[e.g.][]{desy3-3x2,Kids:1000}. 

In particular, a common approach is to combine three two-point functions of the galaxy density field $\delta_{g}$ and the weak lensing shear field $\gamma$: galaxy clustering $\langle \delta_{g} \delta_{g} \rangle$, galaxy-galaxy lensing $\langle \delta_{g} \gamma \rangle$ and cosmic shear $\langle \gamma \gamma \rangle$. In these measurements, we usually refer to the galaxy sample used for $\delta_{g}$ as the {\it lens galaxies}, and the sample used for weak lensing as the \textit{source galaxies}. These two-point statistics capture the Gaussian information in the matter field, which is sensitive to cosmological parameters that describe the history and content of the Universe. The main advantage of combining the three probes in a coherent analysis is that since each probe depends on  cosmological and nuisance parameters in a different way, combining them allows us to effectively break the degeneracies between the parameters and tighten the overall cosmological constraints, as well as constraints on nuisance parameters. Following the community's convention, we refer to this combination of three two-point correlation functions as the 3$\times$2pt probe. 

Building on the success of the Stage-III\footnote{The Stage-III and Stage-IV classification was introduced in the Dark Energy
Task Force report \citep{Albrecht2006}, where Stage-III refers to the ongoing dark energy experiments that started in the 2010s and Stage-IV refers to those that start in the 2020s.} galaxy surveys the Dark Energy Survey \citep[DES,][]{Flaugher2005}, the Kilo-Degree Survey \citep[KiDS,][]{deJong2013} and the Hyper Suprime-Cam survey \citep[HSC,][]{Aihara2018}, we are at the very beginning of Stage-IV galaxy surveys with the Rubin Observatory Legacy Survey of Space and Time \citep[LSST,][]{Ivezic2019}, the ESA satellite Euclid \citep{Euclid2011} and the NASA's Nancy Grace Roman Space Telescope \citep{Akeson2019} ramping up their activities, and DESI \citep{Levi2019} already operating. This paper in particular focuses on LSST, which is scheduled to begin its 10 year survey in 2024. The 3$\times$2pt analysis is one of the baseline pillars of the LSST cosmology analysis  and is expected to deliver a Dark Energy Task Force Figure of Merit (DETF FoM) $\sim 60$ \citep{DETF2006}. Together with supernovae, galaxy clusters and strong lensing, LSST is expected to deliver a DETF FoM $\sim 500$ \citep{DESC2018}. 

To achieve these projections, one of the key factors will be our ability to control for systematic effects. The significantly lower statistical uncertainties in LSST compared to current data places stringent requirements on the level of systematic effects that we can tolerate. At this point, the community collectively has extensive experience from Stage-III surveys where systematic effects that were previously ignored are becoming relevant, and this will only be more evident in LSST. Some examples for these ``new'' systematic effects include: the coupling of shear calibration bias and redshift uncertainties through the blending of object images \citep{MacCrann2022}; biases in photometric redshift distributions from incompleteness of spectroscopic training samples \citep{Hartley2020}; selection effects in galaxy samples used for clustering measurements \citep{Pandey2021}, to name only a few.  

To this end, we cannot necessarily predict the next new systematic effect that will emerge, but we \textit{can} learn from our Stage-III experiences and make the upcoming LSST cosmology analysis process smoother by building a framework to 1) minimize the human errors in carrying out the analysis and 2) enable the users to easily find and test systematic effects that could emerge. This is the main idea behind the creation of \textsc{TXPipe}\footnote{\url{https://github.com/LSSTDESC/TXPipe}}. In \textsc{TXPipe} we focus on the \textit{measurement} side of the 3$\times$2pt cosmology analysis, but note that its general structure could be applied to other analyses as well. We also note that similar concepts exist from the theory side in a number of common cosmology tools such as CosmoSIS \citep{Zuntz2015} but to our knowledge it has not been developed at a similar level on the measurement side, which is equally important. In an earlier work in \citet{Chang2019}, we explored a prototype measurement pipeline \textsc{WLPipe}\footnote{\url{https://github.com/pegasus-isi/pegasus-wlpipe}} and applied it on four precursor datasets. The work demonstrated the value of such a framework by exposing a number of issues in earlier measurements. In a similar spirit, \citet{Phillips-Longley2022} recently reanalyzed the DES Y1, HSC-Y1 and KiDS-1000 cosmic shear analyses using \textsc{TXPipe}, finding a few additional issues. 

The goal of this paper is to first present the structure and design of \textsc{TXPipe} and then validate the basic functionalities using mock galaxy catalogs. Here we target the pipeline requirement for the 3$\times$2pt cosmology analysis using the first year (Y1) of LSST data as it is the first near-term goal we expect from LSST. To perform the validation we use two sets of complementary mock galaxy catalogs: first, a set of simple Gaussian simulations to test that the basic measurements of the estimators do indeed recover the input signal; second, the DESC \textsc{CosmoDC2} catalog presented and validated in \citet{Korytov2019} and \citet{Kovacs2021} to test the various functionalities in \textsc{TXPipe} that involve more realistic galaxy properties. We compare the measurements and the theory prediction both at the data vector level and at the level of the cosmological constraints. Whenever possible, we also validate the 3$\times$2pt-related components of other DESC software packages with this exercise -- these include \textsc{fireCrown}\footnote{\url{https://github.com/LSSTDESC/firecrown}} (likelihood), \textsc{TJPCov}\footnote{\url{https://github.com/LSSTDESC/TJPCov}} (covariance), the Core Cosmology Library \citep[CCL\footnote{\url{https://github.com/LSSTDESC/CCL}},][]{Chisari2019} and others. All together we are able to demonstrate that the performance of core measurement components is sufficient for an LSST Y1 3$\times$2pt cosmology analysis. We fully expect that the baseline analysis that will be adopted in later years of LSST will evolve depending on what we find, and so have designed \textsc{TXPipe} to be adaptable to future changes. We keep the constituent parts of the pipeline only loosely coupled to each other, so that changes can be isolated and their impacts carefully assessed. We store all the pipeline's outputs in one place, so that they can be compared one-to-one. And we maintain pipeline configurations under version control, and propagate this information to output metadata, so that the nature of changes made is never lost or unclear.

The paper is organized as follows. In Section~\ref{sec:background} we introduce the basic background formalism used in a 3$\times$2pt cosmology analysis. This includes both the theory prediction and the estimators used in this work. In Section~\ref{sec:txpipe} we present the design and functionalities of \textsc{TXPipe}.  In Section~\ref{sec:validation} we validate \textsc{TXPipe} at both the data vector and the cosmology level using the LSST Y1 like Gaussian simulations and in Section~\ref{sec:cosmodc2} we apply it to \textsc{CosmoDC2}. We discuss lessons learned and future steps in Section~\ref{sec:discsussion} and summarize in Section~\ref{sec:summary}.

\section{Modeling: 3$\times$2pt cosmology}
\label{sec:background}

The $3\times 2$pt probes are the autocorrelation of the positions of galaxies, the cross-correlation of galaxy shapes and galaxy positions and the autocorrelation of galaxy shapes respectively. These measurements can be done both in real space and in harmonic space. We use two samples to perform these measurements, a \textit{lens} sample for which we have position information and a \textit{source} for which we have both position and shape information. We also use (true) redshift information for both samples.
In this section we will describe how we define and model these two-point correlation functions. 

Unless explicitly stated, during this study we use a cosmology close to the best-fit cosmological parameters from WMAP-7~\citep{Komatsu2011}: $\Omega_{c}=0.22, \Omega_{b}=0.048, h=0.71, n_{s}=0.963, \sigma_{8}=0.8$, which is the assumed cosmology in the \textsc{CosmoDC2} simulation. 

\subsection{Background theory} \label{sec:background_theory}

Considering a real signal $a(\mathbf{n})$ on the unit sphere $S^2$, it is equivalently described in terms of its spherical harmonic coefficients $\{a_{\ell m}\}$ with $\ell\in \mathbb{N} $ and $m \in \left\{-\ell , ..., \ell\right\}$. The angular power spectrum $\{C^{ab}_{\ell}\}$ between two signals $a$ and $b$ is defined as:
\begin{equation}
  \langle a_{\ell m} b^*_{\ell' m'}\rangle \equiv C^{ab}_\ell\,\delta_{\ell\ell'}\,\delta_{mm'},
\end{equation}
and can be obtained using the following estimator assuming coverage over the full-sky:
\begin{equation}
    C_{\ell} = \sum_{m=-\ell}^\ell \frac{ | a_{\ell m}| ^2}{2 \ell + 1}.
\end{equation}
The angular power spectrum corresponds to the average power in fluctuations on scales of the order of $\pi/\ell$ on the sphere. If we assume that $a$ is a random field, the power spectrum can be interpreted as a compression technique, and used to perform statistical inference on physical models of the field. In particular, if $a$ is an isotropic Gaussian random field, the power spectrum is a sufficient statistic containing all the relevant information in the realization $a$. In our case, $a$ will be the galaxy number overdensity for galaxy clustering, and the convergence (or shear) field for weak lensing observables. 

\subsection{Harmonic and real space model}
\label{sec:fourier}

Under the Limber approximation \citep{Limber53, Limber_LoVerde2008} and assuming a flat Universe cosmology, we obtain the weak lensing \textit{shear power spectrum} as a projection of the 3D matter power spectrum $P_{\rm mm}$:
\begin{equation}\label{eq:Cgammagamma}
C^{ij}_{\gamma \gamma} (\ell)= \int d\chi \frac{ q_s^i(\chi)\,  q_s^j(\chi)}{\chi^2} P_{\rm mm} \left(k = \frac{\ell+1/2}{\chi},z(\chi)\right),
\end{equation}
between two source redshift bins $i$, $j$ where $q_s(\chi)$ are the window functions of the given source populations of galaxies, $k$ is the 3D wavenumber, $\ell$ is the 2D multipole moment and $\chi$ is the comoving distance. We can also model the cross-correlation between the lens and source samples $C_{\delta_{g}\gamma}$ and express it as a projection of the 3D galaxy-matter power spectrum $P_{\rm gm}$. For a lens redshift bin $i$ with a window function $N_l(\chi)$ and a source redshift bin $j$, 
\begin{equation}\label{eq:C_gkappa}
    C_{\delta_{g}\gamma}^{ij}(\ell) = \int d\chi \frac{N_l^i(\chi)\, q_s^j(\chi)}{\chi^2}P_{\rm gm}\left(k = \frac{\ell+1/2}{\chi},z(\chi)\right)\,.
\end{equation}
Finally the galaxy clustering harmonic space correlation function is a projection of the 3D galaxy power spectrum $P_{\rm gg}$: 
\begin{equation}\label{eq:C_gg}
    C_{\delta_{g}\delta_{g}}^{ij}(\ell) = \int d\chi \frac{N_l^i(\chi)\, N_l^j(\chi)}{\chi^2}P_{\rm gg}\left(k = \frac{\ell+1/2}{\chi},z(\chi)\right)\,.
\end{equation}
Limber's approximation holds if the 3D galaxy overdensity field of the lenses and the 3D matter overdensity field at the redshift of the source galaxies vary on length scales much smaller than the typical length scale of their respective window functions in the line of sight direction. The lens window function is defined as:
\begin{equation}
N_l^i(\chi) = \frac{n^i_l\,(z)}{\bar{n}^i_l}\frac{dz}{d\chi}, 
\end{equation}
where $n^i_l$ is the lens redshift distribution and $\bar{n}^i_l$ is the mean number density of the lens galaxies. The weak lensing window function of the source galaxies is:
\begin{equation} \label{eq:lensing_window}
q_s^j(\chi) = \frac{3H^2_0 \Omega_m }{2c^2} p(\ell) \frac{\chi}{a(\chi)} g(\chi),
\end{equation}
where $a$ is the scale factor, $p(\ell)$ is the $\ell$-dependent prefactor in the lensing observables due to the spin-2 nature\footnote{Here we use $p(\ell) = \ell^2/(\ell + 1/2)^2$, which corresponds to the 1st order extended Limber flat projection (\texttt{ExtL1Fl}), as defined in table 1 from \citet{Kilbinger2017}.} of the shear and $g(\chi)$ is the lensing efficiency kernel: 
\begin{equation}
 g(\chi) = \int_\chi^{\chi_\text{h}} d \chi'  \frac{n^j_s\,(z) }{\bar{n}^j_s} \frac{dz}{d\chi'}\frac{\chi'- \chi}{\chi'},
\end{equation}
with $n_s^j(z)$ being the redshift distribution of the source galaxies, $\bar{n}^j_s$ the mean number density of the source galaxies and $\chi_\text{h}$ is the comoving distance to the horizon.

We use \textsc{CCL} \citep{Chisari2019} to obtain the model for the two-point correlation functions. We compute the non-linear matter power spectrum using the \citet{Takahashi_2012} version of \textsc{Halofit}. For the  linear power spectrum,   we use  the \textsc{CAMB} algorithm \citep{Lewis:2002ah}.

\subsubsection{Galaxy bias model}
 In our fiducial model we assume that lens galaxies trace the mass distribution following a simple linear biasing model ($\delta_g = b \:\delta_m$, where $b$ is modeled as a constant for each redshift bin), so the galaxy power spectrum and the galaxy-matter power spectrum relate to the matter power spectrum by different factors of the galaxy bias:
\begin{align}
\label{eq:linearbias}
 P_{\rm gg} &= b^2  P_{\rm mm}, \\
 P_{\rm gm} &= b \,  P_{\rm mm}.
\end{align} 

\subsubsection{Intrinsic alignment model}

Intrinsic alignments (IAs) are due to correlations between the intrinsic galaxy shapes. IAs contribute to the total observed cosmic shear angular power spectra $C^{ij}_{GG} (\ell)$ with unknown additive terms of the form:
\begin{equation}
  C^{ij}_{GG} (\ell)= C^{ij}_{\gamma \gamma} (\ell) + C^{ij}_{II} (\ell) + C^{ij}_{\gamma I} (\ell) + C^{ji}_{\gamma I } (\ell) \ , 
\end{equation}
and to the total galaxy-shear angular spectra $C_{\delta_{g}G}^{ij}(\ell)$ with:
\begin{equation}
   C_{\delta_{g}G}^{ij}(\ell)  =  C_{\delta_{g} \gamma}^{ij}(\ell)  +  C_{\delta_{g}I}^{ij}(\ell), 
\end{equation}
where $G$ represents the total observed shape of the galaxies that include the cosmological shear $\gamma$ and the intrinsic shape $I$. In Eqs.~(\ref{eq:Cgammagamma}) and (\ref{eq:C_gkappa}) we have defined the projections with respect to the true shear involving the matter power spectrum, here we define the rest of projections:
\begin{equation}
    C^{ij}_{II} (\ell) = \int d\chi \frac{ N_s^i(\chi)\,  N_s^j(\chi)}{\chi^2} P_{\rm II} \left(k = \frac{\ell+1/2}{\chi},z(\chi)\right)
\end{equation}
\begin{equation}
    C^{ij}_{\gamma I} (\ell) = \int d\chi \frac{ q_s^i(\chi)\,  N_s^j(\chi)}{\chi^2} P_{\rm \gamma
    I} \left(k = \frac{\ell+1/2}{\chi},z(\chi)\right)
\end{equation}
\begin{equation}
    C^{ij}_{\delta_{g} I} (\ell) = \int d\chi \frac{ N_l^i(\chi)\,  N_s^j(\chi)}{\chi^2} P_{\rm \gamma  I} \left(k = \frac{\ell+1/2}{\chi},z(\chi)\right)
\end{equation}
where we have introduced the source window function $N_s^i(\chi)$ which is analogous to the lens one:
\begin{equation}
N_s^i(\chi) = \frac{n^i_s\,(z)}{\bar{n}^i_s}\frac{dz}{d\chi}. 
\end{equation}
The $P_{\rm I I}$ and the $P_{\rm \gamma I}$ power spectra are generic. Usually they are assumed to be 
linearly related to the local tidal field and to be of the same shape as the matter power spectrum except from a redshift dependent scaling:
\begin{equation}
    P_{\rm I I} (k,z) = A(z)^2 P_{mm} (k,z) \, , 
\end{equation}
\begin{equation}
    P_{\rm \gamma I} (k,z) = A(z) P_{mm} (k,z).
\end{equation}
In this work we choose to use the two-parameter nonlinear alignment (NLA) model \citep{Bridle2007}, which defines the amplitude parameter as:
\begin{equation}
    A(z) = -A_{IA} \bar{C}_1 \frac{3H_0^2\Omega_m}{8\pi G} D^{-1} \left( \frac{1+z}{1+z_0}\right)^{\eta_{IA}}. 
\end{equation}
$A_{IA}$ is one of the two free parameters of the NLA model and it is a dimensionless amplitude that  governs the strength of the IA contamination. Here $G$ is the gravitational constant and $D(z)$ is
the linear growth factor. The normalization constant $\bar{C}_1$ is typically
fixed at a value obtained from the SuperCOSMOS Sky Survey
\citep{Brown2002} of  $\bar{C}_1 = 5 \times 10^{-14} M_\odot^{-1} h^{-2} Mpc^{3}$. $\eta_{IA}$ is the other free parameter which controls the redshift scaling. $z_0$ is the pivot redshift which we fix to 0.62, which is a common choice.

\subsubsection{Real space projection}

Finally, each real-space two-point correlation function is related to the total angular power spectrum for cosmic shear, galaxy-galaxy lensing and galaxy clustering via 
    \begin{equation}
    w^{i,j}(\theta) = \int \frac{d \ell \ell}{2\pi} C_{\delta_{g}\delta_{g}}^{i,j}J_{0}(\ell \theta),
    \label{eq:wtheta}
    \end{equation}
    \begin{equation}
    \gamma_{t}^{i,j}(\theta) = \int \frac{d\ell \ell}{2\pi} C_{\delta_{g}G}^{i,j} J_{2}(\ell \theta),
    \end{equation}
and
    \begin{equation}
    \xi^{i,j}_{+/-}(\theta) = \int \frac{d\ell \ell}{2\pi} C_{GG}^{i,j} J_{0/4}(\ell \theta)
    \label{eq:xi}
    \end{equation}
under the flat-sky approximation, where the $J_\alpha$ represent the Bessel functions of the first kind. We have tested that the flat-sky approximation is good enough for $w(\theta)$ given the LSST Y1 footprint and $\xi_{+/-}$ and $\gamma_t$ are expected to be much less impacted by this effect \citep{Kilbinger2017}.

\subsubsection{Redshift and shear marginalization}

When obtaining cosmological constraints we marginalize over some observational systematics such as  redshift and shear calibration effects to obtain more realistic posterior uncertainties. In particular, we marginalize over a shift in the mean redshift $\Delta z^i$  both for the lens and source input redshift distributions $n_{\mathrm{input}}^i$:
\begin{equation}
   n^i (z) = n_{\mathrm{input}}^i (z - \Delta z^i).
\end{equation}
For the shear, we marginalize over a multiplicative shear bias $m$ per each source bin, which modifies the shear and galaxy-shear angular power spectra in the following way:
\begin{equation}
     C_{\delta_{g}G}^{ij}(\ell) = (1+m^j)  C^{ij}_{\delta_{g}G, {\mathrm{input}}} (\ell)
\end{equation}
\begin{equation}
    C^{ij}_{GG} (\ell)= (1+m^i)(1+m^j) \ C^{ij}_{GG, {\mathrm{input}}} (\ell)
\end{equation}
Note that in this first validation of the DESC software we do not consider some effects found significant in recent data analyses such as magnification \citep*{Elvin-Poole2022} or redshift space distortions \citep{Krause2022} (although these are now implemented in CCL).

\section{\textsc{TXPipe}}
\label{sec:txpipe}

\begin{figure*}
\begin{center}
\includegraphics[width=1.\textwidth]{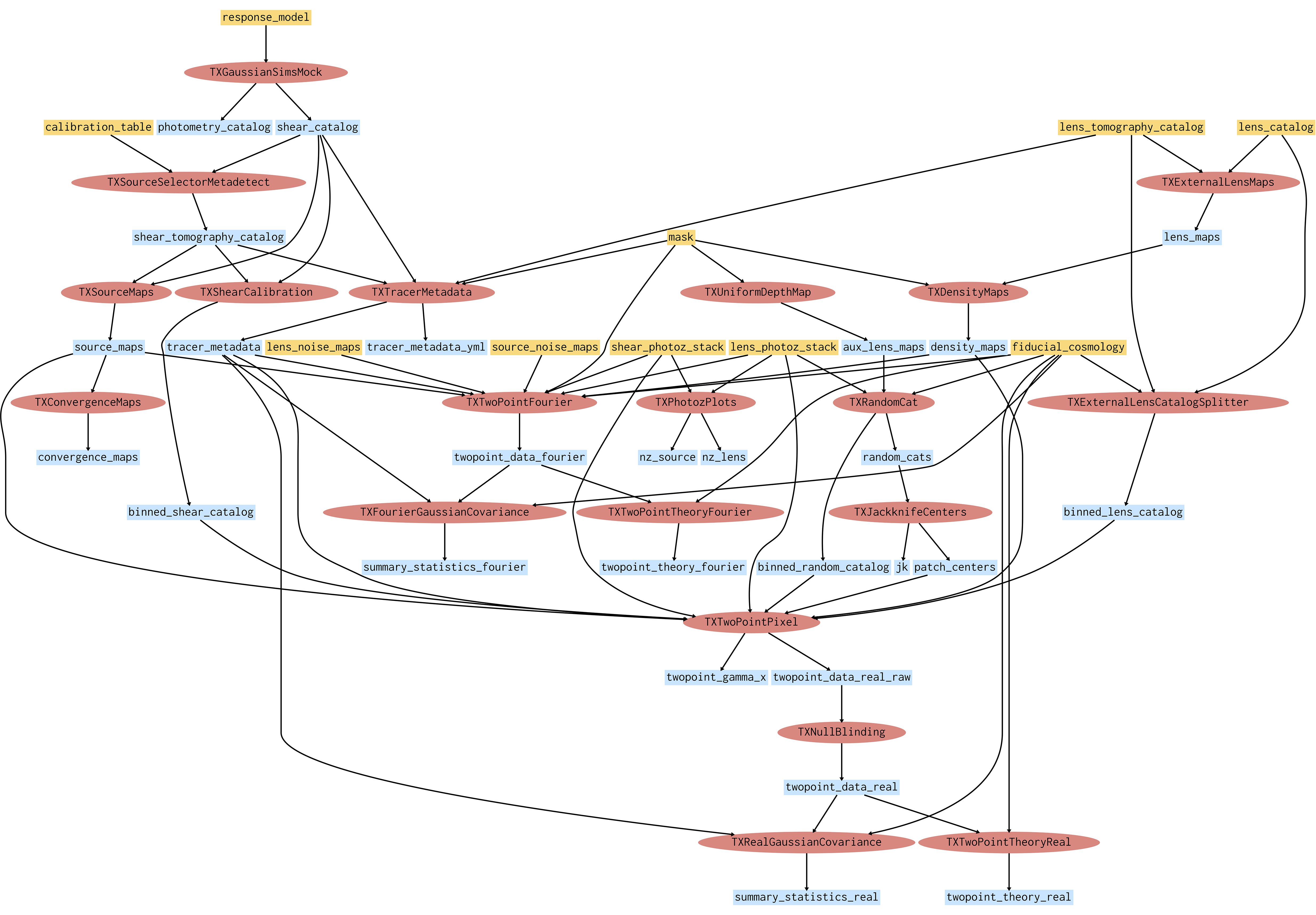}
\caption{TXPipe flowchart for the Gaussian simulation pipeline. The inputs are represented by yellow boxes and are typically \textsc{HDF5} or \textsc{YAML} files, the \textsc{TXPipe} stages by red ellipses and each of the outputs by blue boxes, which are typically \textsc{HDF5}, \textsc{SACC}, \textsc{PNG} or \textsc{PDF} files.  The input and output formats can be defined by the user.}
\label{fig:flowchart}
\end{center}
\end{figure*}

\textsc{TXPipe} is DESC's implementation of the pipeline that ingests catalog data and performs all the calculations required to obtain quality-assured two-point measurements of lensing and clustering, 
and their covariances, with the metadata necessary for parameter estimation.  The code is designed to collect and formalize the many calculations and analysis stages that in previous surveys have often been manually connected.

The goal of the project is that the complete pipeline, from the output of the LSST Science Pipelines\footnote{\url{https://pipelines.lsst.io/}} to the inputs to cosmological parameter estimation, can be run and re-run in a single operation. Such unification is convenient, but it also permits provenance tracking of the pipeline to be essentially complete, so that the collaboration can be sure of precisely what code was run to generate data products. 

We use continuous integration features to test the pipeline as changes are made, using \textit{Github Actions}\footnote{\url{https://docs.github.com/en/actions}}. This system automatically runs a unit test suite whenever changes are proposed or made, and also a set of pipelines on 1 deg${}^2$ of data. This is not large enough to ensure that the pipeline is numerically correct, but does ensure that core functions work.

In Figure~\ref{fig:flowchart} we show a flowchart that illustrates the \textsc{TXPipe} pipeline in its most basic functionality. This flowchart is automatically generated by \textsc{TXpipe}. The Figure shows the stages that have been used for the Gaussian simulation analysis which we present in Sec.~\ref{sec:validation}. Besides the most basic stages presented and tested in this initial paper, in Section~\ref{sec:data-runs} we summarize  additional features which are implemented within \textsc{TXPipe} that will be needed for the data analysis. 

Much of \textsc{TXPipe} functionality involves connecting together external codes that perform measurements. The package wraps each of these tools as a Python class, specifying input and output files required for each, and launching them using the \textsc{Ceci}\footnote{\url{https://github.com/LSSTDESC/Ceci}} library and executable which automatically interfaces them to workflow management frameworks. \textsc{TXPipe} also performs various calculations internally, again using a Python class to embody each pipeline stage. The outputs of the pipeline can be automatically organized as a web page. It is also important to highlight that in all these cases \textsc{TXPipe} is optimized for the large data volumes of LSST. Wherever possible we use parallel and online algorithms, which do not require all data to be loaded in memory at once and are therefore scalable. All correlation function outputs are saved using the  \textsc{SACC}\footnote{\url{https://github.com/LSSTDESC/sacc}} format, a dedicated DESC software for storing measurements, covariances and redshift distributions in a unified way. Below we describe the \textsc{TXPipe} pipeline stages grouped in different blocks:
\begin{itemize}
    \item \textbf{Data ingestion, shear and redshift calibration:} The \textsc{TXPipe} pipeline starts with the ingestion of the galaxy catalogs which are prepared into suitable formats (e.g. HDF5 files for catalogs). \textsc{TXPipe} then splits the catalogs into different tomographic bins, typically using criteria described below in Section \ref{sec:cosmodc2}. Next, it builds the redshift distributions\footnote{In this work we use true redshift distributions throughout, making this step a simple histogram. However, implementations are undergoing to make this step realistic. See the additional functionalities section in \ref{sec:data-runs} for more details.}, performs the requested shear calibration using the methods of \textsc{Metadetection} and \textsc{Metacalibration} described in \citet{Sheldon2020}, \citet{Sheldon2017} and \textsc{lensfit} described in \citet{Miller2007}, depending on the catalog type.
    \item \textbf{Two-point measurements:} We use the harmonic-space and real-space estimators described in Appendix~\ref{sec:estimators}. In essence, for the harmonic-space estimator, we first create shear and density maps in \texttt{HEALPix}\footnote{\url{https://healpix.sourceforge.io/}} formats. Then we use the \textsc{NaMaster}\footnote{\url{https://github.com/LSSTDESC/NaMaster}} \citep{namaster} pseudo-$C_{\ell}$ code to measure the spectra. For the real-space estimators, we invoke the fast tree code \textsc{TreeCorr}\footnote{\url{https://github.com/rmjarvis/TreeCorr}}  \citep{Treecorr} to perform the measurements. A mask is needed for the harmonic space estimator, which we generate directly using a depth map obtained from the lens galaxies (using a resolution much coarser than the average separation between galaxies). This is a simplistic approach that will need to be updated once we test on more complex simulations or data. We also generate a random catalog from the depth maps to be used in the real-space estimator (see Equations~\ref{eq:gt} and~\ref{eq:wtheta_estimator}). Typically, the number of random points is set to be at least 20 times larger than that of the lens galaxies \citep{Prat2022}. This puts a significant memory load on the real-space measurements given the number of lenses expected for LSST Y1. As a result, we also implement a pixel-based estimator described in Sec.~\ref{sec:pixel_estimators} -- our validation tests for real-space galaxy clustering and galaxy-galaxy lensing measurements were performed using this pixel-based estimator. 
    \item \textbf{Two-point predictions:} \textsc{TXPipe} interfaces with \textsc{CCL} to obtain a theory prediction for the corresponding measured two-point correlation functions, automatically using the same angular binning as the measurements, i.e. generating a \textit{theory data vector} that we use to compare with the measurements. The same theory predictions are also used in the covariance matrix calculation described below.
    \item \textbf{Covariance matrix:} \textsc{TXPipe} obtains a covariance matrix for the corresponding data vector using one of several codes with a unified interface in the DESC \textsc{TJPCov} package. Currently \textsc{TJPCov} includes Gaussian covariances with different treatments of the mask. We compare the different options in Section~\ref{sec:cov} and Appendix~\ref{app:mask_effects}.
    \item \textbf{Extra diagnostics:} \textsc{TXPipe} generates useful diagnostic plots and metadata throughout the pipeline, including plots of maps, catalog histograms, masks, jackknife patches, $n(z)$ distributions as well as quantities such as galaxy number densities, shape noise values, mask area etc\footnote{The full list of currently implemented diagnostics can be found in \url{https://txpipe.readthedocs.io/en/latest/stages/Diagnostics.html}}.
\end{itemize}

\begin{figure}
\begin{center}
\includegraphics[width=0.4\textwidth]{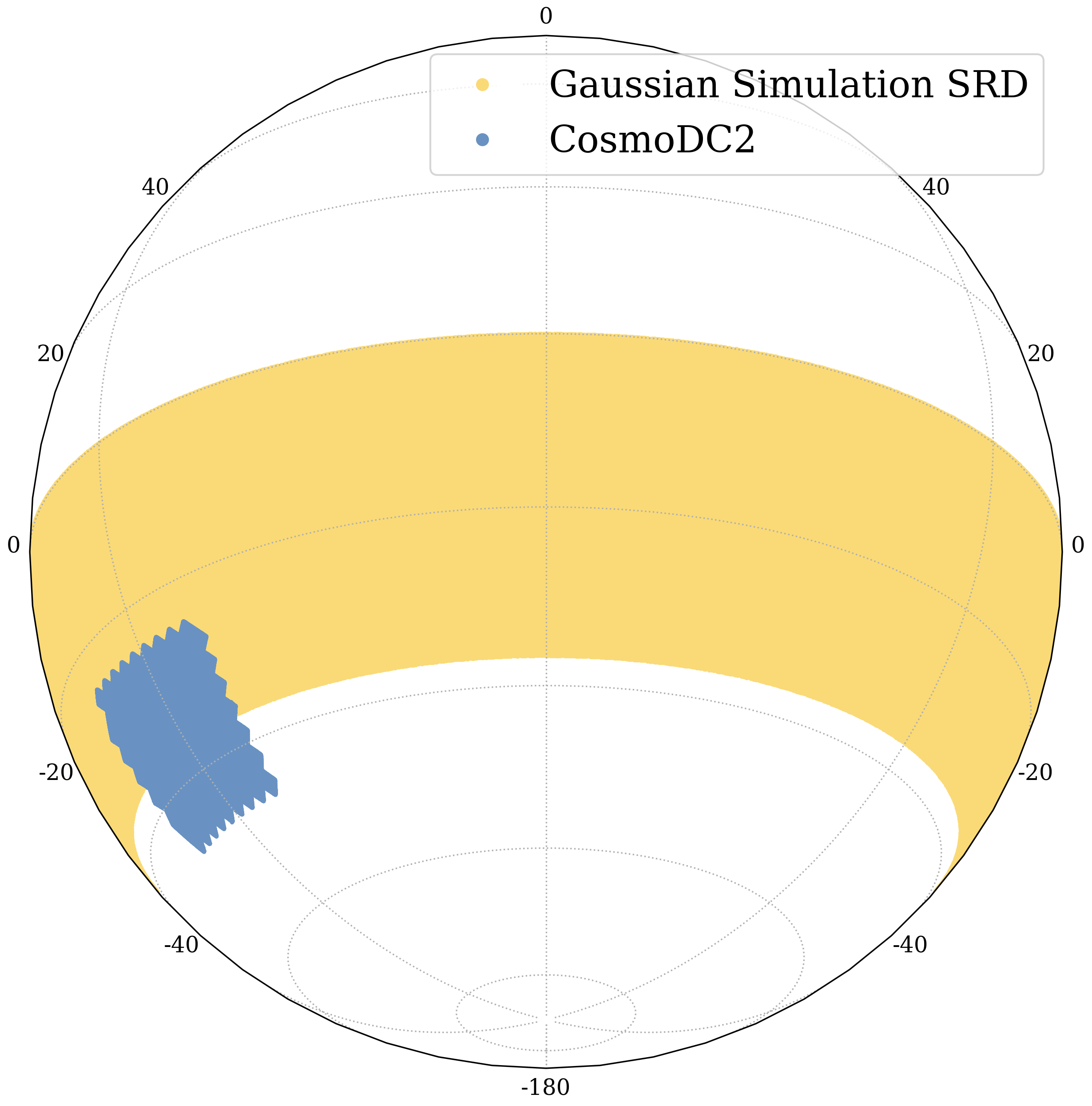}
\caption{Sky coverage or masks of the two simulations we use in this work: a Gaussian simulation mimicking a general LSST-Y1 like mask with 12,300 deg$^2$ and \textsc{CosmoDC2} with 440 deg$^2$.}
\label{fig:masks}
\end{center}
\end{figure}

\begin{figure}
\begin{center}
\includegraphics[width=0.48\textwidth]{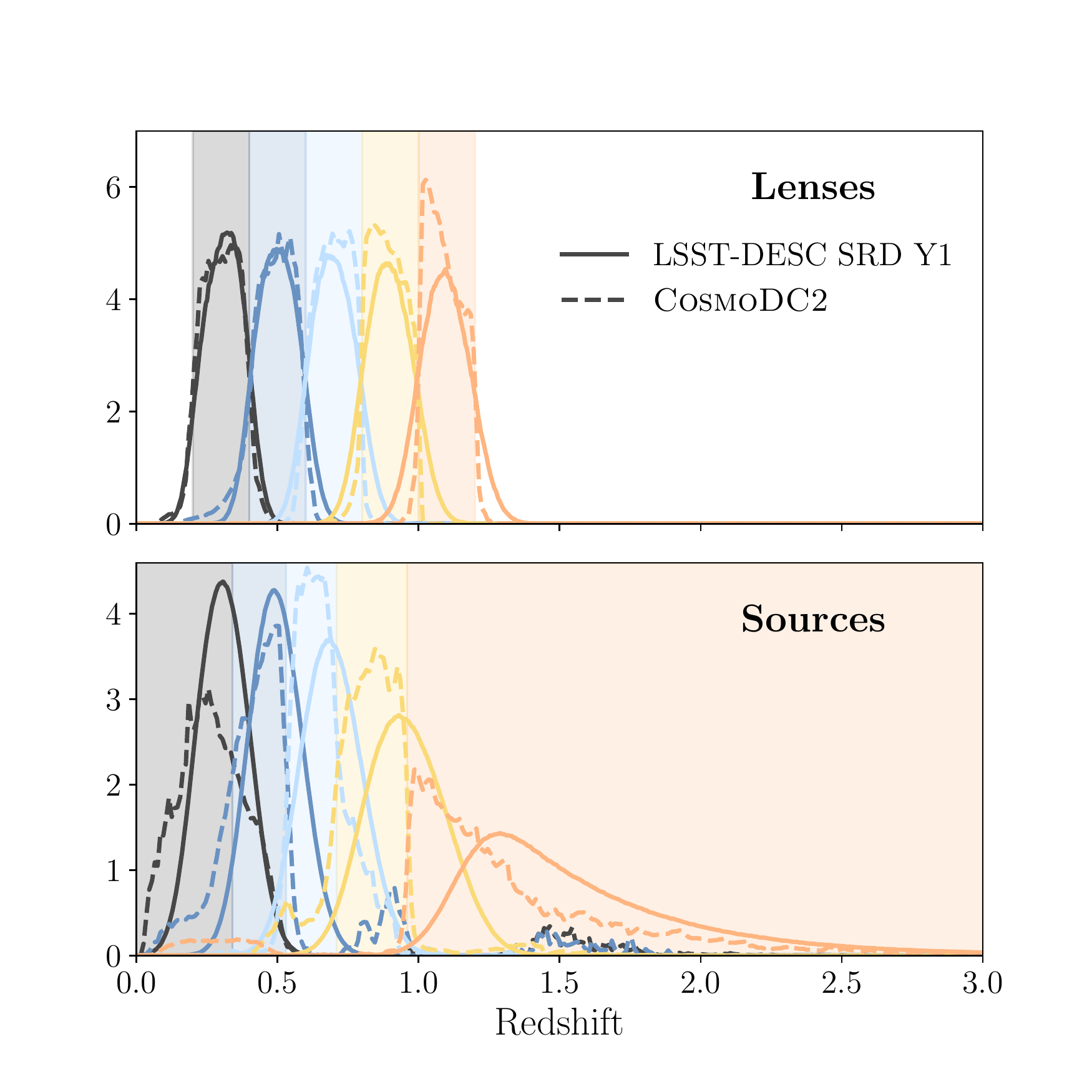}
\caption{Redshift distributions of the tomographic bins for the lens (top) and source (bottom) samples as specified in the LSST-DESC SRD for the first year of LSST operation and for the samples we have defined on the \textsc{CosmoDC2} simulation. The shaded regions represent the bin edges for the \textsc{CosmoDC2} samples, which are designed to 1) match the SRD binning for the lenses and 2) produce a similar number density in each source bin.}
\label{fig:nzs_srd}
\end{center}
\end{figure}

\subsection{Additional functionalities} \label{sec:data-runs}

Most of the validation tests performed in this paper assume an idealized LSST Y1 dataset. We also test \textsc{TXPipe} on the more realistic mock catalog \textsc{CosmoDC2} in Section~\ref{sec:cosmodc2}, but still with idealized conditions. Thus, we only test the core functionalities of \textsc{TXPipe}. There are additional functionalities that have been developed for more realistic analyses and that will continue to be developed and tested in future work. We briefly describe them here: 
\begin{itemize}
    \item \textbf{Redshift distribution estimation:} \textsc{TXPipe} is integrated with the DESC photometric redshift code \textsc{RAIL} tool\footnote{\url{https://github.com/LSSTDESC/RAIL}}, which contains a suite of algorithms to estimate both point-estimate redshifts per galaxy\footnote{Including BPZ, Delight, FlexZBoost, and PZFlow \citep{Benitez2000,Coe2006,LeistedtHogg17,IzbickiLee17,Dalmasso20,Crenshaw20}} and redshift distribution estimates for ensembles of galaxies\footnote{Including self-organizing maps, direct calibration, and variational-inference based models \citep[e.g.][]{CarrascoKind14,Wright19,Rau2021}}. The former are sometimes used to define tomographic bins \citep[although, see][for alternative binning approaches]{Zuntz2021O}, while the latter are used for making theory predictions for cosmological inference.
    \item \textbf{Null tests:} In most Stage-III surveys, the measurement of the data vectors are accompanied by a large suite of null tests to ensure that the data vector is not significantly contaminated by systematic effects \citep[see, for example][]{Gatti2021b,Giblin2021,Li2022,RodriguezMonroy2022}. These tests include correlations of shear and density with survey property maps and catalogs, testing for B-mode leakage, calculating mean shear in bins of size and signal-to-noise, etc. In the idealized simulations used in this paper, we do not put in any systematic effects, thus these functionalities will not be rigorously tested here and we leave this for future work. 
   \item \textbf{Blinding:} In the era of precision cosmology, it is important to build in some mechanism so that we do not base our analysis choices on the outcome of our measurements. These so-called ``observer biases'' can be minimized by ``blinding'' the analysis and only ``unblinding'' the results once all analysis choices are frozen. In \textsc{TXPipe}, we currently implement the methodology described in \citet{Muir2020}, where the data vectors are shifted to a slightly different cosmology in a way that preserves the relation between all parts of the data vector. 
   \item \textbf{Convergence maps:} \textsc{TXPipe} interfaces with the \textsc{WLMassMap} tools\footnote{\url{https://github.com/LSSTDESC/WLMassMap}}, where the weak lensing shear catalogs are converted to convergence maps using a range of methods \citep[e.g.][]{Jeffrey2021}. These maps are typically used for higher-order statistics such as peak counts \citep{Liu2015,Zucher2022} and moments \citep{Gatti2020, Gatti2021}. This functionality highlights the flexibility of \textsc{TXPipe}, where data vectors beyond 3$\times$2pt can be incorporated into the framework and share the same infrastructure.  \\
\end{itemize}
\noindent In general, it is straightforward to build on the existing \textsc{TXPipe} structure and incorporate new functionalities thanks to its modular and transparent design. We expect \textsc{TXPipe} to grow and become more complete/mature as the LSST data arrives in the coming years.

\subsection{Performance requirements}

\begin{table*}
    \centering
    \begin{tabular}{cccccc}
    \hline
    \textsc{TXPipe} Stage & Time & $N_\mathrm{n}$ & $N_\mathrm{p}$  & $N_\mathrm{t}$ & Description\\
    \hline
    \texttt{TXConvergenceMaps} & 38 min & 1 & 1 & 32 & Make a convergence map from a source map the using Kaiser-Squires method. \\
    \texttt{TXDensityMaps} & 1 min & 1 & 1 & 1 & Convert galaxy count maps to overdensity maps.\\
    \texttt{TXFourierGaussianCovariance} & 2 min & 1 & 1 & 64 & Compute a Gaussian Fourier-space covariance with \texttt{TJPCov} using $f_{\text{sky}}$ only.\\
    \texttt{TXGaussianSimsMock} & 49 min & 1 & 1 & 20 & Simulate mock photometry from Gaussian simulations.\\
    \texttt{TXJackknifeCenters} & 8 min & 1 & 1 & 1 & Generate jackknife centers from random catalogs.\\
    \texttt{TXExternalLensMaps} & 132 min & 2 & 10 & 1 & Make tomographic lens number count maps from external lens catalog. \\
    \texttt{TXRealGaussianCovariance} & 19 min & 1 & 1 & 64 & Compute a Gaussian real-space covariance with  \texttt{TJPCov} using $f_{\text{sky}}$ only. \\ 
    \texttt{TXShearCalibration} & 7 min & 1 & 7 & 1 & Split the shear catalog into calibrated bins.\\ 
    \texttt{TXSourceMaps} & 51 min & 4 & 4 & 10 & Make tomographic shear maps from shear catalogs and tomography.\\ 
    \texttt{TXSourceSelectorMetadetect} & 11 min & 2 & 64 & 1 & Source selection and tomography for \texttt{metadetect} catalogs.\\ 
    \texttt{TXTwoPointFourier} & 32 min & 4 & 4 & 20 & Make Fourier space 3$\times$2pt measurements using \textsc{NaMaster}. \\ 
    \texttt{TXTwoPointPixel} & 780 min & 6 & 6 & 32 & Compute pixelated versions of the 3$\times$2pt real space correlation functions.\\ 
    \texttt{TXTwoPointTheoryFourier} & 1 min & 1 & 1 & 1 & Compute theory predictions for Fourier-space 3$\times$2pt measurements.\\ 
    \texttt{TXTwoPointTheoryReal} & 0.2 min & 1 & 1 & 1  &  Compute theory predictions for real-space 3$\times$2pt measurements.\\ 
    \texttt{TXUniformDepthMap} & 0.1 min & 1 & 1 & 1 & Generate a uniform depth map from the mask.\\ 
    \hline
    \end{tabular}
    \caption{Brief description of each stage for Gaussian simulations pipeline and time it takes to run. $N_\mathrm{n}$ is the number of nodes, $N_\mathrm{p}$ is the total number of processes and $N_\mathrm{t}$ the number of threads per process. See \url{https://txpipe.readthedocs.io/en/latest/stages.html} for the latest documentation of each stage.}
    \label{tab:times}
\end{table*}

Because of its depth and area, LSST data volumes will be very large: the final 10-year data release will contain $O(10^{10})$ galaxies. This size, and the corresponding increases in required systematic accuracy, mean that many methods and algorithms must be re-designed for LSST.  DESC's primary computing facility is the National Energy Research Scientific Computing Center (NERSC), and we have targeted the parallelization features to deal with this size at the systems they host. 

Our primary limitation is memory; many algorithms begin by loading entire catalog columns before performing operations on them, or on subsets on them. At full LSST scale a single column can be more than 50 GB in size. Although loading these is possible on high-memory systems, depending on the number of columns needed, doing so limits the number of processes that can simultaneously operate on data.  We minimize such patterns in \textsc{TXPipe} wherever possible, preferring to model our data tables as streams and processing them in parallel. Since many operations involve core descriptive statistics on data columns (means, standard deviations, etc), we use a DESC library\footnote{\url{https://github.com/LSSTDESC/parallel_statistics/}} of tools to implement them where possible.

Given the large data set size, processing speed is also greatly important. In many cases I/O is the bottleneck in \textsc{TXPipe} stages. The complete analysis described here iterates through the complete input catalogs multiple times; some algorithmic steps in the pipeline inherently require multi-pass runs, although sometimes it is a choice. We make heavy use of NERSC's LUSTRE file system's parallel I/O facilities, which \emph{stripe} chunks of data files across multiple storage targets, and enable fast access to different parts of the data from different processes in parallel. Storing data in the HDF5 format makes it straightforward to both read and write different data chunks in parallel through MPI, provided that data chunks are large and contiguous.

We use a hybrid of Message Passing Interface (MPI) process and OpenMP (thread) parallelism paradigms. For example, when assigning objects to tomographic bins in \texttt{TXSourceSelector}, a low-CPU and thus I/O-dominated operation, we use pure MPI, creating as many processes as there are CPUs on a node, and splitting data among them. In cases where I/O is not the bottleneck, or where we can pre-reduce data down to multiple smaller subsets, such as when computing correlation functions in \texttt{TXTwoPoint}, we typically make use of thread parallelism with \textsc{OpenMP}, creating only a single process per node with many threads.

 A step further in this case is the use of GPU acceleration with \textsc{TXPipe} on for example the NERSC environment. GPU acceleration can speed up numerical calculations that involve a large throughput. We tested here an implementation of the  TXNoiseMaps using the Google Jax library stage of \textsc{TXPipe} so that it can benefit from the speed up allowed by GPU acceleration offered by the NERSC Perlmutter system.  The TXNoiseMaps stage generates random rotations and applies them to galaxy shape catalogs. By default, the stage evaluates rotations of galaxy shears, 100,000 items at a time. We find that for large enough catalogs, the speedup from using GPU acceleration is significant (of the order of a factor of 10). 

There is sometimes a trade-off to be made between wider algorithmic speed and code flexibility and modularity. We can try to minimize the total number of I/O passes of the data, at the cost of requiring much greater coordination and unified behavior between stages. We avoid this where stages are not conceptually connected in \textsc{TXPipe}, which also enables us to remain flexible regarding stage input and output data. For example, we have a pipeline stage that does diagnostic plots on the source sample (such as shear as a function of object size, and similar null tests), and another stage that assigns objects to tomographic bins. Both use the same input data - the shear catalog - so could be combined in a single step. We choose not to do so, however, so that we can modify and verify them independently.  However, we do not go to the extreme length of splitting the diagnostics stage into multiple stages, each for different null tests, since the small gain in clarity would not be useful enough to offset the cost of the time taken for added loops through the data.

We specify the computation time and resources we use for each  \textsc{TXPipe} stage when running on the  LSST-Y1 like Gaussian simulation in Table~\ref{tab:times}. Note that this catalog actually has a much larger effective number density for the lens sample due to reasons specified in Sec.~\ref{sec:sims}. Therefore, the times detailed in this table are expected to be an upper limit for the data runs.

\section{Validation of \textsc{TXPipe}}
\label{sec:validation}

In this section we validate the core functionalities of \textsc{TXPipe} to show that it meets the accuracy/precision requirement needed for the first year of LSST data as specified by the LSST DESC Science Requirement Document \citep[hereafter the DESC SRD,][]{DESC2018}. First we introduce the mock galaxy catalog used for the validation (Section~\ref{sec:sims}). Next we describe our covariance matrix estimation (Section~\ref{sec:cov}). We then validate the measurement of the data vectors (Section~\ref{sec:dv}) and the recovered cosmological constraints (Section~\ref{sec:cosmo}). All validations are done for the 3$\times$2pt probes in both harmonic and real space. 
 
\subsection{Gaussian mock galaxy catalog}
\label{sec:sims}

\begin{table}
    \centering
    \begin{tabular}{cccc}
    \hline
    Lens bin & $\langle z \rangle$ & Number density & Galaxy bias \\
    1 & 0.30 & 2.25 &  1.229\\
    2 & 0.50 & 3.11 &  1.362\\
    3 & 0.70 & 3.09 &  1.502\\
    4 & 0.90 & 2.61 &  1.648\\
    5 & 1.10 & 2.00 &  1.799\\
    \hline
    Source bin  & $\langle z \rangle$ & Number density & Shape noise $\sigma_e$ \\
    1 & 0.31 &  \multirow{5}{*}{1.78} & \multirow{5}{*}{0.26} \\
    2 & 0.49 &   &  \\
    3 & 0.69 &   &  \\
    4 & 0.96 &   &  \\
    5 & 1.59 &   &  \\
    \hline
    \end{tabular}
    \caption{LSST Y1 DESC SRD sample specifications. The number densities are in 1/arcmin$^2$ and  $\sigma_e$ is defined in Eq.~\ref{eq:sigma_e}. The same number density and shape noise is assumed in all source redshift bins.}
    \label{tab:srd_sample}
\end{table}

We generate a set of idealized Gaussian simulations following the galaxy sample specifications described  in the LSST DESC SRD to stress-test \textsc{TXPipe}. Such a setup with perfectly known inputs allows us to validate whether the pipeline is capable of recovering our input signal to the precision required for LSST-Y1 analyses. Once this baseline is validated, we can then move on to more realistic mock galaxy catalogs from e.g. $N$-body simulations (see Section~\ref{sec:cosmodc2}). We show the sky coverage for each simulation in Fig.~\ref{fig:masks} using the \textsc{cartosky}\footnote{\url{https://github.com/kadrlica/cartosky}} code.

To generate the mock catalog, we use the redshift distributions shown in Figure~\ref{fig:nzs_srd} with five lens redshift bins and five source redshift bins, which follow the definitions given in the DESC SRD. We also assume the number densities listed in that document, which are 10 gal/arcmin$^2$ for the source catalog and 18 gal/arcmin$^2$ for the lens catalog, but note that these are the total number densities if one were to integrate over the full redshift distributions ignoring the tomographic bins. We use the redshift distributions to obtain  the binned number densities which are listed in Table~\ref{tab:srd_sample}.

Assuming the cosmology from Section~\ref{sec:background}, we generate theory predictions of all the 3$\times$2pt auto and cross $C_{\ell}$ data vectors in harmonic space using \textsc{CCL} up to $\ell_\mathrm{max}=16384$. We use  the galaxy bias values listed in Table~\ref{tab:srd_sample}. From the $C_{\ell}$'s we generate correlated, noiseless Gaussian maps at a  \textsc{HEALPix} resolution $N_\mathrm{side}=8192$\footnote{Note that even though later we only use large scales in the analysis we still need a high $N_\mathrm{side}$ value because of the  following reasoning: 
Since the simulations were designed to work both for harmonic space and real-space mimicking the process we would apply to real data, we needed to use a high $\ell_\mathrm{max}$ in order for the real space input theory to also be accurate. We found that in this case there is a significant aliasing of power to lower $\ell$, which produces a significant mismatch of the resulting measurements with the input theory, unless the resolution is high enough.}
for both spin-0 fields (density, $\delta_{m}$) and spin-2 fields (shear, $\gamma$) using the approach described in \cite{Giannantonio2008}.  
To turn the noiseless maps into galaxy catalogs, we employ the following process:
\begin{itemize}
    \item We apply a simple mask to define the survey footprint, illustrated in Fig.~\ref{fig:masks}. We apply a declination cut of $-36.61^{\circ}<{\rm Dec}<0^{\circ}$, which results in an area of 12,300 deg$^{2}$, consistent with the DESC SRD specification for the LSST Y1 data for the 3$\times$2pt analysis. Note that the actual LSST footprint will cover a larger range of declinations but maintain the same area due to regions of high Galactic dust as detailed in \citet{Lochner2021, Lochner2018}.
    \item To generate source galaxies, we randomly sample points inside the survey mask with a number density according to the DESC SRD as listed in Table~\ref{tab:srd_sample}. For each galaxy, we obtain a true shear value $\gamma = \gamma_{1} + i \gamma_{2}$ corresponding to the pixel in the shear maps it falls in. Then, this catalog  is ingested into \textsc{TXPipe}, which adds shape noise to it. To do that, we draw from a Gaussian distribution with a width $\sigma=0.26$ to add a random ``intrinsic shape'' component. We draw independent values for $\gamma_{{\rm int},1}$, $\gamma_{{\rm int},2}$,  which represent the intrinsic shape of each galaxy $\gamma_{\rm int} =\gamma_{{\rm int},1} + i\gamma_{{\rm int},2}$. We then add $\gamma_{\rm int}$ to the true shear $\gamma$ to form the observed ellipticity values $e$ \citep{Seitz1997}:
    \begin{equation}
    e=\frac{\gamma_{\rm int}+\gamma}{1+\gamma^{*}\gamma_{\rm int}}, 
    \end{equation}
    from which we extract each ellipticity component as $e_1 = \Re[e]$, $e_2 = \Im [e]$.
    %\begin{equation}
    %    \gamma_{{\rm obs},i} = \gamma_{i} + \gamma_{{\rm int},i}, \;\; i=1,2.
    %\end{equation}
    \item To generate lens galaxies, we Poisson sample the density map with the galaxy number density listed in Table~\ref{tab:srd_sample}. Since a density field cannot have values smaller than -1, the Gaussian field will have tails that cannot be Poisson sampled. To circumvent this, we scale the density maps for each tomographic bin by factors  1/[2.458, 2.043, 1.878, 2.060, 2.249] such that the fraction of pixels with values $<-1$ is small ($<1\%$), and verify that the imposed cut does not induce differences in the power spectra. Specifically, these numbers come from imposing that $b \, \delta_m > -1 $ for each redshift bin. The galaxy bias $b$ is larger for the higher redshift bins, while the matter fluctuations $\delta_m$ are  larger for the lower redshift bins, thus compensating each other and yielding a similar scaling factor in all redshift bins. The field is then Poisson sampled with the desired galaxy number density  to form the lens sample. Due to this scaling, the galaxy clustering and galaxy-galaxy lensing data vectors measured from this catalog will be artificially low and will need to be rescaled by the same factor to recover the correct amplitude (in the case of galaxy clustering by the square of these factors), as has been done before in e.g. \citet{Elvin-Poole2018}. Note that the target number density (shown in Table.~\ref{tab:srd_sample}) needs to be multiplied by the scaling factor squared when sampling from the field,  otherwise the resulting shot noise levels would be much higher due to this needed scaling of the data vectors. This results in approximately [603 M, 576M, 482 M, 490M, 449M] galaxies per each lens bin and $20$ times more random points  (for comparison the source catalog has $\simeq78$M objects per source bin). 
    \item After the galaxy catalogs are generated, they are fed to \textsc{TXPipe} in a similar way as we would input real data catalogs. Then, we use \textsc{TXPipe} stages to generate the galaxy and shear maps that are used to compute the two-point correlation functions using the estimators detailed in Appendix.~\ref{sec:estimators}. We use $N_\mathrm{side}=4096$ for all the maps generated within \textsc{TXPipe}.
\end{itemize}

\subsection{Covariance matrix}
\label{sec:cov}

\begin{figure*}
\begin{center}
\includegraphics[width=1.\textwidth]{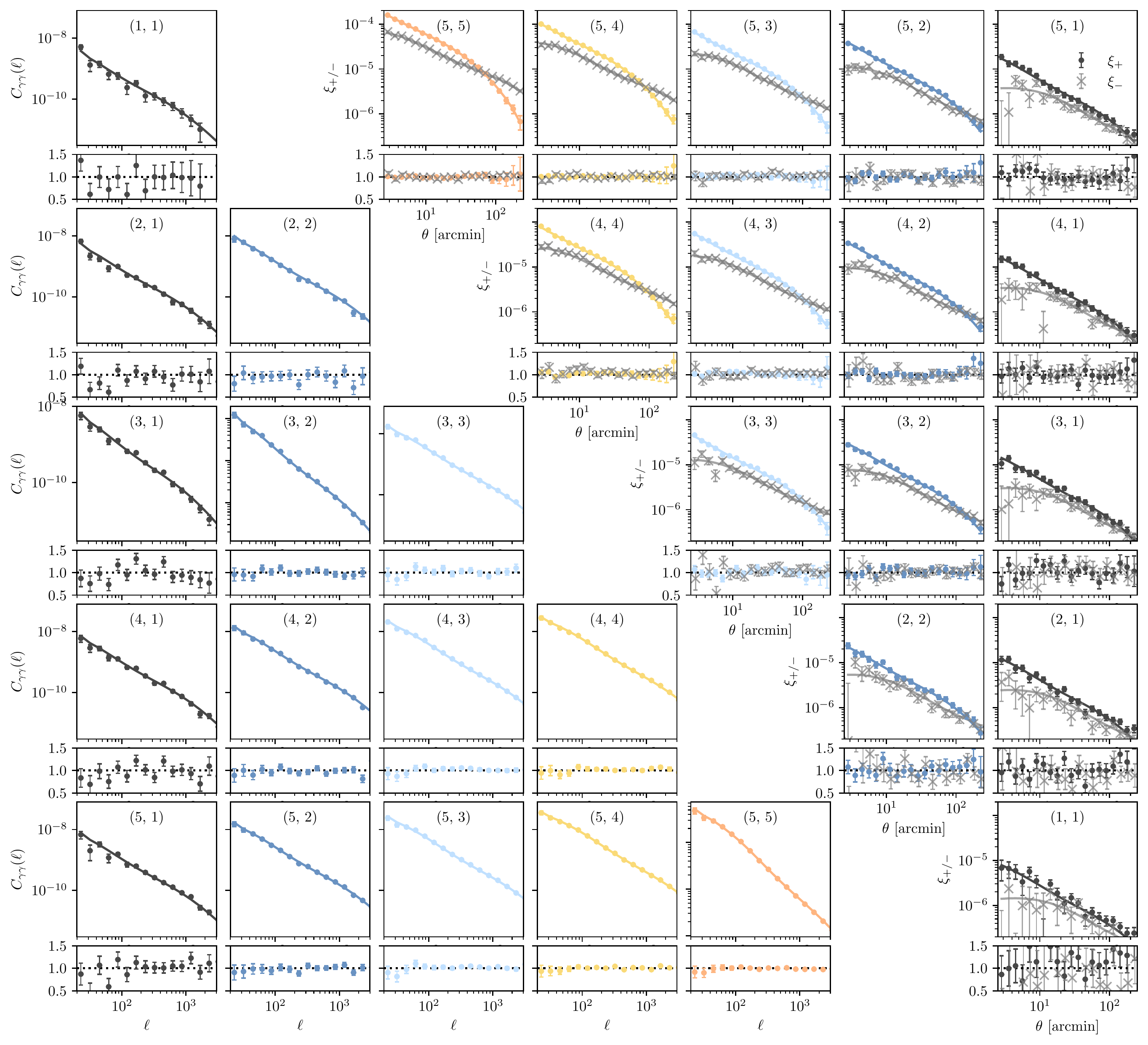}
\caption{Cosmic shear measurements (correlation between galaxy weak lensing and galaxy weak lensing) from \textsc{TXPipe} in harmonic space $(C_{\gamma\gamma}$, lower left) and real space $(\xi_+, \xi_-$, upper right). Each panel corresponds to a (source, source) redshift bin combination. For each combination, we show the measurements with point markers from our Gaussian simulations and the input theory with lines (upper panels) and the ratio of the measurement to the theory (lower panels). The error bars are the square root of the diagonal covariance.} 
\label{fig:cosmicshear}
\end{center}
\end{figure*}

\begin{figure*}
\begin{center}
\includegraphics[width=0.8\textwidth]{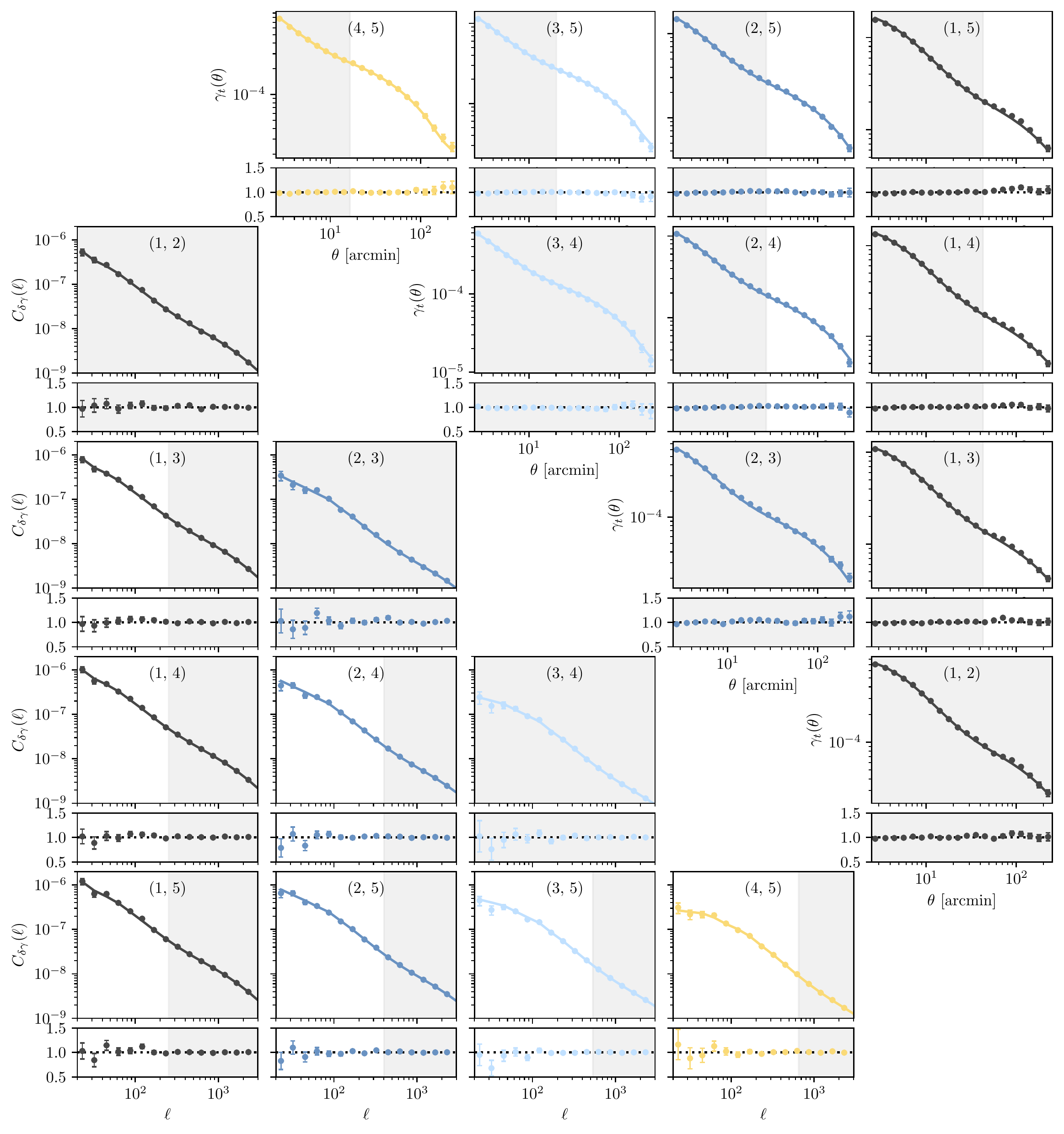}
\caption{Galaxy-galaxy lensing measurements (correlation between galaxy position and galaxy weak lensing) from \textsc{TXPipe} in harmonic space (lower left) and real space measurements (upper right). Each panel corresponds to a (lens, source) redshift bin combination. For each lens-source bin pair, we show the measurements from our Gaussian simulations and the input theory (upper panels) and the ratio of the measurement to the theory (lower panels). The error bars are the square root of the diagonal covariance. For simplicity, we do not show all the bin pairs that are measured, but only the highest signal-to-noise ones, which are the ones included in the cosmological analysis, following the DESC SRD. In addition, we show the SRD scale cuts represented with the gray shaded bands for scales we do not use -- some of the bins are entirely removed.}
\label{fig:ggl}
\end{center}
\end{figure*}

To evaluate the accuracy of \textsc{TXPipe} against theoretical predictions, and to do inference, we need an estimate of the covariance matrix. In this work we use a Gaussian covariance matrix. The Gaussian component coming from the 2-point functions is dominant with respect to the non-Gaussian parts related to using higher order information from the bispectrum. In harmonic space, at fixed cosmology, the Gaussian component is approximately described by \citep{Schneider2002,Crocce2011}
\begin{equation}
    \mathcal{C}[C_{X}^{ij} (\ell), C_{X^{\prime}}^{i^{\prime}j^{\prime}}(\ell^{\prime}))]  = \delta_{\ell \ell^{\prime}} \frac{ D_X^{ii^{\prime}}(\ell) D_{X}^{jj^{\prime}}(\ell)+D_{X}^{ij^{\prime}}(\ell) D_{X}^{ji^{\prime}}(\ell)}{(2 \ell + 1) f_{\rm sky}},
\label{eq:cov_cl}
\end{equation}
where $i,j$ and $i^{\prime},j^{\prime}$ denotes the redshift bin pairs associated with the two considered power spectra; $X$ is either $\gamma \gamma$, $\delta_{g} \gamma$ or $\delta_{g} \delta_{g}$. $D_{X}^{ij}(\ell)\equiv C_{X}^{ij}(\ell)+N_X^{ij}(\ell)$ is the sum of the signal $C_{X}^{ij}$ (described in Section~\ref{sec:fourier}) and noise power spectra $N_X^{ij}$, $\delta_{\ell \ell^{\prime}}$ is the Kronecker delta function and $f_{\rm sky}$ is the fractional sky coverage. The noise power spectra is $1/n_{\rm gal}$ for $C_{\delta_{g} \delta_{g}}$ and $\sigma_{e}^2/n_{\rm gal}$ for $C_{\gamma\gamma}$ and are assumed to be zero for cross-correlations between different redshift bins and for the galaxy-shear cross-correlation. $\sigma_{e}$ is the shape noise per component defined as
\begin{equation}\label{eq:sigma_e}
     \sigma_{e} \equiv \sqrt{\frac{1}{2}  \left( \frac{\mathrm{var}(e_1)}{R^2_{11}}+\frac{\mathrm{var}(e_2)}{R^2_{22}} \right)}
\end{equation}
within the \textsc{Metacalibration} framework for a diagonal response matrix. $R_{ii}$ is the response factor for the ellipticity component $i$. For the Gaussian simulations, we assume an identity response matrix and thus we recover $\sigma_e = 0.26$, the same value  we input as the standard deviation of the ellipticity per component.

In practice, the covariance is non-Gaussian and cosmology-dependent. There are plans to implement these improvements for the analysis with future LSST data. However, for the analysis presented here these approximations are sufficient -- see figure~3 from \citet{Friedrich2021} which shows that the non-Gaussian terms represent less than 10\% of the diagonal elements of the covariance for both cosmic shear and angular galaxy clustering. The validation of these terms with DESC software is left for future work.

To convert Equation~\ref{eq:cov_cl} into a real-space covariance we  apply the Hankel transform operator to the Fourier space covariance, implementing the approach in \citet{Singh2021} and  \textsc{Skylens}\footnote{\url{https://github.com/sukhdeep2/Skylens_public}}.

In Equation~\ref{eq:cov_cl} and also in the real space transformation we have made the assumption that the precise geometry of the footprint of the dataset does not have a big effect on the covariance, and we use the simple $f_{\rm sky}$ factor to account for the size of the footprint. As was shown in \citet{Troxel2018}, this can lead to biases in the covariance when the footprint is small. However, we check in Appendix~\ref{sec:mask_cov} that the effect is small for our setup for the LSST Y1-like Gaussian simulation. In particular we find the diagonal errorbars without including this effect are slightly smaller, thus making our validation tests if anything more stringent. For the \textsc{CosmoDC2} simulation this  effect becomes important, especially at large scales. Thus, for the tests in \textsc{CosmoDC2} we include the mask effects in the harmonic space covariance using the DESC \textsc{TJPCov} package, see Appendix~\ref{sec:mask_cov} and Section~\ref{sec:cosmodc2} for more details.

\begin{figure*}
  \centering
  \includegraphics[width=0.95\textwidth]{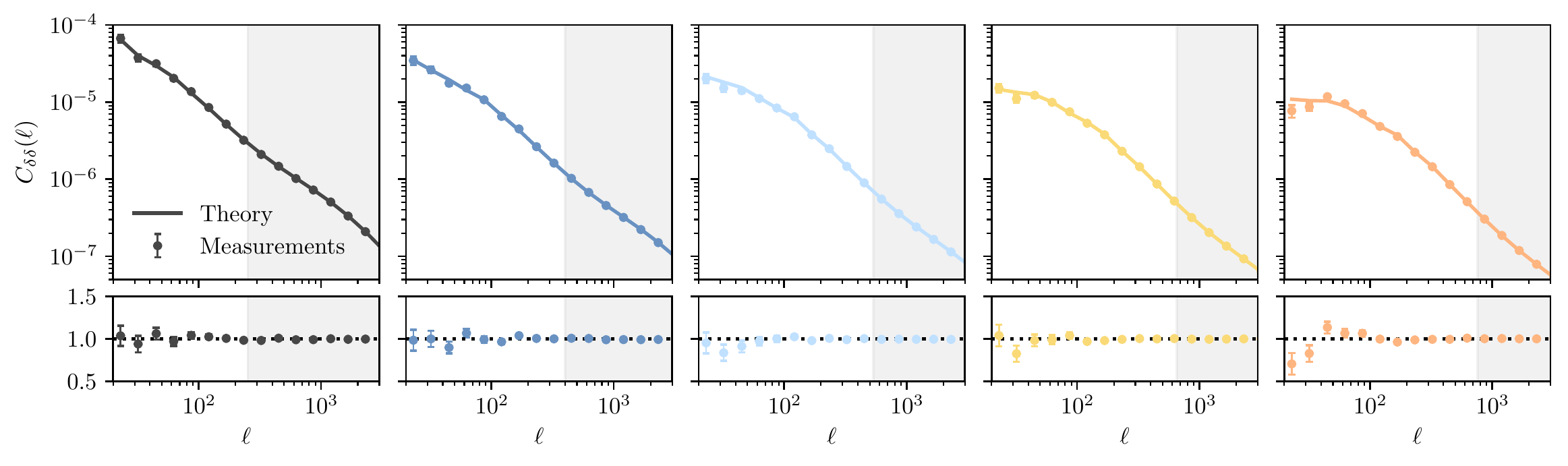} \\
   \includegraphics[width=0.95\textwidth]{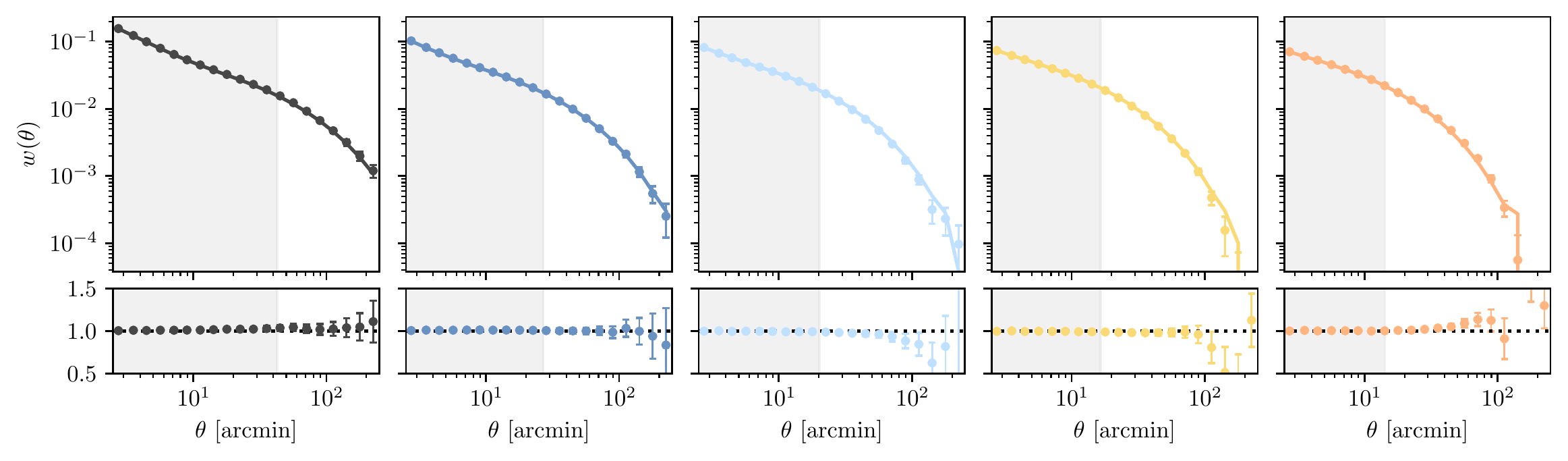}
   \caption{Galaxy clustering measurements from \textsc{TXPipe} in harmonic (top row) and real space (bottom row). We only show the  auto-correlation measurements, which are the ones included in the cosmological analyses, following the DESC SRD specifications. For each bin, we show the measurements from our Gaussian simulations and the input theory (upper panels) and the ratio of the measurement to the theory (lower panels). The error bars are the square root of the diagonal covariance. The gray shaded bands represent the SRD scale cuts.} \label{fig:clustering}
 \end{figure*}

\subsection{Data vector}
\label{sec:dv}

\begin{table}
    \centering
    \begin{tabular}{ccc}
     & Harmonic space & Real space \\ 
    \hline
    Lens bin &  &  \\ 
    1 & $\ell < 253$ & $\theta>42.81$ arcmin\\
    2 & $\ell < 402$ & $\theta>26.93$ arcmin \\
    3 & $\ell < 535$ & $\theta>20.21$ arcmin \\
    4 & $\ell < 654$ & $\theta>16.52$ arcmin \\
    5 & $\ell < 761$ & $\theta>14.21$ arcmin \\
  \hline
    Source bin &  &  \\
    1--5 & $\ell < 3000$ & $\theta>2.5$ arcmin\\
    \hline
    \end{tabular}
    \caption{Scale cuts in real space and harmonic space, that in the case of the lens sample  correspond to $k_\text{max}=0.3 h$/Mpc, accordingly with the DESC SRD specifications. }
    \label{tab:srd_scale}
\end{table}

\begin{table*}
\centering 
\begin{tabular}{llllll }

  & Probe & $\chi^2 / \nu$ (r1) & PTE (r1) & $\chi^2 / \nu$ (r2) & PTE (r2) \\ 
  \hline
\multirow{5}{*}{Real space} & $\xi_{+}$ & 259.9/300 = 0.87 & 0.95  & 260.8/300 = 0.87 & 0.95\\
 & $\xi_{-}$& 283.7/300 = 0.95 & 0.74 & 284.7/300 = 0.95 & 0.73\\
  & $\gamma_{t}$& 91.5/67 = 1.37 & 0.02 & 60.8/67 = 0.91 & 0.69\\
  & $w$& 57.8/53 = 1.09 & 0.30 & 71.3/53 = 1.35 & 0.05\\
   & 3$\times$2pt &  722.6/720 = 1.00  & 0.47 & 680.3/720 = 0.94 & 0.85 \\
\hline
\multirow{4}{*}{Harmonic space} &  $C_{\gamma \gamma}$& 232.8/225 = 1.03 & 0.35 & 223.9/225 = 1.00 & 0.51 \\
  & $C_{\delta_{g} \gamma}$ & 75.3/63 = 1.20 & 0.14 &56.3/63 = 0.89 & 0.71 \\
  & $C_{\delta_{g} \delta_{g}}$ & 55.5/49 = 1.13 & 0.24 &  40.2/49 = 0.82 & 0.81\\
   & 3$\times$2pt & 383.2/337  = 1.14 & 0.04 & 339.7/337 = 1.01 & 0.45\\
  \hline
\end{tabular}
\caption{$\chi^2$ per degree of freedom and the Probability-To-Exceed (PTE), after scale cuts, when comparing \textsc{TXPipe} output and input for the DESC SRD LSST Y1-like Gaussian simulation. We list the results for two different realizations: r1 and r2. For the rest of the paper we always use the datavectors from r1.  }
\label{tab:chi2}
\end{table*}

In Figures~\ref{fig:cosmicshear}, \ref{fig:ggl} and~\ref{fig:clustering} we show the data vector outputs of \textsc{TXPipe} for cosmic shear, galaxy-galaxy lensing and galaxy clustering, in real and harmonic space. In all the figures we also show the input theoretical prediction as well as the ratio of our measurements to the theory. For cosmic shear we show all the bin combinations. For galaxy-galaxy lensing and galaxy clustering, we only show a subset of the bin combinations for simplicity, in particular the ones where the signal-to-noise is larger and that are included in our analysis. The error bars in these figures are the square root of the diagonal of the covariance matrix described in Section~\ref{sec:cov}. Our measurements are performed on scales $2.5<\theta<250$ arcmin for real space (with 20 logarithmic bins) and $20<\ell<3722$ for harmonic space (with 17 logarithmic bins, to match the scales defined in the DESC SRD). In all ongoing galaxy surveys, the modeling of the 3$\times$2pt data vector is uncertain on small angular scales due to either uncertainty in the matter power spectrum $P_{\rm mm}$ or the uncertainty in nonlinear galaxy bias, which affects both $P_{\rm gg}$ and $P_{\rm gm}$. These scale cuts are currently one of the determining factors of the constraining power of a given dataset. We adopt the DESC SRD scale cuts listed in Table~\ref{tab:srd_scale}, with $k_{\rm max}=0.3\,  h \,  \rm{Mpc}^{-1}= 0.213 \,  \rm{Mpc}^{-1}$ for the lenses used in galaxy clustering and galaxy-galaxy lensing and $\ell<3000$ for cosmic shear. We convert from the 3D quantity $k$ to the projected $\ell$ using:
\begin{equation}
\ell_\text{max} = k_{\rm max}  \chi (z_l) - 0.5.
\end{equation}
For the galaxy-galaxy lensing and galaxy clustering probes, we convert to a real space scale cut using the approximation  $\theta_{\rm{min}} \sim \pi/\ell_\text{max}$. Moreover, for galaxy-galaxy lensing we only use the lens-source redshift bin combinations that are indicated in the DESC SRD, which only include: $z_l^1 - z_s^3$, $z_l^1 - z_s^4$, $z_l^1 - z_s^5$, $z_l^2 - z_s^4$, $z_l^2 - z_s^5$, $z_l^3 - z_s^5$ and $z_l^4 - z_s^5$, corresponding to the combinations with higher signal-to-noise and where the lenses and the sources barely overlap in redshift. For galaxy clustering we only include the auto-correlations, as specified in the SRD.
Note the choice of redshift combinations and of all the scale cuts might need to be revised for the analysis with data. In particular: a) overlapping bins between the lenses and sources will probably need to be included in future analyses since they are useful to self-calibrate the intrinsic alignment parameters, as it was done in e.g. the DES Y1 and in the DES Y3 3$\times$2pt analyses \citep{DES3x2, desy3-3x2}, and b) clustering cross-correlations between redshift bins are useful to self-calibrate the redshift distributions of the lens sample \citep{Nicola2020, Hang2021}. For cosmic shear, we include all the scales we measure in real space and we cut at $\ell<3000$ in harmonic space, following the SRD. In general, within the scale cuts, we find excellent agreement between the measurements and the input. To quantify this agreement, we evaluate the $\chi^2$ per degree of freedom $\nu$ defined as
\begin{equation}
    \chi^2/\nu = \frac{1}{\nu} (D_{\rm TXPipe}-D_{\rm theory})^{t}\mathcal{C}^{-1}(D_{\rm TXPipe}-D_{\rm theory}),
\end{equation}
where $\nu$ is the number of data points, $D_{\rm TXPipe}$ is the data vector as measured by \textsc{TXPipe}, $D_{\rm theory}$ is the theory data vector evaluated at the same angular scales as the measurements\footnote{In harmonic space we weight the scales using the bandpower window functions within the \textsc{Namaster} formalism, as detailed in Appendix~\ref{sec:estimators}. In real space we evaluate the theory at the mean angular scale value, as measured in the data using the \textit{meanlogr} \texttt{TreeCorr} function.} and $\mathcal{C}$ is the covariance matrix. We list the $\chi^2/\nu$ values in Table~\ref{tab:chi2}. We also report the Probability-To-Exceed (PTE) or also sometimes called p-value, defined as:
\begin{equation}
   \mathrm{PTE} = 1 - \int_0^{\chi_D^2} \chi^2(x, \nu_D ) dx \, , 
\end{equation}
where the $\chi^2$ function is integrated until the given $\chi^2_D$ value for the data set $D$.

Note that we find some of the  PTE values to be below 0.05 or above 0.95. This is expected given a distribution of values but we still want to check that they are indeed due to noise and not due to an identifiable issue. In order to do so we run a second realization of the Gaussian simulations, for which we also list the results in Table.~\ref{tab:chi2}. We find that most of the deviations that are below 0.05 or above 0.95 in the first realization yield a good agreement (values between 0.16 and 0.84) in the second realization, except for the PTE 0.95 value for $\xi_+$ that remains the same in both realizations, pointing to a possible overestimation of the covariance for this probe and also potentially for $\xi_-$. Besides this case, we conclude that the rest of the anomalies are due to statistical fluctuations.

\subsection{Cosmology inference}
\label{sec:cosmo}

\begin{table}
\centering 
\begin{tabular}{cc}
\hline
Parameter & Prior   \\ 
\hline
\multicolumn{2}{l}{\textit{Cosmological parameters}}  \\
$\Omega_{\rm c} $  &$\mathcal{N}[0.220, 0.2]$  \\
$\Omega_{\rm{b}}$  &$\mathcal{N}[0.0448, 0.006]$   \\
$\sigma_{8}$  &$\mathcal{N}[0.8, 0.14]$  \\
$A_s \times 10^{-9}$ & $\mathcal{N}[2.16, 0.378]$\\
$h$  &$\mathcal{N}[0.71, 0.063]$ \\
$n_{\rm{s}}$  &$\mathcal{N}[0.963, 0.08]$  \\
$w_{0}$& $\mathcal{N}[-1, 0.8]$   \\ 
$w_{a}$& $\mathcal{N}[0.0, 2.0]$   \\ 
\hline
 \multicolumn{2}{l}{\textit{Astrophysical nuisance parameters}} \\
$b^{1\cdots 5}$  & $\mathcal{N}[1.229, 0.9]$, $\mathcal{N}[1.362, 0.9]$, $\mathcal{N}[1.502,0.9]$,\\
& $\mathcal{N}[1.648, 0.9]$,  $\mathcal{N}[1.799, 0.9]$  \\
$A_{IA}$ &  $\mathcal{N}[0.0, 3.9]$\\
$\eta_{IA}$ &  $\mathcal{N}[0.0, 2.3]$\\
\hline
 \multicolumn{2}{l}{\textit{Observational nuisance parameters}} \\
$m^{1\cdots 5}$  & $\mathcal{N}[0.0, 0.013]$  \\
$\Delta z_{l}^{1\cdots 5} \times 10^{3}$ &  $\mathcal{N}[0.0, 6.6]$, $\mathcal{N}[0.0, 7.5]$, $\mathcal{N}[0.0, 8.5]$,\\
 &  $\mathcal{N}[0.0, 9.5]$, $\mathcal{N}[0.0, 10.5]$\\
$\Delta z_{s}^{1\cdots 5}\times 10^{3}$ &  $\mathcal{N}[0.0, 2.6]$, $\mathcal{N}[0.0, 3.0]$, $\mathcal{N}[0.0, 3.4]$,\\
 &  $\mathcal{N}[0.0, 3.9]$, $\mathcal{N}[0.0, 5.2]$\\
\hline
\end{tabular}
\caption{Priors for the parameters of our model. $\mathcal{N(\mu, \sigma})$ is a Gaussian distribution with a mean $\mu$ and a width $\sigma$. The priors mostly follow those specified in the LSST DESC SRD, but centered on the true values of our simulations.}
\label{table:prior}
\end{table}

\begin{figure*}
\begin{center}
\includegraphics[width=0.91\textwidth]{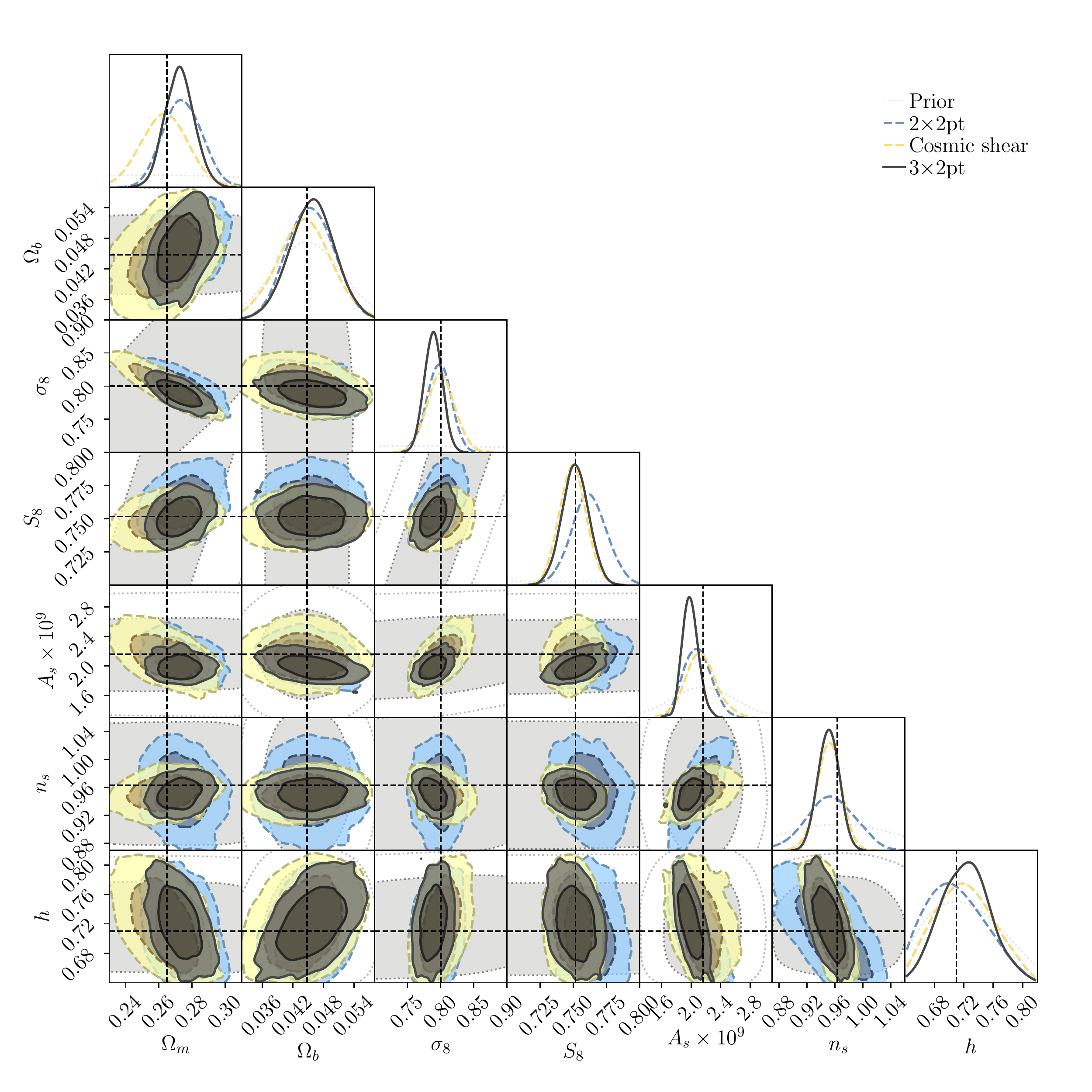}
\caption{Cosmological contours assuming the $\Lambda$CDM model for the LSST Y1-like Gaussian mock catalogs compared with the input values and the DESC SRD priors. We show the results for the harmonic space  2$\times$2pt analysis (galaxy clustering and galaxy-galaxy lensing, \textit{dashed blue}), cosmic shear only (\textit{dashed yellow}) and their combination 3$\times$2pt (\textit{solid black}). }
\label{fig:lcdm-cs-2x2}
\end{center}
\end{figure*}

We now take the data vectors measured from \textsc{TXPipe} and propagate them into cosmological constraints to validate that we can recover the input cosmology to the required precision and accuracy. To this end, we use the DESC likelihood package \textsc{fireCrown}\footnote{\url{https://github.com/LSSTDESC/firecrown}}. In this work we assume a Gaussian likelihood throughout. The core of \textsc{fireCrown} uses \textsc{CCL} as its theory prediction and it is designed to interface with established cosmological inference code such as \textsc{Cobaya} and \textsc{CosmoSIS} so that we can take advantage of the fast sampling and caching techniques implemented therein. This is the first time \textsc{fireCrown} is used for a scientific project, and is a suitable test case given the idealized settings. We use the \textsc{Emcee} sampler \citep{Foreman-Mackey2013}, and obtain 640,0000 samples for each chain, for which we apply a burn-in of 150,000 samples in all cases.  We use \textsc{ChainConsumer} \citep{Hinton2016} with a Kernel Density Estimation (KDE) smoothing value of 2.0 for the plots and to obtain the mean and 1-$\sigma$ values shown in the tables. 

We fit the measured data vectors to a $\Lambda$CDM model and a $w_0-w_a$CDM model with the priors on the cosmological parameters and nuisance parameters listed in Table~\ref{table:prior}, which closely follows that used in the DESC SRD -- see Appendix~\ref{app:srd_comparison} for a detailed description of the differences between our analysis and DESC SRD. In Fig.~\ref{fig:lcdm-cs-2x2} we show the  cosmological results from \textsc{fireCrown} for the harmonic space analysis using the data vectors derived from \textsc{TXPipe} in Section~\ref{sec:dv}. We show the cosmic shear only results, the galaxy-galaxy lensing and galaxy clustering (also called 2$\times$2pt) results and their combination. In Figures~\ref{fig:lcdm} and~\ref{fig:wcdm} we also compare the constraints coming from the real and harmonic space estimators for a subset of the parameters, which we find to be in very good agreement. We show the rest of the parameters for these cases in  Appendix~\ref{sec:full_posteriors}, together with the nuisance parameter posteriors. We also present a summary of the results in Table~\ref{tab:model_params_lcdm} for $\Lambda$CDM and in Table~\ref{tab:model_params_wcdm} for $w_0-w_a$CDM. In all cases, we recover the true input cosmological parameters within the 1$\sigma$ contours. 

We use the \textsc{apriori} sampler provided by \textsc{CosmoSIS} within \textsc{fireCrown} to sample the prior and understand which parameters we are able to constrain with respect to the prior. We add the \textsc{apriori} samples in all the contour plots as the gray dotted lines. Overall we find that with LSST Y1 precision all the cosmological parameters are significantly constrained with respect to the prior in the 3$\times$2pt combination. Also, as shown in Fig.~\ref{fig:lcdm-cs-2x2} we find that cosmic shear gives more precise results for $n_s$ and $S_8$, while the 2$\times$2pt combination is able to constraint $\Omega_m$ and $\Omega_b$ more tightly.

Moreover, using the priors stated in Table~\ref{table:prior} we find that  both the $\Lambda$CDM and the $w_0-w_a$CDM LSST-Y1 analyses are systematics limited, meaning that  the constraints on $S_8$ improve significantly when fixing  observational systematics. Specifically, we find the constraints on $S_8$ are $\sim$5 times more precise in $\Lambda$CDM and  $\sim$1.8  more precise in $w_0-w_a$CDM times when fixing the shear and redshift calibration parameters, while the parameters describing the equation of state of dark energy do not vary as much and yield similar constraints. In Fig.~\ref{fig:gs_cosmo_full} we display this comparison.

Then, we also compare our results on the LSST Y1 patch simulated using the SRD specifications with the forecast on cosmological parameters from the DESC SRD. Generally we find the constraining power of our posteriors is similar to what was predicted in the SRD. This is an important cross-check given that the software tools used in the two studies are significantly different. For more details we refer the reader to Appendix~\ref{app:srd_comparison}. 

We also compare the results from this work with the most constraining Stage-III posteriors to date, the 3$\times$2pt constraints from the first three years (Y3) of DES data \citep{desy3-3x2, desy3-extensions}. We find that the LSST Y1 results on the Gaussians simulation provide a constraint in $S_8$ that is $\simeq$2 times tighter than DES Y3 in $\Lambda$CDM and $\simeq$2.5 times in $w_0-w_a$CDM, as well as a significant improvement in the rest of the cosmological parameters. Note that this comparison is not straightforward because of two main reasons: 1) The priors from the DES Y3 analysis are uninformative (flat) while we use Gaussian priors. 2) The IA model is not the same for the DES Y3 $\Lambda$CDM and our work (we marginalize over a the 2-parameter NLA model and the DES Y3 $\Lambda$CDM analysis marginalizes over a 5-parameter TATT model). The IA models \textit{are} the same in both analyses for $w_0-w_a$CDM, which provides a cleaner comparison that we show in Figure~\ref{fig:srd_comparison}. Therefore, even with these caveats in mind, we expect that the LSST Y1 data set will  be able to shed light into the current $S_8$ tension, see e.g. \citet{DiValentino2021}.

Overall, the work presented here shows that the DESC tools that are needed to perform an end-to-end 3$\times$2pt analysis are sufficiently accurate in their most basic functionalities for LSST Y1 precision. This includes analyses both in harmonic space and in real space, from the data vector measurement to the parameter inference.

\begin{table*}
    \centering
    \begin{tabular}{lcccc}
        \hline
		$\qquad  \qquad \qquad \Lambda$CDM & $\sigma_8$ & $\Omega_m$ & $S_8$ & FoM \\ 
		\hline
		Harmonic Gauss. Sims. LSST Y1 & $0.789^{+0.013}_{-0.014}$ & $ 0.2724^{+0.0086}_{-0.0092}$ & $0.7511^{+0.0104}_{-0.0097}$ &  12,317  \\ 
		Harmonic Gauss. Sims. LSST Y1  fixed obs. sys. &  $0.7920^{+0.0092}_{-0.0089}$  & $ 0.2715^{+0.0062}_{-0.0067}$  &  $0.7532^{+0.0019}_{-0.0017}$  & 91,760  \\ 
		Real Gauss. Sims. LSST Y1  & $0.797^{+0.012}_{-0.014}$ & $ 0.2711^{+0.0091}_{-0.0081}$ & $0.7582^{+0.0092}_{-0.0095} $ & 13,604 \\ 
		Real Gauss. Sims. LSST Y1 fixed obs. sys. & $0.7935^{+0.0079}_{-0.0091}$ & $0.2721^{+0.0062}_{-0.0058}$ & $0.7554^{+0.0020}_{-0.0019}$ & 89,928 \\ 
		Harmonic $\textsc{CosmoDC2}$ & $0.767\pm 0.031$ & - & $0.737\pm 0.016$  & - \\ 
		Real $\textsc{CosmoDC2}$ &  $0.771^{+0.031}_{-0.029}$ & - &  $0.740^{+0.016}_{-0.018}$ & - \\
		\hline
    \end{tabular}
    \caption{Results for the LSST Y1  Gaussian simulation and  \textsc{CosmoDC2}  under the $\Lambda$CDM model, shown also in Fig.~\ref{fig:lcdm}. We also add the results fixing the observational systematics parameters, which include multiplicative shear biases and photometric redshifts (see Appendix C). The Figure of Merit is defined as $\rm{FoM} =\det \left[ \mathcal{C}\, (\Omega_m,\sigma_8)\right]^{- 1/2}$, where $\mathcal{C}$ is the covariance of the posterior parameters.}
     \label{tab:model_params_lcdm}
\end{table*}

\begin{table*}
    \centering
    \begin{tabular}{lcccccc}
        \hline
		$\qquad  \qquad \qquad  w$CDM & $\sigma_8$ & $\Omega_m$ & $S_8$ & $w_0$ & $w_a$ & FoM \\ 
		\hline
        Harm. Gauss. Sims. LSST Y1  & $0.793^{+0.020}_{-0.022}$ & $0.267^{+0.017}_{-0.015}$ & $0.7500^{+0.0093}_{-0.0108}$ & $-1.08^{+0.15}_{-0.17}$ & $0.37^{+0.32}_{-0.47}$ & 33.1 \\ 
		Harm. Gauss. Sims. LSST Y1  fixed obs. sys.& $0.804^{+0.012}_{-0.014}$ & $0.259^{+0.011}_{-0.010}$ & $0.7470\pm 0.0057 $ & $-1.20^{+0.14}_{-0.12}$ & $0.66^{+0.28}_{-0.48}$ & 49.4 \\ 
		Real Gauss. Sims. LSST Y1  & $0.800^{+0.022}_{-0.023}$ & $0.267^{+0.019}_{-0.018}$ & $0.756^{+0.012}_{-0.013}$ & $-1.05^{+0.17}_{-0.19}$ & $0.23^{+0.50}_{-0.46}$ & 27.2 \\ 
		Real Gauss. Sims. LSST Y1  fixed obs. sys. & $0.801^{+0.014}_{-0.013}$ & $0.265\pm 0.012$ & $0.7531^{+0.0067}_{-0.0061}$ & $-1.14^{+0.17}_{-0.12}$ & $0.59^{+0.31}_{-0.54}$ & 42.0\\ 
	
		Harmonic $\textsc{CosmoDC2}$ & $0.766^{+0.033}_{-0.031}$ & -  & $0.736\pm 0.020$ & - & - & -\\
		Real $\textsc{CosmoDC2}$ &   $0.753^{+0.038}_{-0.031}$ & - & $0.740^{+0.019}_{-0.017}$ & - & - & -\\ 
		Forecasted in DESC SRD LSST Y1 & $0.831\pm 0.020$ & $0.315\pm 0.020$  & $0.852\pm 0.010$  & $-1.00\pm 0.21$ & $0.00\pm 0.69$ & 17.5\\
		\hline
    \end{tabular}
    \caption{Analogous results for the $w_0-w_a$CDM model, shown also in Fig.~\ref{fig:wcdm}. In the last row we also add the forecast produced by the DESC SRD assuming the same area as our Gaussian simulation. The center values of that forecast were different but the uncertainties can be compared. The Figure of Merit is defined as $\rm{FoM} =\det [\mathcal{C}(w_0,w_a)]^{-1/2}$, where $\mathcal{C}$ is the covariance of the posterior parameters.}
    \label{tab:model_params_wcdm}
\end{table*}

\section{Application to \textsc{CosmoDC2}}
\label{sec:cosmodc2}

\begin{table}
    \centering
    \begin{tabular}{cccc}
    \hline 
    Lens bin & $\langle z \rangle$ & Number density & Galaxy bias \\
    1 & 0.30 & 2.48 & 0.87 \\
    2 & 0.50 & 3.51 &  1.02\\
    3 & 0.70 & 4.11 &  1.19\\
    4 & 0.90 & 4.12 &  1.30\\
    5 & 1.10 & 2.49 &  1.54\\
    \hline
    Source bin  & $\langle z \rangle$ & Number density & Shape noise $\sigma_e$\\
    1 & 0.37 &  2.85 & 0.288 \\
    2 & 0.52 &  2.87 & 0.317 \\
    3 & 0.66 &  3.35 & 0.305 \\
    4 & 0.83 &  5.00 & 0.335 \\
    5 & 1.29 &  7.79 & 0.353 \\
    \hline
    \end{tabular}
    \caption{CosmoDC2 sample specification. The number densities are in 1/arcmin$^2$ and $\sigma_e$ is defined in Eq.~\ref{eq:sigma_e}. We also list the galaxy bias values that we input to the covariance. We check they are consistent with the posteriors in Figure~\ref{fig:cosmodc2_bias}.}
    \label{tab:cosmodc2_sample}
\end{table}

\begin{figure}
\begin{center}
\includegraphics[width=0.49\textwidth]{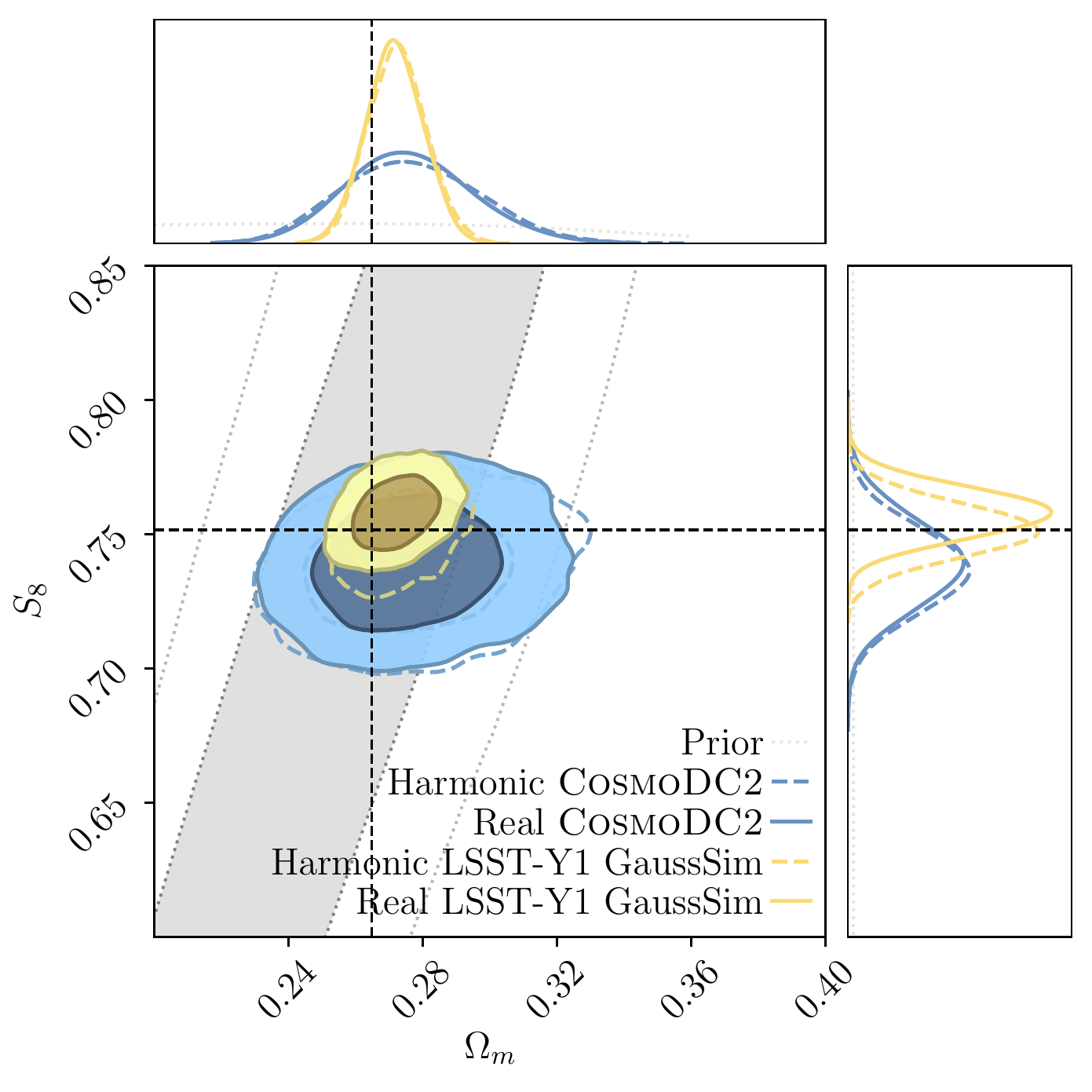}
\caption{Cosmological contours assuming the $\Lambda$CDM model for the \textsc{CosmoDC2} and LSST Y1-like Gaussian mock catalogs  compared with the input values and the DESC SRD priors that we use in both cases.  }
\label{fig:lcdm}
\end{center}
\end{figure}

\begin{figure*}
\begin{center}
\includegraphics[width=0.7\textwidth]{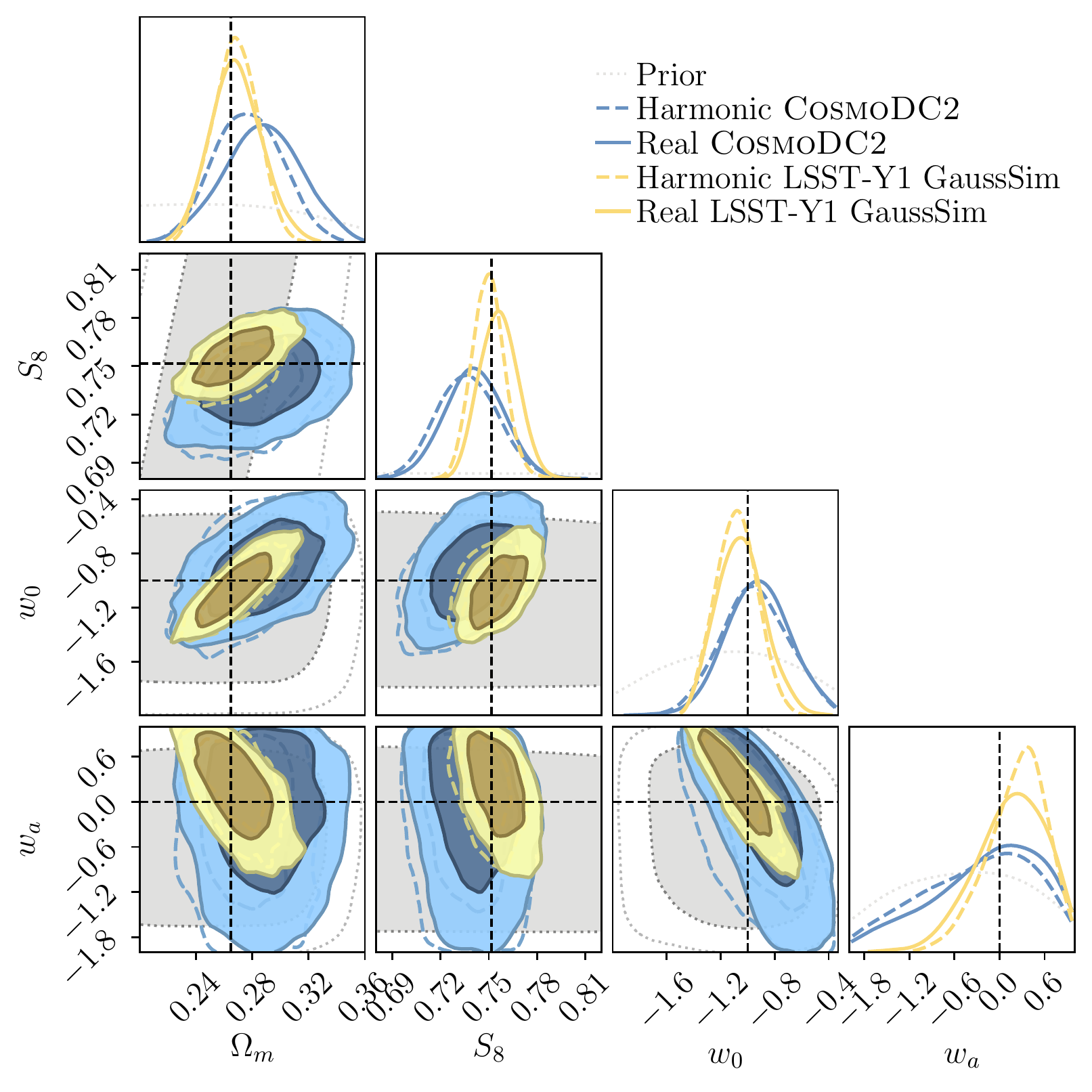}
\caption{Cosmological contours assuming the $w_0-w_a$CDM model for the \textsc{CosmoDC2} and LSST Y1-like Gaussian mock catalogs  compared with the input values and the DESC SRD priors that we use in both cases.}
\label{fig:wcdm}
\end{center}
\end{figure*}

\begin{figure}
\begin{center}
\includegraphics[width=0.49\textwidth]{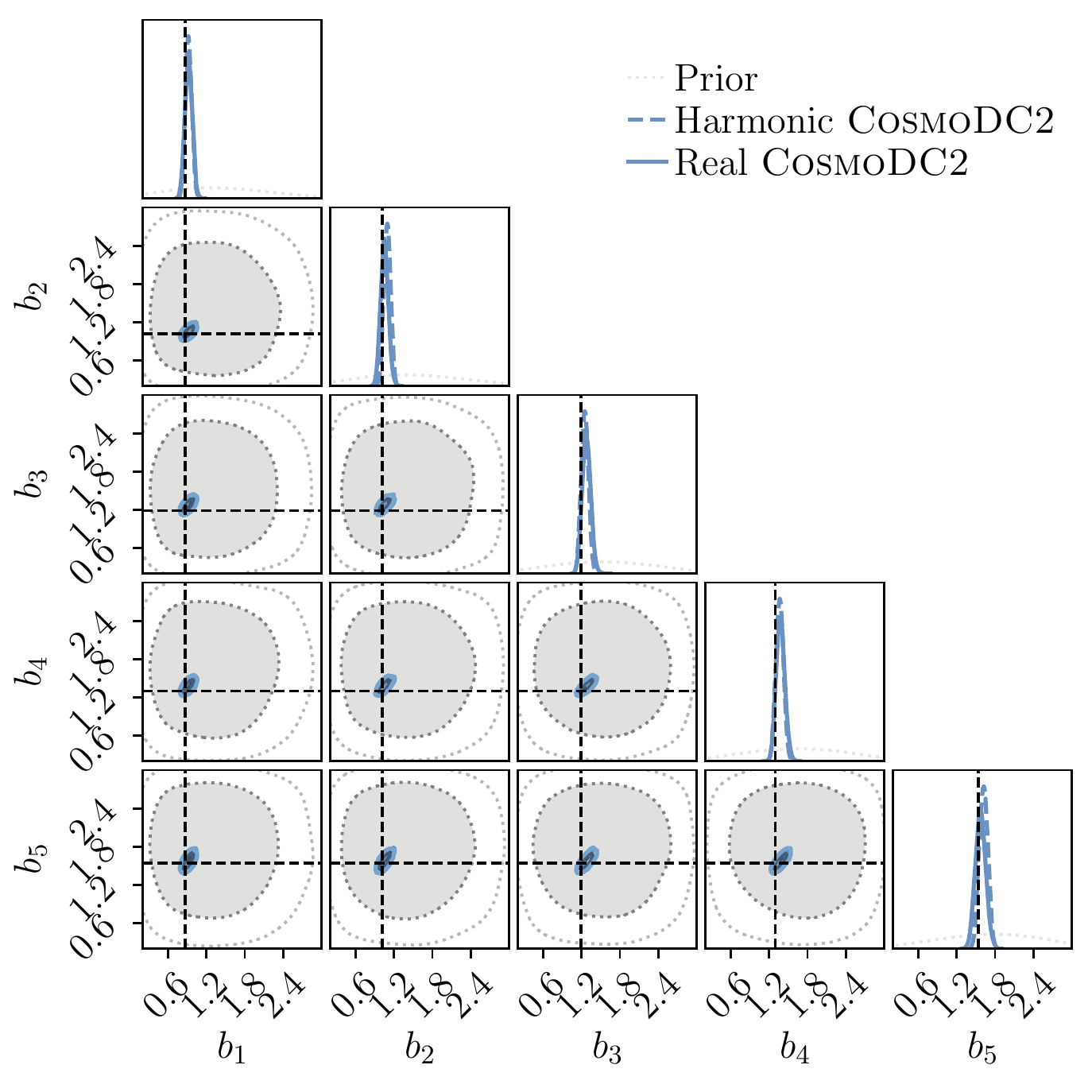}
\caption{Linear galaxy bias for the \textsc{CosmoDC2} lens catalog. The dashed lines mark the values we have used as input to the covariance matrix and for the $\chi^2$ values.}
\label{fig:cosmodc2_bias}
\end{center}
\end{figure}

Now that we have validated the core functionalities of \textsc{TXPipe} at scale, we describe its application to a more complex mock galaxy catalog as a first step towards applying the pipeline to data (we note that in \citealt{Phillips-Longley2022}, \textsc{TXPipe} was used with Stage-III data, but that paper focused on testing cosmic shear in real space, while we test the full 3$\times$2pt data vector in both real and harmonic space here). In particular, we run \textsc{TXPipe} on the LSST DESC simulation suite \textsc{CosmoDC2} \citep{Korytov2019,Kovacs2021}. 

\textsc{CosmoDC2} is a mock galaxy catalog based on the Outer Rim N-body simulation \citep{Heitmann2019}. It covers $\sim440$ deg$^{2}$ area to redshift and depth much beyond that expected for LSST. Each galaxy is assigned observable quantities from photometry to morphology based on a number of semi-analytical models. The galaxies also carry cosmological information from both their position on the sky and ray-traced weak lensing properties. This provides a fairly realistic test bed for \textsc{TXPipe} while still having a known true input cosmology to the simulations.

We add noise to the \textsc{CosmoDC2} observed magnitudes following the methodology described in \citet{Ivezic2010}, which accounts for the width of the point spread function (PSF), sky brightness, instrumental noise, and pixelization\footnote{Future noise simulations will be done with updated values of the parameters in that document, which may be found at \url{https://smtn-002.lsst.io/}}. We use 16 visits per band per pixel, approximating an LSST Y1 survey. We also include the noise and estimator response on the shear truth values, using a constant shape noise per component $i$ of $\sigma = 0.26$ and shear estimator response values \citep{Sheldon2020}. We sample the responses from a multivariate normal with the mean and standard deviation from a spline in size and signal-to-noise fitted to the DES-Year 1 catalog \citep{Zuntz2018}, and a correlation coefficient $\rho=0.2$ between the size and shape response, as measured from the DES-Year 1 shape catalog. We do not vary any noise properties across the field. Using Eq.~\ref{eq:sigma_e}, we obtain the shape noise values per component $\sigma_e$ displayed in Table~\ref{tab:cosmodc2_sample}.

We construct a similar lens sample as that used in Section~\ref{sec:sims} based on the DESC SRD, using all galaxies with $i < 24.1$.  For source galaxies we select objects on the $riz$ signal-to-noise and size as measured by the trace of the moments matrix compared to that of the PSF: $\mathrm{SNR}>10,\, T/T_\mathrm{psf} > 0.5$ (see \citealt{Zuntz2018} for more details). For both samples we use a random forest algorithm to assign galaxies to tomographic redshift bins, training a classifier on an ideal spectroscopic sample. For the lens sample we use $grizy$ for the training while for the source sample the training can only be done in the $riz$ bands because of the requirements of metacalibration - for more details on both the algorithm and this limitation see \citet{Zuntz2021O}.

The resulting redshift distribution and galaxy characteristics are shown in Figure~\ref{fig:nzs_srd} and Table~\ref{tab:cosmodc2_sample}. We note that this is not identical to the specification in DESC SRD, while it is  consistent with the findings in \citet{Zhang2022}.  We next run \textsc{TXPipe} with this sample. \textsc{CosmoDC2} offers us additional opportunities to test \textsc{TXPipe} as well as the simulations. For instance, we are able to test effects such as non-unity \textsc{Metadetection} responses (see Section~\ref{sec:responses}), non-linear effects in the matter power spectrum on small scales, generating tomographic bins based on the galaxy colors, and many more.

Then, we estimate the linear galaxy bias of the lens sample performing a direct fit to the large scales for galaxy-galaxy lensing and galaxy clustering. We report these values in Table~\ref{tab:cosmodc2_sample}. Using these galaxy bias values, we obtain a good agreement between the measurements and the theory data vectors both in real and harmonic space after applying the scale cuts detailed in Table~\ref{tab:srd_scale}. We display the resulting $\chi^2$ values in Table~\ref{tab:chi2cosmodc2}. We also show the two-point measurements in Appendix~\ref{app:cosmodc2_datavectors}. 

We use these bias values to infer the Gaussian covariance matrix and use \textsc{fireCrown} to perform parameter inference. Then, we check the galaxy bias posteriors are consistent with the input values, which we demonstrate in Figure~\ref{fig:cosmodc2_bias}. Since they are found consistent, we do not update our initial estimates in the covariance. Note that the difference between the galaxy bias we find in \textsc{CosmoDC2} and what we input for the Gaussian sims might have an impact on the relative constraining power of each simulation. However, since the area difference is so large, we expect this will not be a dominant effect.

The final cosmological constraints are shown in Figure~\ref{fig:lcdm} for the $\Lambda$CDM model and in Figure~\ref{fig:wcdm} for $w_0-w_a$CDM. Comparing with the priors, we find we are only able to constraint  $\sigma_8$ and $S_8$, which we list in Tables~\ref{tab:model_params_lcdm} and \ref{tab:model_params_wcdm} -- the rest of the parameters are prior-dominated, as shown in Figure~\ref{fig:srd_comparison}. In the latter, we also compare them with the DES Y3 3$\times$2pt results assuming  the same $w_0-w_a$CDM model. 

\begin{table}
\centering 
\begin{tabular}{llll}

  & Probe & $\chi^2 / \nu$ & PTE \\ 
  \hline
\multirow{5}{*}{Real space} & $\xi_{+}$ &  295.7/300 = 0.99 & 0.56 \\
 & $\xi_{-}$& 297.5/300 = 0.99 & 0.53 \\
  & $\gamma_{t}$& 67.4/67  = 1.01 & 0.46\\
  & $w$& 48.8/53 = 0.92 &  0.64\\
   & 3$\times$2pt & 716.8/720 = 1.00 & 0.53 \\
\hline
\multirow{4}{*}{Harmonic space} &  $C_{\gamma \gamma}$&  200.4/225 = 0.89 & 0.88\\
  & $C_{\delta_{g} \gamma}$ & 47.9/63 = 0.76 & 0.92\\
  & $C_{\delta_{g} \delta_{g}}$ & 45.3/49 = 0.92 & 0.62 \\
   & 3$\times$2pt & 296.4/337 = 0.88 & 0.95 \\
  \hline
\end{tabular}
\caption{$\chi^2$ values when comparing \textsc{TXPipe} output and input for \textsc{CosmoDC2}. For CosmoDC2 input we use the galaxy bias values listed in Table~\ref{tab:cosmodc2_sample} and the input cosmology to the simulations. }
\label{tab:chi2cosmodc2}
\end{table}

\section{Discussion: lessons learned and future work}
\label{sec:discsussion}

\paragraph{Lessons learned} Several practical lessons can be learnt from this validation exercise relevant for LSST-scale datasets, some of which would not have been obvious in Stage-III datasets:
\begin{itemize}
 
    \item Monitoring intermediate steps is critical for multiple parts of the pipeline, and devising a mechanism to do this automatically requires careful choices. In many cases, such as when validating maps generated in our pipeline, a key test is the general question ``does anything look unusual in this map?'' rather than a purely numerical comparison that is not easily automated. We also have a number of null tests that we can run automatically on data, to ensure that, for example, galaxy shear is not correlated with signal-to-noise. But we have enough of these tests that we do expect some to randomly fail at the $2\sigma$, even on correct data, so we cannot automate this process entirely.
    \item The complex web of external software dependencies in umbrella frameworks like TXPipe makes testing challenging; continuous integration is a relatively small part of addressing this problem, and more complete regression tests on large data sets were needed. Changes in code or packaging of other DESC projects like \textsc{RAIL}  and general packages like \textsc{numpy} have required careful version management.  We found that a containerization approach using Docker \citep{docker} is effective for management, using two images, one for experimentation and one more stable, and both built automatically using Github Actions.
    \item Having redundancy in the pipeline allows for very powerful cross-checks and debugging and we should incorporate this as much as possible into our pipelines. This includes multiple estimators, multiple implementations of covariance and noise estimations, etc. For example, as shown in Fig.~\ref{fig:noise} and detailed in Appendix~\ref{eq:harmonic_estimator} we have implemented two ways of estimating the noise power spectrum needed for the Fourier space measurements. This comparison  revealed some subtle issues related to handling maps with weights per object and a partial mask.
    \item Similarly, comparing different parts of the \textsc{TXPipe} output to external  codes is an essential exercise that often reveals problems that would otherwise be hard to track. Even though this kind of tests are especially needed for validating new code, periodic comparisons can expose unexpected behaviors from different revisions of the code. In this work, we carried out several comparisons of different outputs such as: (i) Real space two-point measurements and theory data vectors, comparing them with another DESC code \textsc{DESCQA} (see Appendix~\ref{app:descqa} for examples of the aftermath of such a comparison); (ii) harmonic space two-point measurements and theory data vectors, comparing them with a personal code (iii) real space analytical covariance with the external code \textsc{CosmoCov}\footnote{\url{https://github.com/CosmoLike/CosmoCov}} \citep{Krause2017, Fang2020} and with the \textsc{TXPipe}'s jackknife covariance (iv) harmonic space analytical covariance with a covariance obtained from several gaussian simulation realizations and with an external personal code from one of the authors. All these comparisons exposed issues in each of the pieces and helped to disentangle errors in such a large pipeline when the PTE validation tests were not meeting our criteria. 
    \item It is extremely valuable to validate pipelines on simple simulations such as the Gaussian simulation used in this work before embarking on tests with more realistic N-body simulations, in order to be able to disentangle between different issues in an easier way. Even this nominally simple set-up has allowed us to find multiple issues and inaccurate approximations. In Appendix~\ref{app:descqa}  we list some examples of this, to illustrate such a process.
     \item Once the pipeline has been tested in the simplest setup it becomes equally essential to test it with a more realistic scenario that mimics the data better. For example,  non-unity lens and source galaxy weights and non-unity response functions are prone to introduce many bugs if not thoroughly tested.
    \item The usage of real-space randoms to account for mask effects becomes unsustainable at LSST scale, given the memory load it needs. Instead, pixel-based estimators are much more efficient in this regime. More generally, the right memory vs speed trade-offs may change significantly at LSST scale vs. what has been found in Stage III galaxy surveys.
\end{itemize}

\paragraph{Future work} We have introduced a number of simplifications in this work in order to carry out clean and unambiguous tests of the basic pipeline. Moving forward, we will need to improve on several areas described below in order to bring the DESC infrastructure to the readiness level required for the LSST Y1 data: 
\begin{itemize}
    \item \textit{Sample selection}: We have used the specifications from DESC SRD as a guide for the sample characteristics (number density, redshift distribution, associated nuisance parameters) as well as many of the analysis choices (scale cuts, models of nuisance parameters, priors on cosmological parameters). Some of these specifications should be updated given our current knowledge from Stage-III experiments. 
    \item \textit{$n(z)$:} We have used the true redshift information for the source and lens samples in the modeling of this work to decouple the effect of photometric redshift estimation. Future analyses will employ realistic photometric redshift algorithms to determine the impact of potential biases in the estimated redshift distributions. 
    \item \textit{Mask:} We use a very simplistic LSST Y1 mask, without holes and without accounting for the Milky Way region or Rubin's survey strategy. Future analyses should use more realistic masks such as the one from \citet{Lochner2018} or an updated version of that when available. Moreover, we do not account for  generally spatially varying systematics, which would usually be corrected with LSS weights. Future analyses should implement and test  methods to correct for these effects. 
    \item \textit{Covariance matrix:} Here we rely on a Gaussian covariance and only include mask effects in harmonic space for the \textsc{CosmoDC2} simulation. Future analyses will include non-Gaussian terms and mask effects both in harmonic and real space.  
    \item \textit{Estimators:} In this analysis we have implemented the standard estimators that have already been used in Stage-III 3$\times$2pt analyses. As an exception, we did not include lens-source clustering effects in the galaxy-galaxy lensing estimator, usually corrected for with the so-called ``boost factors'', which future analyses should account for. Moreover, a whole suite of alternative estimators optimized for different goals exist in the literature,  such as  the KNN's  \citep{Banerjee2021} sensitive to higher order information in the galaxy density field; the COSEBIs  \citep{Schneider2010}, an alternative E/B mode decomposition of the cosmic shear information; or the $\Delta \Sigma$ estimator \citep{Sheldon2004} for galaxy-galaxy lensing, which is often used to perform galaxy-halo connection studies. \textsc{TXPipe}'s flexible framework will make it possible to implement and test these estimators in future analyses.
    \item  \textit{Model}: To perform a 3$\times$2pt analysis with the LSST Y1 data set the model will need to be extended in several aspects to \textit{at least} consider: non-Limber terms, Tidal Alignment Tidal Torque (TATT) IA model, redshift-space distortions and magnification effects. 
    \item \textit{Cosmology inference code}: This is the first analysis where \textsc{fireCrown} has been used, and as such we have only tested the most basic implementation of it. Future analyses will be able to test more advanced features such as different matter power spectrum models or nuisance parameter marginalization schemes.  
\end{itemize}

\section{Summary}
\label{sec:summary}

In this paper we perform a rigorous validation test on the software pipeline in LSST DESC to be used for a cosmology analysis with three two-point functions: galaxy clustering, galaxy-galaxy lensing, and cosmic shear, or the 3$\times$2pt analysis. The core of the validation surrounds the software package \textsc{TXPipe}, which carries out the measurement of the 3$\times$2pt data vector in both real and harmonic space. But \textsc{TXPipe} also interfaces with several other DESC software packages -- \textsc{CCL} (for the theoretical modeling), \textsc{fireCrown} (for the cosmological inference), \textsc{TJPCov} (for the covariance matrix calculation), \textsc{SACC} (for storing data vectors and relevant metadata) -- which are also tested via this process. 

We first validate the pipeline using a  Gaussian simulation that is designed to mimic the galaxy sample used in the LSST DESC Science Requirements Document (DESC SRD). In particular, we test the pipeline with the statistical power expected for the first year (Y1) of LSST data. We find that \textsc{TXPipe} can properly recover the input theory to the LSST Y1 precision at both the data vector level and the cosmological constraints, with both the real-space and harmonic-space estimators. This first validation of the pipeline assumes a number of simplifications that will need to be revisited when considering the first-year data analysis. We leave for future work a validation of the pipeline including additional effects.

We also find that in this set up, the systematics contribute to a significant part of the uncertainties, particularly in the $S_8$ parameter, which we find it would be $\sim 5$(1.8) times more tightly constrained under the $\Lambda$($w_0-w_a$)CDM model if we fixed shear and redshift calibration parameters. On the other hand we find these systematics only contribute  slightly to the uncertainties in $w_0$ or $w_a$, the parameters describing the equation of state of dark energy. 

We then apply this pipeline to \textsc{CosmoDC2}, a DESC mock galaxy catalog built on an $N$-body simulation, designed to be sufficiently deep for LSST and have more realistic galaxy characteristics. We find good agreement between the measurements and theory when we only consider scales where we supply the correct theory. We are also able to recover the input cosmological parameters, in particular in the $\sigma_8$ and $S_8$ parameters, which we find are the only parameters that are not prior-dominated in this setup.

Overall, we have developed and validated to LSST-Y1 precision a catalog-to-cosmology framework with DESC tools to obtain cosmological information from weak lensing and galaxy clustering measurements. In this paper we have focused on two-point weak lensing and galaxy clustering estimators, but the scheme we have developed can easily be  extended to include  higher order estimators and also other probes. Therefore, we have accomplished the first milestone in the roadmap to perform cosmological analyses with LSST data.

\section*{Acknowledgments}
This paper has undergone internal review in the LSST Dark Energy Science Collaboration. 
The internal reviewers were David Alonso, Eric Gawiser, François Lanusse.
% Standard papers only: author contribution statements. For examples, see http://blogs.nature.com/nautilus/2007/11/post_12.html
% This work used TBD kindly provided by Not-A-DESC Member and benefitted from comments by Another Non-DESC person.
% Standard papers only: A.B.C. acknowledges support from grant 1234 from ...
The DESC acknowledges ongoing support from the Institut National de Physique Nucl\'eaire et de Physique des Particules in France; the Science \& Technology Facilities Council in the United Kingdom; and the Department of Energy, the National Science Foundation, and the LSST Corporation in the United States.  DESC uses resources of the IN2P3 Computing Center (CC-IN2P3--Lyon/Villeurbanne - France) funded by the Centre National de la Recherche Scientifique; the National Energy Research Scientific Computing Center, a DOE Office of Science User Facility supported by the Office of Science of the U.S.\ Department of Energy under Contract No.\ DE-AC02-05CH11231; STFC DiRAC HPC Facilities, funded by UK BEIS National E-infrastructure capital grants; and the UK particle physics grid, supported by the GridPP Collaboration.  This work was performed in part under DOE Contract DE-AC02-76SF00515. Work at Argonne National Laboratory was supported under the U.S. Department of Energy contract DE-AC02-06CH11357. This research used resources of the Argonne Leadership Computing Facility, which is supported by DOE/SC under contract DE-AC02-06CH11357.
JP, YO and CC were supported by DOE grant DESC0021949 and the Clare Boothe Luce Foundation. 
TT acknowledges funding from the Swiss National Science Foundation under the Ambizione project PZ00P2\_193352.
CGG acknowledges support from European Research Council Grant No:  693024 and the Beecroft Trust.
DA is supported by the Science and Technology Facilities Council through an Ernest Rutherford Fellowship, grant reference ST/P004474.
EG was supported by the U.S. Department of Energy, Office of Science, Office of High Energy Physics Cosmic Frontier Research program under Award No. DE-SC0010008.

\input{author_contributions}

\input{affiliations}

\bibliographystyle{mnras_2author}
\bibliography{references.bib}

\appendix

\section{Two-point estimators} \label{sec:estimators}

\subsection{Harmonic space estimator}

We use \textsc{NaMaster} \citep{namaster} to measure the harmonic space power spectra, based on the pseudo-$C_\ell$ method. For the harmonic space measurements we start from a map, either of galaxy density $\delta_g$ or shear $\gamma$. An observed map $\Tilde{a}$ is the product of the true map $a$ and a weight map $w$: $\Tilde{a} = w a$. The weights map can either be a binary mask or an inverse-variance local weight, down-weighting regions of high noise. 

The spherical harmonic coefficients of the observed map  $\Tilde{a}$ are a convolution of the spherical harmonic coefficients of $a$ and $w$, given the convolution theorem. This leads to mode-coupling between the power spectra of the observed and true fields and prevents us from directly measuring the desired $C_\ell$. We can instead define first the coupled pseudo $\Tilde{C}^{ab}_\ell$ between fields $a$, $b$ with weights $v, w$ respectively:
\begin{equation}
    \Tilde{C}^{ab}_\ell = \frac{1}{2\ell + 1} \sum^{\ell}_{m=-\ell} \Tilde{a}_{\ell m} \Tilde{b}^*_{\ell m},
\end{equation}
which is related to the true power spectrum via the mode-coupling matrix (MCM):
\begin{equation}
    \left< \Tilde{C}^{ab}_\ell \right> = \sum_{\ell'} M^{v w}_{\ell \ell '} C^{ab}_\ell,
\end{equation}
where the averaging $\left< \, \right>$ is over different realizations of $a$ and $b$ of the initial Gaussian fields. The coupling matrix depends only on the weights of the two fields being correlated. Explicit expressions for the coupling matrix of different combinations of spin-0 and 2 fields can be found in \citet{namaster}. If the data covers all sky, then the coupling matrix is the identity. When we are only able to observe a part of the sky, this produces off-diagonal coupling between neighboring multipoles. To obtain $C^{ab}_\ell$, we need to invert the coupling matrix. However, $M^{v w}_{\ell \ell '}$ is non-invertible in many  scenarios. This is usually solved by binning the power spectra in \textit{bandpowers} $B$, containing weighted sums of different multipoles:
\begin{equation}
\Tilde{C}^{ab}_q = \sum_{l \in q } B^\ell_q \Tilde{C}^{ab}_\ell ,
\end{equation}
in which case the binned coupling matrix is defined as: 
\begin{equation}
    \mathcal{M}^{v w}_{q q '} = \sum_{l \in q } \sum_{l' \in q' } M^{v w}_{\ell \ell '},
\end{equation}
and is usually invertible for sufficiently broad bandpowers. Then, the decoupled $\hat{C}^{ab}_q$ for a chosen bandpower binning $q$ is:
\begin{equation}\label{eq:harmonic_estimator}
    \hat{C}^{ab}_q = \sum_{q'} (\mathcal{M}^{v w})^{-1}_{q q'} \, (\Tilde{C}^{ab}_{q'} - \Tilde{N}^{ab}_{q'}),
\end{equation}
where $\Tilde{N}^{ab}_{q'}$ is the (binned) mode-coupled noise power spectrum, which needs to be removed first for autocorrelations. The form of the noise power spectrum depends on the maps being correlated, and their calculation is described below in \ref{app:noise_subtraction}. The results of Equation~(\ref{eq:harmonic_estimator}) are the measurements we plot in Figures~\ref{fig:cosmicshear}, \ref{fig:ggl} and \ref{fig:clustering}.
\begin{figure*}
\begin{center}
\includegraphics[width=0.8\textwidth]{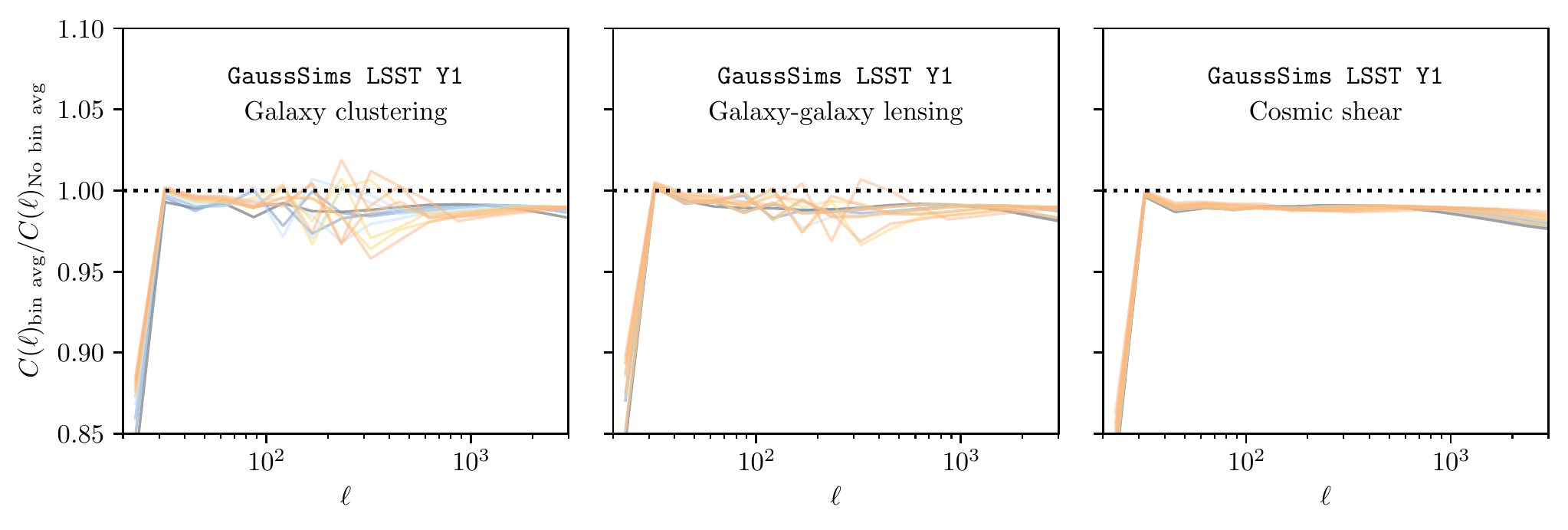}
\caption{Comparison between the theory data vector in harmonic space including proper bin averaging as described in App.~\ref{sec:estimators} using Equation~(\ref{eq:bin_avg}) vs. using the equations described in Section~\ref{sec:background} evaluated at the center $\ell$ of each angular bin. The colors represent different redshift bins, with the darker (black) colors corresponding to lower redshift, and the lighter (orange) colors to higher redshift bin combinations.}
\label{fig:bin_avg}
\end{center}
\end{figure*}

Then, the decoupled $\hat{C}^{ab}_q$ would be an unbiased estimator of the true power spectrum
evaluated at, e.g. the central multipole of each bandpower, if the power spectra was exactly
constant within each bandpower. Since that is not the case, when comparing with theory we need to propagate this averaging with the \textit{bandpower window functions}, which relate the theoretical prediction for the bandpowers $C^{ab}_q$ to the theory power spectrum $C^{ab}_\ell$:
\begin{equation}\label{eq:bin_avg}
    C^{ab}_q = \sum_\ell \mathcal{F}^{vw}_{q\ell} C^{ab}_\ell,
\end{equation}
where 
\begin{equation}
    \mathcal{F}^{vw}_{q\ell} = \sum_{q'} (\mathcal{M}^{v w})^{-1}_{q q'} \, \sum _{\ell'\in q'} w^{\ell'}_{q'} M^{u v}_{\ell \ell'}.  
\end{equation}
In Figure~\ref{fig:bin_avg} we show the impact of this effect with the angular binning that we use in this work for harmonic space, which are 17 logarithmically spaced bandpowers between $\ell=20$ and $\ell=3722$, to match the scales defined in the SRD. We find small $\ell$ values are impacted the most, reaching a $\sim$20\% effect at the largest scale. We use a common \textsc{HEALPix} resolution of $N_{\rm side} = 4096$ for all sky maps.

Finally, we apply a beam correction to account for the window function of \texttt{HEALPix}, which depends on $N_{\rm side}$. The correction is largest for small scales (large $\ell$), reaching $\sim$3\% at $\ell\simeq 3000$.

\subsubsection{Harmonic space noise subtraction} \label{app:noise_subtraction}
In Equation~\ref{eq:harmonic_estimator} we subtract the mode-coupled noise power spectrum $\Tilde{N}^{ab}_{q'}$ from the measured cosmological signal $\Tilde{C}^{ab}_{q'}$. In \textsc{TXPipe} we have implemented two different ways to compute the noise power spectrum,  analytically and using a number of realizations of noise maps. Here we detail these implementations for the galaxy clustering  and  cosmic shear estimators in harmonic space. 

For galaxy clustering, the noise is due to poisson sampling of the fields when we observe galaxies. We can  compute the \textit{coupled} noise analytically by averaging the  mask and dividing by the number density of galaxies $n_g$, in each redshift bin:
\begin{equation}
    \Tilde{N}_\text{coupled, clustering} = \frac{\left< \text{mask}\right>}{n_g},
\end{equation}
where in our case the mask is uniform with unity values within the footprint and with zero values outside of it, i.e.  $\left< \text{mask}\right>$ will approximately be the same as $f_\text{sky}$\footnote{Note that the decoupled noise can be expressed as: $N_\text{decoupled, clustering} \simeq \frac{1}{n_g}$, and that the decoupling operation corresponds to approximately an $f_\text{sky}$ factor.}. 

For the cosmic shear case, the coupled noise can be computed as the \textit{sum of weights} estimator from Equation~(2.24) in \citet{Nicola2021}: 
\begin{equation}
   \Tilde{N}_\text{coupled, cosmic shear} = \left< \sigma^2_{e, \rm pix}\right> A_\text{pix}, \, \rm{for}\  \ell \geq 2,  
\end{equation}
and zero for $\ell<2$ (in general $\Tilde{N} = 0$ for $\ell <$ spin), where $A_\text{pix}$ is the area of each pixel, the averaging $\left<\right>$ happens across all pixels in the sphere and $\sigma^2_{e, \rm pix}$ is the sum of the weighted variance of the calibrated ellipticity $e$ computed in each pixel as
\begin{equation}
    \sigma^2_{e, \rm pix} = \frac{1}{N}\sum_i^N \frac{1}{2} \left(e^2_{1, i} + e^2_{2, i}\right) w^2_i, 
\end{equation}
where $w_i$ are the weights associated with the source sample for each object $i$ in a given pixel that has a total of $N$ galaxies.  $\sigma^2_{e, \rm pix}=0$ outside the footprint. For \textsc{CosmoDC2}, we calibrate this quantity under the \textsc{Metacalibration} framework with $\sigma^2_{e, \rm pix} = R_{\rm inv}^2  \sigma^2_{e\, (\rm{uncalibrated}), \rm pix}$, where in this case we only use the diagonal part of the mean of the response matrix. 

\begin{figure*}
\begin{center}
\includegraphics[width=0.9\textwidth]{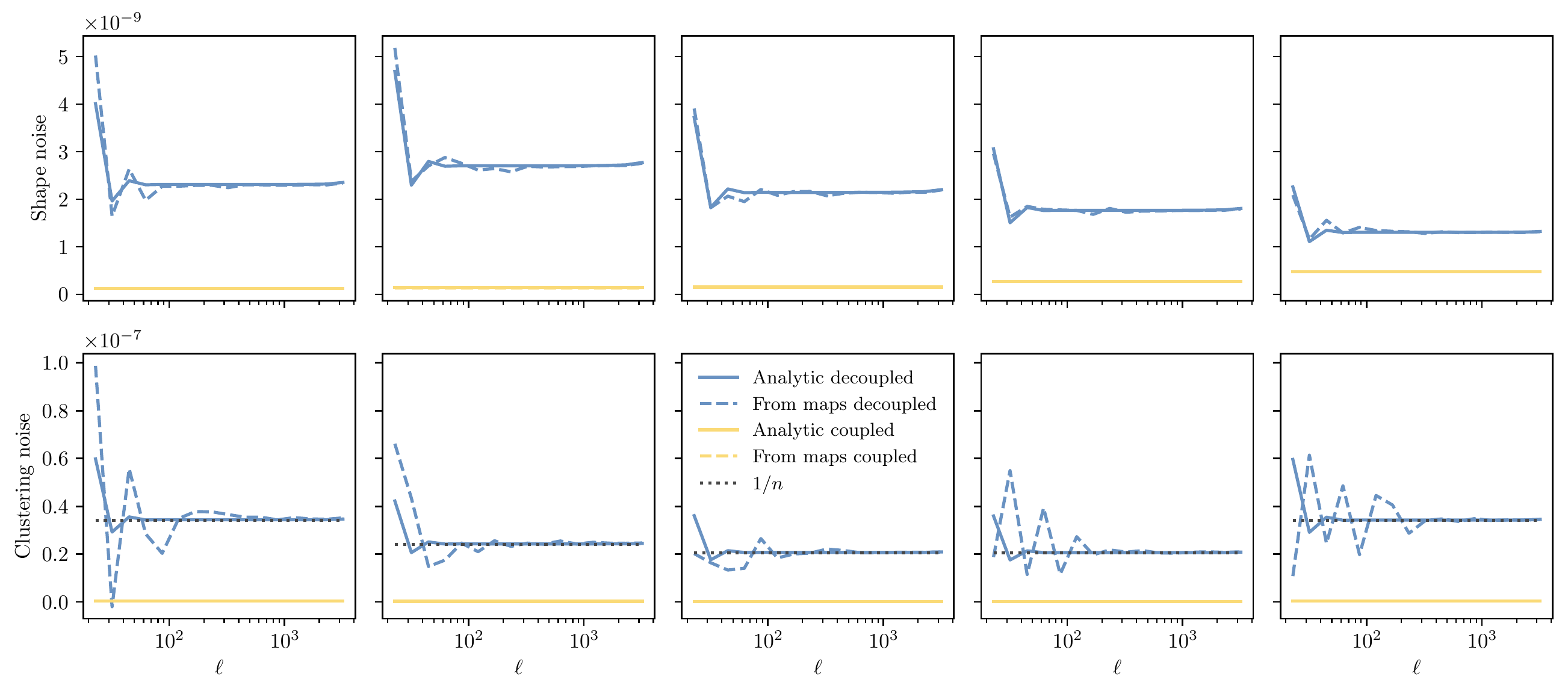}
\caption{Noise power spectrum for \textsc{CosmoDC2}, computed by \textsc{TXPipe} as described in Section~\ref{app:noise_subtraction}.}
\label{fig:noise}
\end{center}
\end{figure*}

We cross-check the above estimates with an alternative method based on generating a given number of noise realizations. For this work we use 30 realizations of the noise maps. For cosmic shear, we generate maps with random rotations of the ellipticities, and then compute the pseudo-$C_\ell$ in each noise realization $r$, and estimate the noise as the mean across all of them, following the method from \citet{Nicola2021}. For galaxy clustering, we generate gaussian realization of the density field, and then for each noise map we measure the pseudo-$C_\ell$ for half the galaxies in the map, which we label as $h1$ and $h2$. Then, we subtract the measurements from each half to null the signal and be left with the noise part, which we divide by 4 to account for the fact that the number density for each split is $1/2$ of the total one:
\begin{equation}
    \Tilde{N}_\text{coupled, clustering, maps} = \frac{1}{4}\left< \Tilde{C}^{\rm{h1}}_\ell - \Tilde{C}^{\rm{h2}}_\ell \right>_r. 
\end{equation}
We find agreement between the analytic and the noise maps realizations methods, which we show in  Figure~\ref{fig:noise} for the \textsc{CosmoDC2} simulation. 

\subsection{Real space estimator}
For the real space measurements we start from galaxy or shear catalogs and not from maps as in the harmonic space case. We use $\xi_{\pm}$ as the estimators for cosmic shear, defined in terms of the tangential ($t$) and cross ($\times$) components of the ellipticity $\hat{e}$ defined along the line that connects each pair of galaxies $a,b$. Correlating galaxies in a pair of redshift bins $(i,j)$ we define,
\begin{equation}
\xi_{\pm}^{ij}(\theta)=
\frac{\underset{ab}{\sum} w_a w_b \left(\hat{e}_{t,a}^{i}\,\hat{e}_{t,b}^{j}
\pm
\hat{e}_{\times,a}^{i}\,\hat{e}_{\times,b}^{j}\right)}
{\underset{ab}{\sum}w_{a}w_{b},
},\label{eq:xipm_estimator}
\end{equation} 
\noindent
with inverse variance weighting $w$ and  where the sums run over pairs of galaxies $a,b$, for which the angular separation falls within the range $|\boldsymbol{\theta}-\Delta\boldsymbol{\theta}|$ and $|\boldsymbol{\theta}+\Delta\boldsymbol{\theta}|$. The ellipticities that enter Equation \eqref{eq:xipm_estimator} are corrected for residual mean shear, such that $\hat{e}^i_k\equiv e^i_k - \langle e_k \rangle_i$ for components $k\in (1,2)$ and redshift bin $i$. Also, they have already been corrected by the \textsc{Metacalibration} response factors as described below. 

For galaxy-galaxy lensing, the mean tangential shear estimator is usually expressed as:
\begin{equation}
\hat{\gamma}^{ij}_t(\theta)  = \frac{\sum_{LS} w_{LS} \, \hat{e}_{t,LS}^{ij}(\theta)}{\sum_{LS} w_{LS}(\theta)} - \frac{\sum_{RS} w_{RS} \, \hat{e}_{t,RS}^{ij}(\theta)}{\sum_{RS} w_{RS}(\theta)},
\label{eq:gt}
\end{equation}
where $LS$ refers to lens-source galaxy pairs that are separated by a given angular scale that falls within the bin. $RS$ refers to random-source pairs, and this term removes the tangential shear around a sample of random points  to correct for mask effects and reduce the noise in the jackknife covariance\footnote{We do not use the jackknife covariance in this work to infer cosmology but nonetheless we validated its implementation in TXPipe, which is useful for other purposes.} \citep{Singh_2017, Prat2018}.  Note in this work we do not include the effect of lens-source clustering in the tangential shear estimator, usually referred as ``boost factors'',  which is left for future work. In our setup we do not expect them to have a significant impact given the scale cuts we use and the mostly non-overlapping redshift bins  between lenses and sources.

Finally, for the angular 2-point correlation function, $w(\theta)$ the typical estimator is the  Landy-Szalay estimator \citep{Landy1993} which can be written as: 
\begin{equation}
 \hat{w}^{ij}(\theta) = \frac{D^i D^j - D^i R^j - D^j R^i + R^i R^j }{R^i R^j} \, ,
\label{eq:wtheta_estimator}
\end{equation}\\ 
where $DD$, $RR$ and $DR$ are the normalized galaxy-galaxy, random-random and galaxy-random pair counts within an angular bin, respectively. We use 20 log-spaced $\theta$ bins between $2.5$ and $250$ arcminutes for all the probes in real space.

\subsubsection{Pixel based estimators}\label{sec:pixel_estimators}
Both the galaxy-galaxy lensing and the galaxy clustering  estimators written above require a sample of random points. This sample usually needs to be at least 20 times larger than the sample of lens galaxies \citep{Prat2022}, to avoid adding significant extra noise to the measurements\footnote{We tested using less amount of random points, in particular 12 times larger than the lens sample, and found this was insufficient to obtain a good fit to the theoretical predictions.}.  We found that for the Gaussian simulation built following the LSST Y1 SRD  lens number densities and area, it becomes infeasible to apply these estimators due to the high amount of memory they require, even after attempting several memory optimizations. As a result, we also implement a pixel-based estimator which we use for the fiducial measurements presented in Section~\ref{sec:dv}.  The pixel-based version of the Landy-Szalay estimator between two lens redshift bins $i, j$ is 
\begin{equation}\label{eq:wtheta_estimator_pixel}
   \hat{w}^{ij}(\theta) = \sum_{l=1}^{N_\text{pix}} \sum_{k=1}^{N_\text{pix}} \frac{(N^i_l- \bar{N}^i) (N^j_k-\bar{N}^j)}{\bar{N}^i \bar{N}^j} \Theta_{l,k},
\end{equation}
where $N_l$ is the galaxy number density in pixel $l$, $\bar{N}$ is the mean
galaxy number density over all pixels within the footprint and $\Theta_{l,k}$
is a top-hat function which is equal to 1 when pixels $l$ and $k$ are
separated by an angle $\theta$ within the angular bin. In the limits of an infinitely large random catalog and  small enough pixel size this is equivalent to the Landy-Szalay estimator.The fractional coverage of each pixel is taken into account in the calculation of $N_l$
and $\bar{N}$. We also use \textsc{TreeCorr} for this pixel-based estimator. As input for this function, we use the same density contrast maps that are used in the harmonic space two-point measurements. Note this estimator has already been implemented with DES Y3 data in \citet{Monroy2022}. The analogous pixel-based estimator for the mean tangential shear can be written as:
\begin{equation}
     \hat{\gamma}^{ij}_t(\theta) = \sum_{l=1}^{N_\text{pix}} \sum_{k=1}^{N_\text{pix}} \frac{(N^i_l- \bar{N}^i)}{\bar{N}^i}  \gamma^j_{t}\Theta_{l,k}
\end{equation}
where we can effectively use the  correlation  using the density contrast and ellipticity maps, named \texttt{KG} as implemented in the  \texttt{TreeCorr} code. We verify that the difference between this pixel version of the estimator and that using random points is negligible for the angular scales we consider given the ${\rm nside} = 4096$ we use and the scales cuts that are applied for the galaxy clustering and galaxy-galaxy lensing probes.

\subsection{Response factors} \label{sec:responses}

We use response factors  that account for shear and selection biases within the \textsc{Metadetection} framework introduced in \citet{Sheldon2020}. In all the above equations, the shear per each galaxy $\hat{e}$ has already been corrected by the response factors. In particular, we use the full response matrix to correct for the shear. We measure the response matrix in each of the source redshift bins, for the whole ensemble of galaxies in each bin, and then apply this mean response factor to each of the galaxies individually, before using the two-point estimators as described above. We test this implementation in TXPipe with the \textsc{CosmoDC2} measurements, for which we generate a non-unity response factor. While this implementation is already more advanced than what has been used in Stage--III surveys (e.g as in \citealt{Amon2022, Secco2022, Prat2022}), more sophisticated treatments may be required for LSST estimates, such as computing response factors for smaller ensembles or the deep-field approach described in \citet{Zhang2022b}.

\section{Mask effect on the covariance}\label{app:mask_effects}
\label{sec:mask_cov}

\begin{figure*}
\begin{center}
\includegraphics[width=0.95\textwidth]{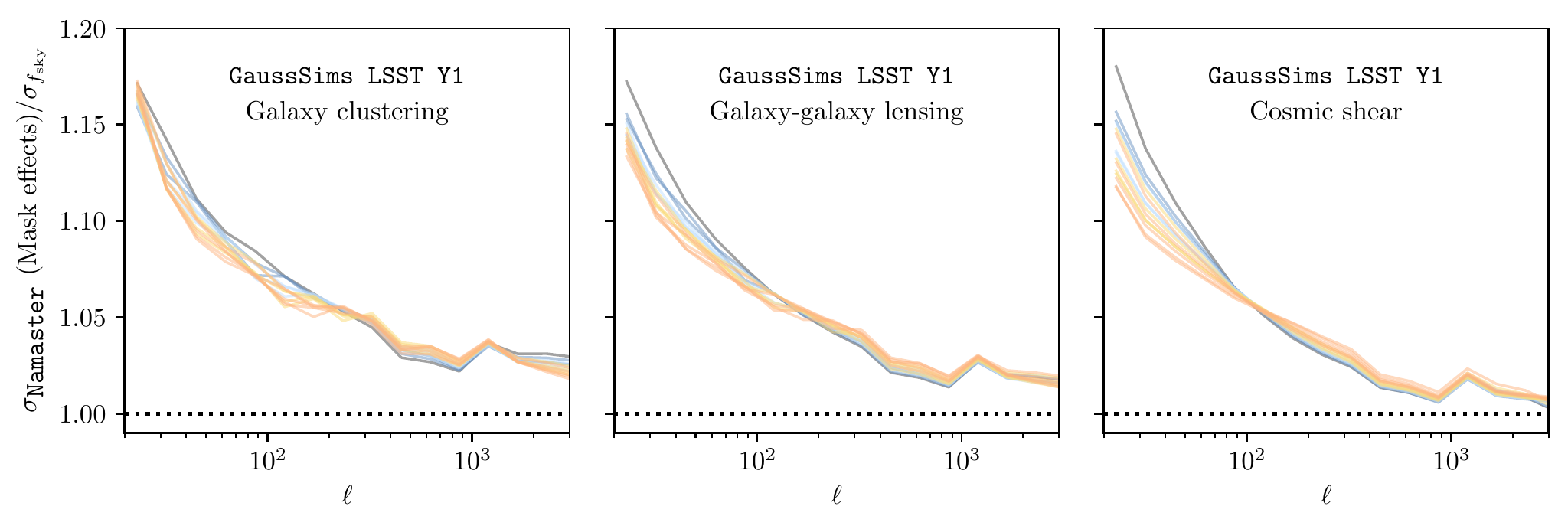}
\includegraphics[width=0.95\textwidth]{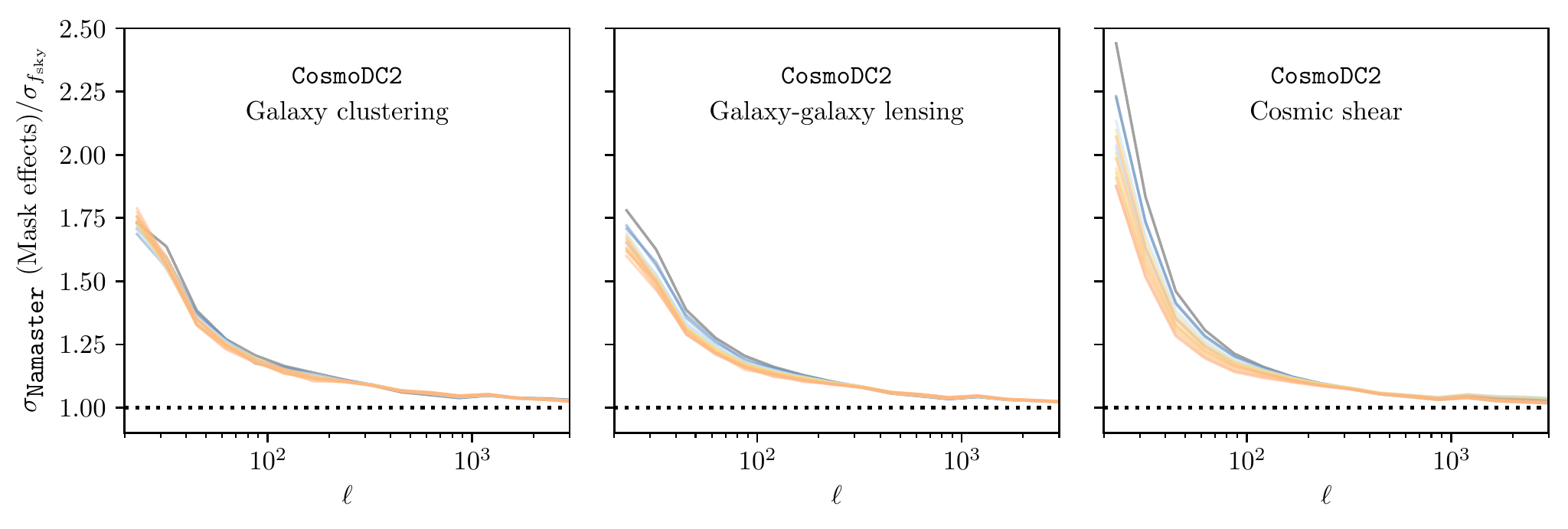}
\caption{Comparison between the covariance including coupling between the $\ell$ modes due to mask effects and the one using the $f_{\text{sky}}$ approximation. The covariance that includes mask effects is computed with the DESC code \texttt{TJPCov} that is based on \texttt{NaMaster}. The colors represent different redshift bins, with the darker (black) colors corresponding to lower redshift, and the lighter (orange) colors to higher redshift bin combinations.}
\label{fig:tjpcov}
\end{center}
\end{figure*}

\begin{figure}
\begin{center}
\includegraphics[width=0.49\textwidth]{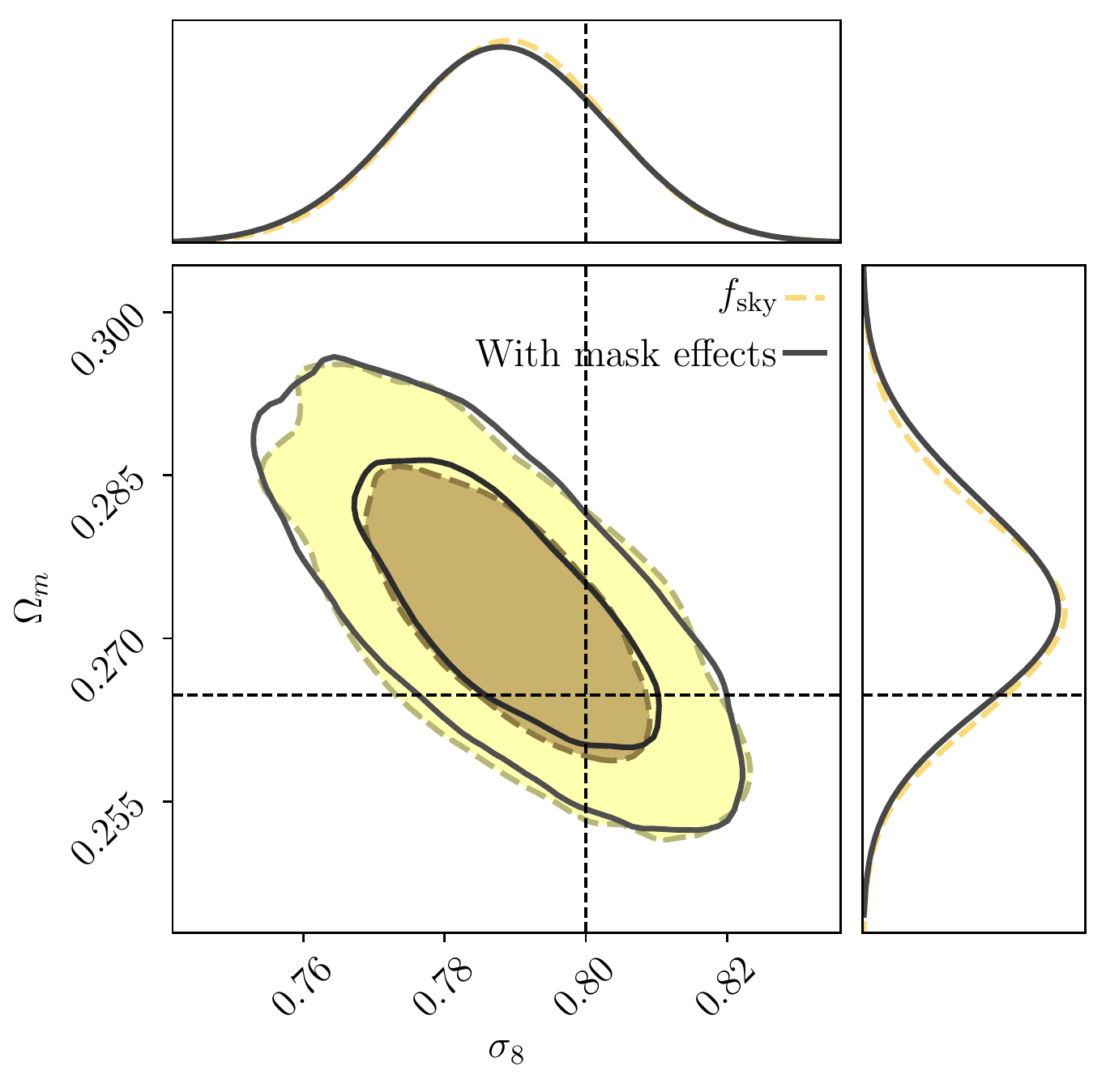}
\caption{Comparison at the cosmological posteriors level for the LSST Y1 like Gaussian simulation in the $\Lambda$CDM model  between using the covariance that includes coupling between the $\ell$ modes due to mask effects and the one using the $f_{\text{sky}}$ approximation, using the harmonic space datavectors. }
\label{fig:tjpcov-cosmo}
\end{center}
\end{figure}

In Figure~\ref{fig:tjpcov} we show the impact of including the mask effects on the covariance for harmonic space. We compare the covariance obtained with the $f_\mathrm{sky}$ approximation with the one where the coupling between modes due to the mask is taken into account. We compute the coupling between modes with the DESC package \textsc{TJPCov}, which calls the \textsc{NaMaster} code. We find the impact is biggest for the largest scales, corresponding to lower $\ell$. In particular, for our simple LSST Y1 mask shown in Figure~\ref{fig:masks} we find that for the largest scale used in this work of $\ell=20$ the effect is $\sim$15\% on the square root of the diagonal elements of the covariance. We find this to be the case for all the probes. At small scales the impact of the mask is $\leq$5\%. We check the impact on the cosmological contours is negligible in Figure~\ref{fig:tjpcov-cosmo}, and therefore do not include it in our fiducial covariance with the LSST Y1 Gaussian simulation.  We note that these conclusions might differ when considering a more realistic mask (e.g. masking bright stars  will affect all angular scales, and not only large scales).

On the other hand, we find this effect to be much larger for the smaller $\sim$440 deg.$^2$ \textsc{CosmoDC2} mask, where we find it can be as large as $\sim$200\% for $\ell=20$. We note however that $\ell=20$ is quite close to the edge of the footprint, which we would normally not use in a typical cosmological analysis. We also find that the difference in the two covariances is most significant for the cosmic shear probe. Given the magnitude of this effect, we use the covariance including mask effects for our fiducial harmonic space results on \textsc{CosmoDC2}. We do not find the impact of the mask to be as important for the real space counterpart, since there we do not use such large scales: for the real space analysis we use scales up to 250 arcmin, which corresponds to about $\ell \sim 45$, using the $\ell \sim \pi/\theta$ approximate scaling.

\section{Comparison with DESC SRD forecasts and Stage-III surveys} \label{app:srd_comparison}

\begin{figure*}
\begin{center}
\includegraphics[width=0.75\textwidth]{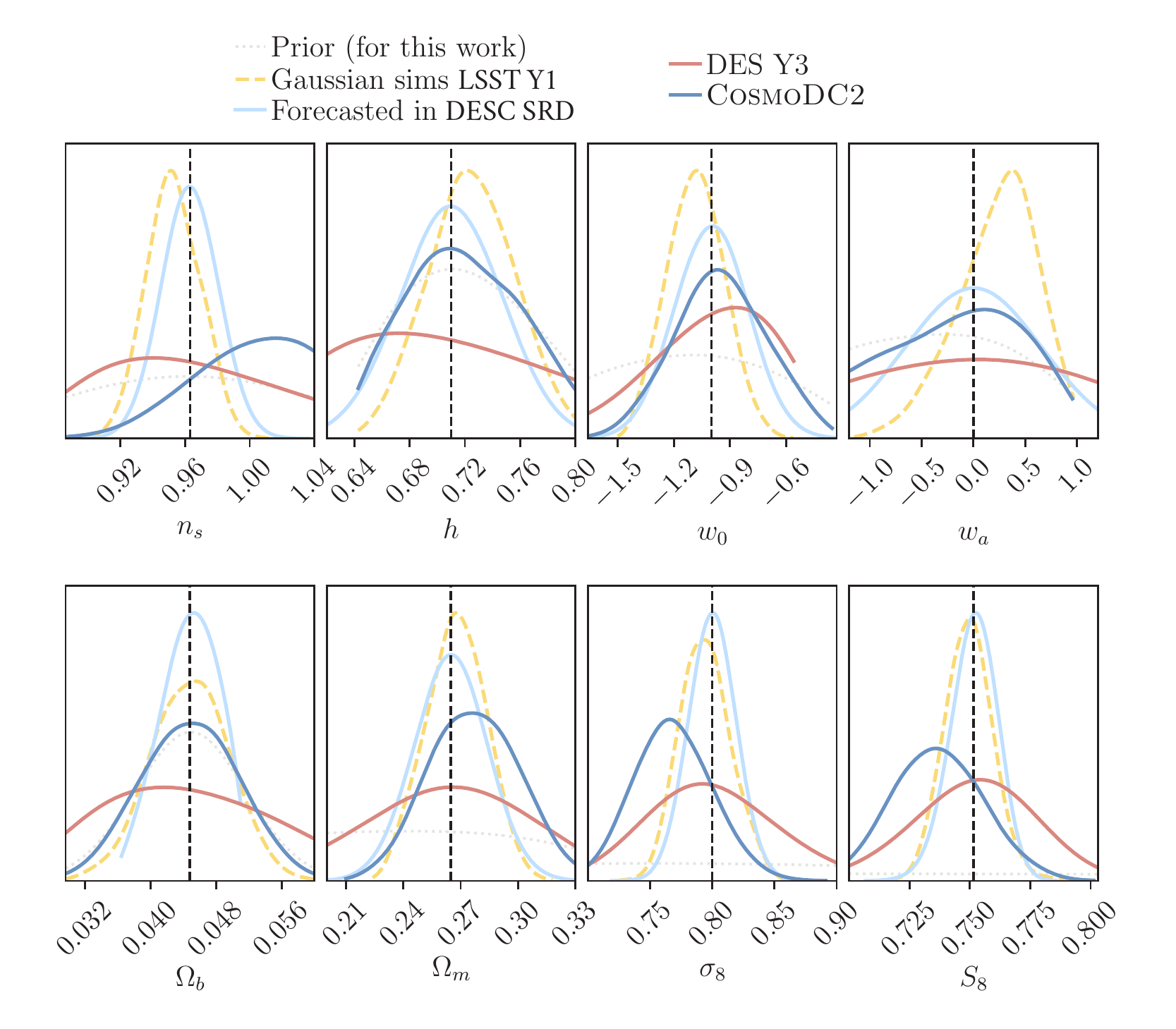}
\caption{Comparison of the 1-D posteriors from different data sets under the $w_0-w_a$CDM  model. First, we compare the results obtained in this paper from  the ~440 deg.$^2$ \textsc{CosmoDC2} simulation and the 12300 deg.$^2$ LSST-Y1 like Gaussian simulation with the prior. Most of the parameters are prior dominated for  \textsc{CosmoDC2}. The vertical dashed lines correspond to the input values for both of these simulations. Then, we compare these results  with the forecasted predictions from the DESC SRD (shifted in this plot to match the same  values as \textsc{CosmoDC2}). We also compare the constraining power to that of stage-III surveys, in particular to the DES Y3 3$\times$2pt $w_0-w_a$CDM results from the extensions work from  \citet{desy3-extensions}, with the caveat that they assume different priors with respect to this work (they assume flat priors in all cosmological parameters). The mean contours of DES Y3 are also shifted to match the input of \textsc{CosmoDC2}. }
\label{fig:srd_comparison}
\end{center}
\end{figure*}

In this appendix we compare our results on the LSST Y1 Gaussian simulation with the forecast from the DESC SRD and from Stage III surveys. However, neither of those are perfect comparisons since there are some differences in the analyses.  First we list the differences between our analysis and the DESC SRD forecast:
\begin{enumerate}
    \item \textbf{Input cosmology:} The width of the priors is matched to the DESC SRD forecast one but not their center values, since we wanted to use the same cosmological  parameters that were input to \textsc{CosmoDC2}.
    \item \textbf{Covariance:} The DESC SRD uses a non-Gaussian covariance, while we only use the Gaussian terms.
    \item \textbf{Noise:} The DESC SRD used noiseless datavectors while we use a noisy realization of the Gaussian simulation. 
    \item \textbf{Intrinsic alignment model:} The DESC SRD used a 4-parameter IA model, with an overall redshift scaling parameter with a Gaussian prior with mean $\mu=5$, $\sigma$ = 3.9 (while we use the same $\sigma$, but center it at zero, since the simulations we use do not include intrinsic alignment effects); a power-law luminosity scaling parameter (which we do not include), a redshift scaling parameter with $\mu = 0$ and $\sigma=2.3$ (identical as ours) and an additional high-redshift scaling parameter (not included here).
\end{enumerate}
Thus, the comparison with the DESC SRD is not completely straightforward but it is still useful as a validation, since each analysis uses very different software tools. Note that we only compare the results in the $w_0-w_a$CDM model, since that is the only one that was considered in the DESC SRD. We display the mean and 1-$\sigma$ uncertainties for both cases in Table~\ref{tab:model_params_wcdm} and plot the 1D posteriors in Figure~\ref{fig:srd_comparison}. Note that for the plot we have shifted the DESC SRD posteriors  to match their mean to the input values of the Gaussian and CosmoDC2 simulations that we use in this work, for an easier visual comparison. Generally we find that the constraining power of our posteriors is similar to what was predicted in SRD in most of the parameters. We only obtain significant differences in the parameters describing the equation of state of dark energy: we find the 1-$\sigma$ uncertainties in $w_0$ are $\sim$1.3 times more constraining than originally predicted and $\sim$2 times more constraining for $w_a$. A potential explanation for these differences is that we assume a simpler IA model (with  2 free parameters instead of the 4 in the SRD). 

Then, we also compare the results from this work with Stage-III posteriors, in particular with the DES Y3 3$\times$2pt $w_0-w_a$CDM analysis from the extensions work in \citet{desy3-extensions}, which assumes the same 2-parameter (NLA) IA model as we do, and allows us to perform a reasonable comparison. However, an important difference between the analyses are the priors: in our work we use the same priors that were used in the DESC SRD, which are Gaussian and are informative in some of the cosmological parameters, such as the Hubble constant, $\Omega_b$ and  $w_a$, as seen in Fig.~\ref{fig:srd_comparison} comparing the yellow and the gray posteriors, while in the DES Y3 analyses all the priors are flat and non-informative in the cosmological parameters (i.e. it would look flat in all the panels from Fig.~\ref{fig:srd_comparison}). With these caveats in mind, we find that  the LSST Y1 results on the Gaussian simulation provide a constraint in $S_8$ that is 2.5 times more constraining than DES Y3 and a significant improvement in the rest of the cosmological parameters. Interestingly, we also find that a LSST Y1 3$\times$2pt analysis will be able to place constraints in the tilt of the power spectrum $n_s$, which was completely unconstrained in Stage-III surveys analyses. 

\section{Full parameter posteriors} \label{sec:full_posteriors}

\begin{figure*}
\begin{center}
\includegraphics[width=0.63\textwidth]{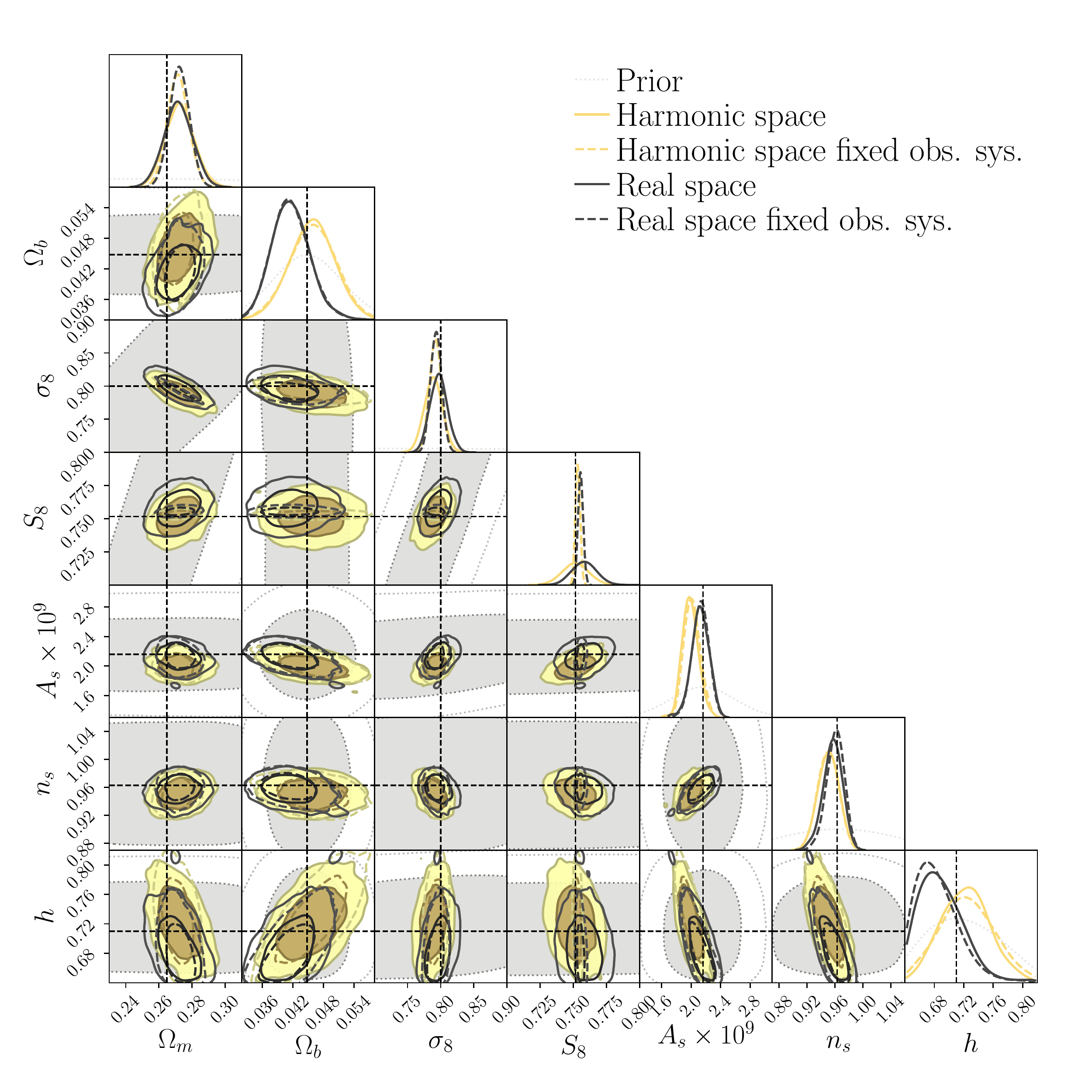}
\includegraphics[width=0.63\textwidth]{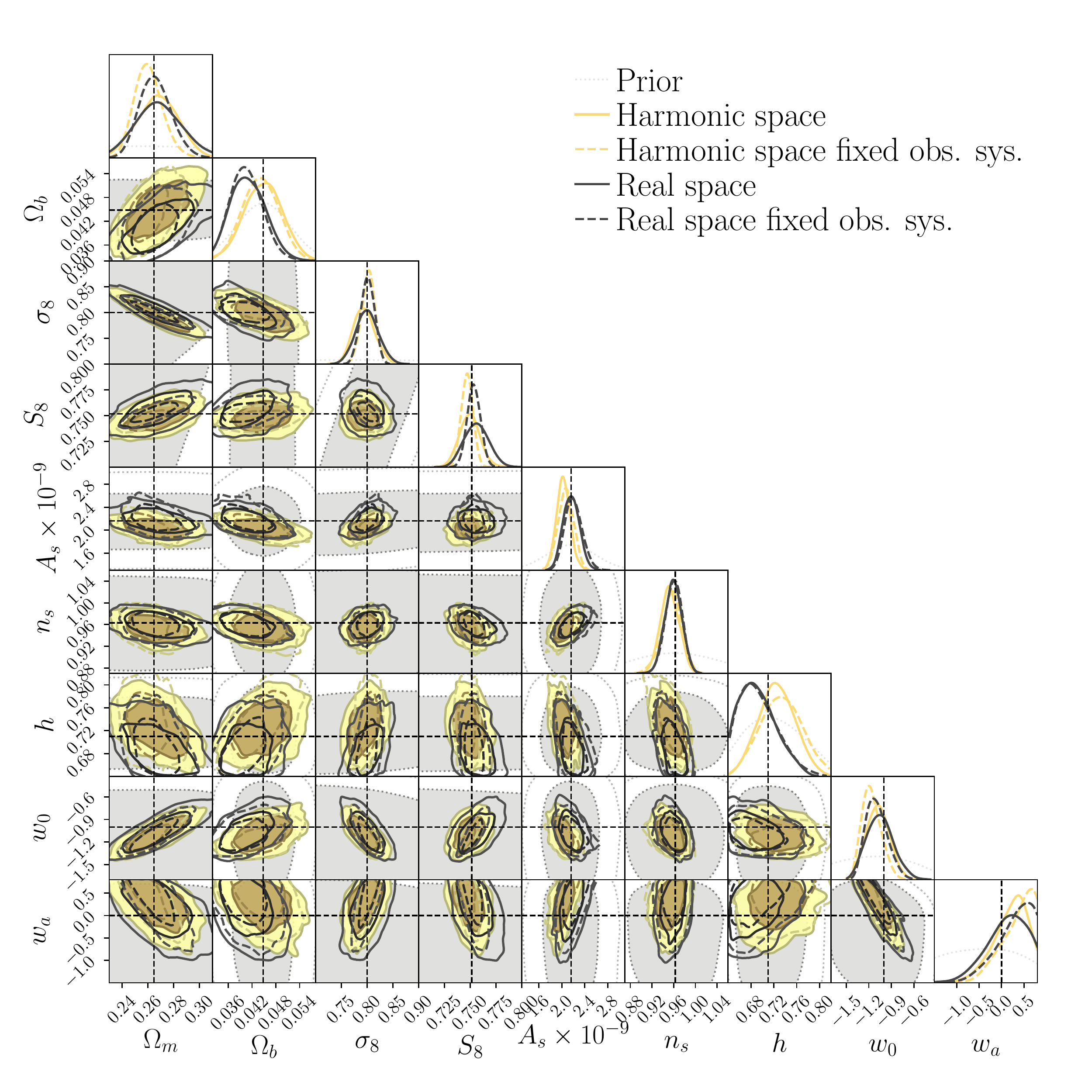}
\caption{$\Lambda$CDM (\textit{top}) and $w_0-w_a$CDM (\textit{bottom}) cosmological contours for the  LSST-Y1 like Gaussian simulation. }
\label{fig:gs_cosmo_full}
\end{center}
\end{figure*}

Here we report the rest of the 2D parameter posteriors for the LSST-Y1 like Gaussian simulation. In Figure~\ref{fig:gs_cosmo_full} we show the cosmological parameters, for $\Lambda$CDM and $w_0-w_a$CDM, in comparison with the prior. We find that all the parameters are constrained with respect to the prior,  not only the usual $\Omega_m$ and $\sigma_8$ parameters. In particular, we obtain a relatively tight constraint on $n_s$, observed for the first time in 3$\times$2pt analyses. 

We also show the difference in the posteriors when we fix the nuisance parameters controlling the redshift and shear calibration parameters (but still varying the intrinsic alignment and galaxy bias parameters).  We find that fixing these systematics results in  $S_8$ constraints that are $\sim$5 times more precise in $\Lambda$CDM, both in real space and harmonic space. In $w_0-w_a$CDM,  $S_8$ uncertainties decrease $\sim$1.8 times when fixing the same observational systematics, $\sim$1.3 times  for $w_0$ and do not change significantly for $w_a$. We note that the DESC SRD only tried to optimize the $w_0$ and $w_a$ parameters in the $w_0-w_a$CDM model and thus it is somewhat expected that we find that $S_8$ is systematics dominated, also especially in the $\Lambda$CDM model which was not considered in the DESC SRD.

\begin{figure*}
\begin{center}
\includegraphics[width=0.65\textwidth]{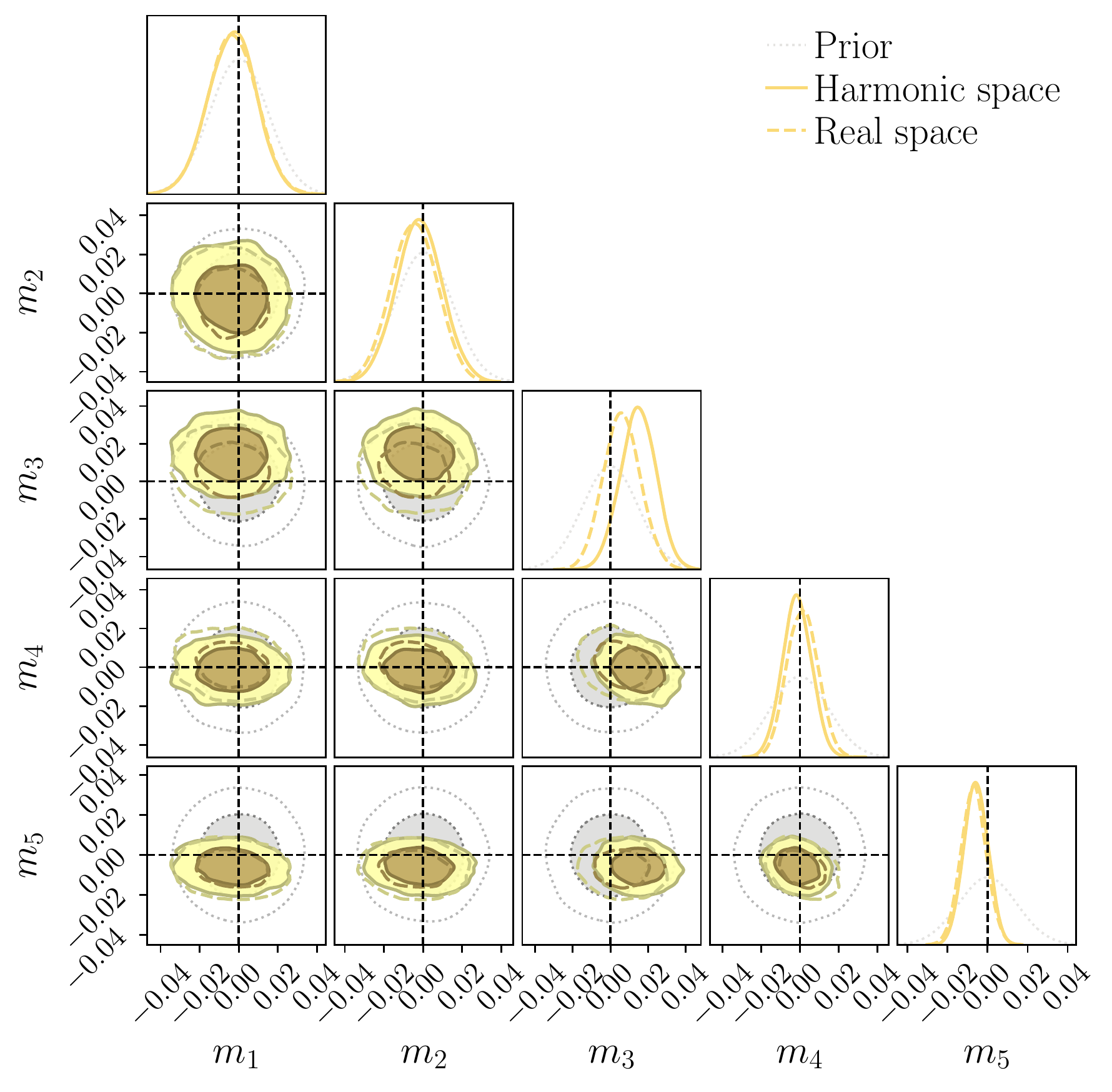}
\includegraphics[width=0.65\textwidth]{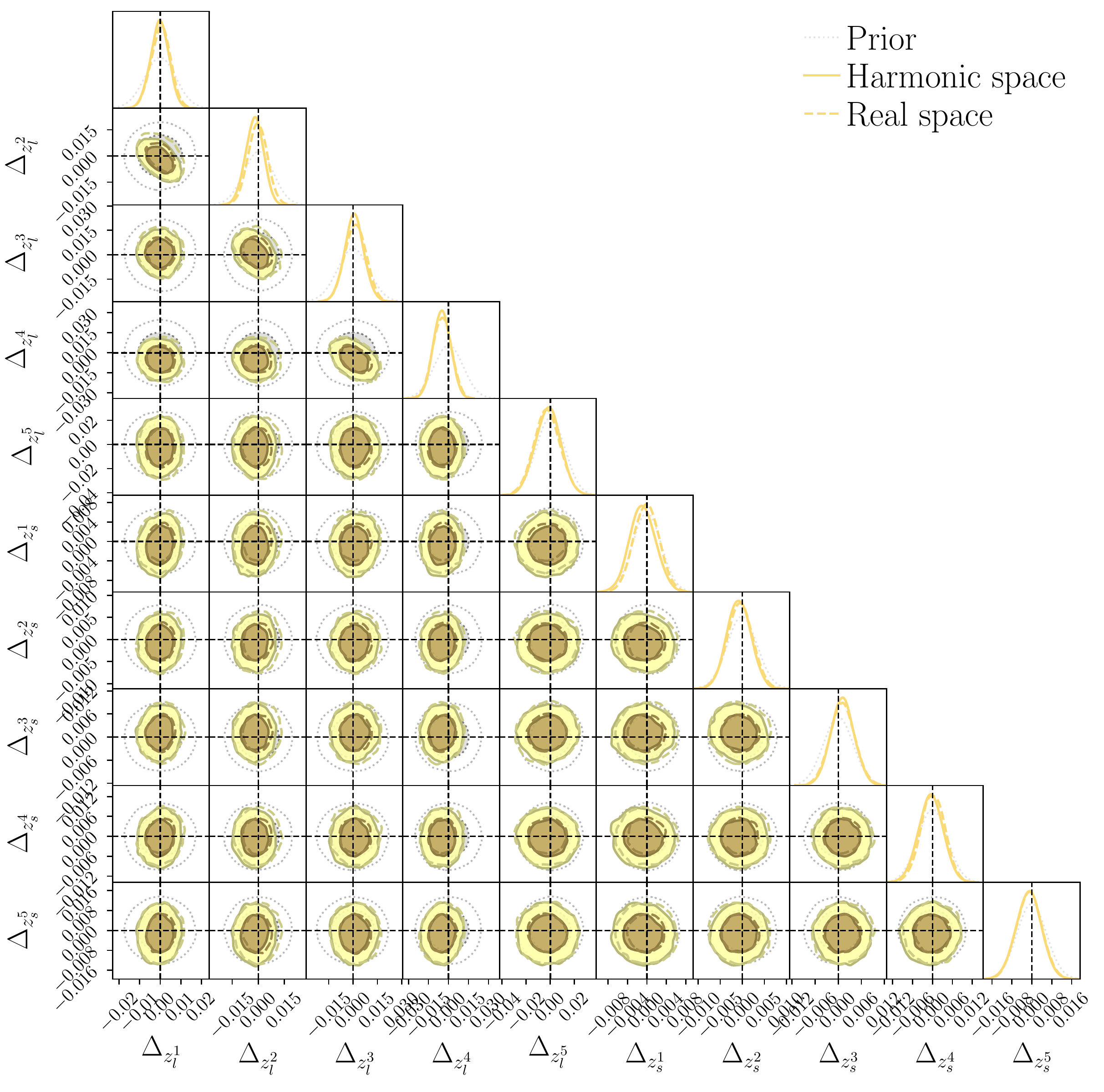}
\caption{Multiplicative shear bias posteriors (top) and redshift calibration parameters (bottom) for the  LSST-Y1 like Gaussian simulation assuming the $\Lambda$CDM model.}
\label{fig:lcdm_deltazs_ms}
\end{center}
\end{figure*}

\begin{figure}
\begin{center}
\includegraphics[width=0.36\textwidth]{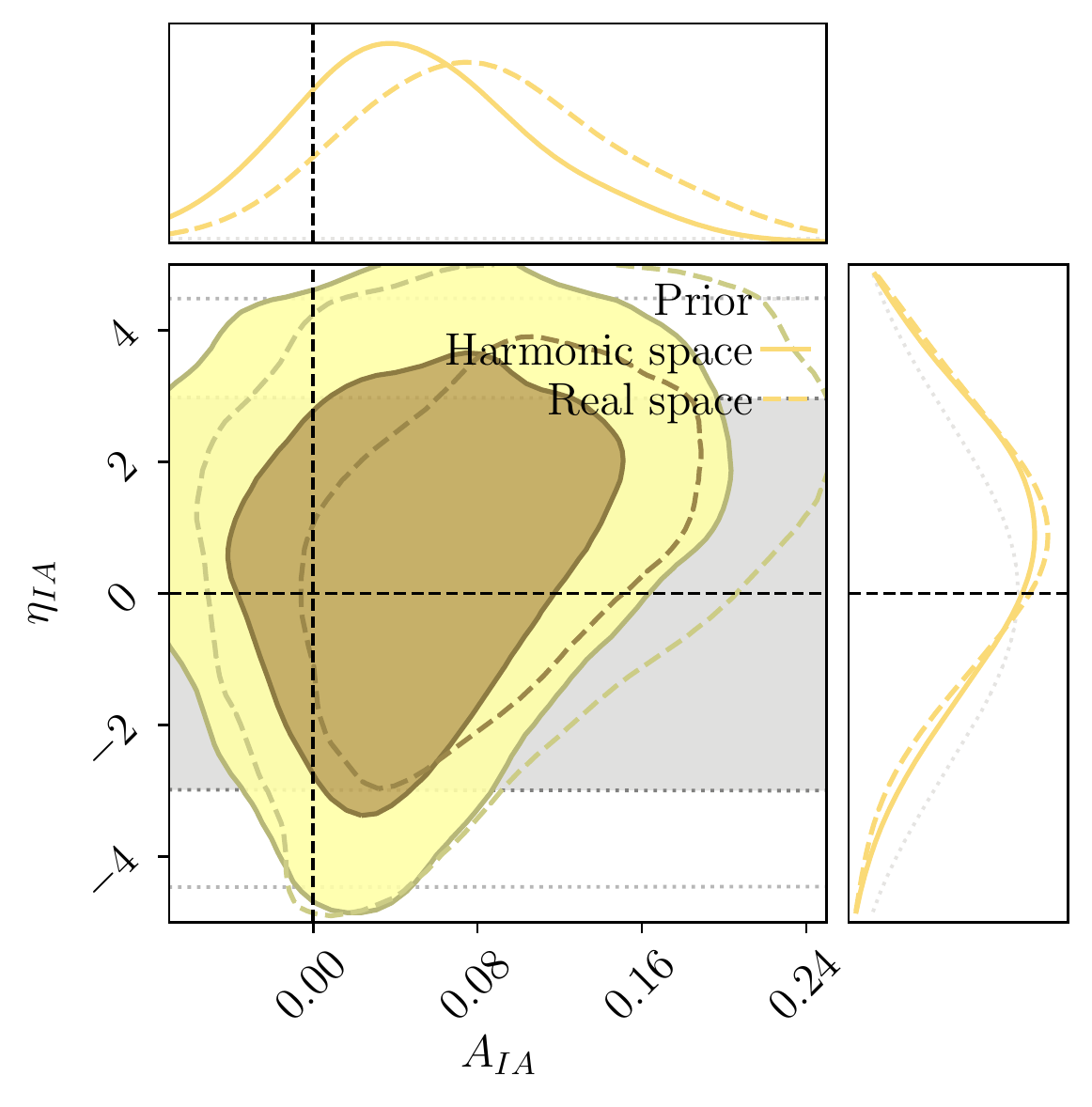}
\caption{Intrinsic alignment parameters posteriors for the  LSST-Y1 like Gaussian simulation assuming the $\Lambda$CDM model.}
\label{fig:lcdm_ia}
\end{center}
\end{figure}

We have also checked the posteriors of the nuisance parameters, for the IA, shear and redshift calibration parameters. In Figure~\ref{fig:lcdm_ia} we show the IA parameter posteriors,  finding that the amplitude of intrinsic alignments has a tight posterior while the parameter controlling its scale dependence is barely constrained. We show the posteriors for the multiplicative shear bias parameters in Figure~\ref{fig:lcdm_deltazs_ms}. We find they are being self calibrated by the data themselves with a substantial improvement with respect to the prior, especially at higher redshift. The redshift calibration parameters are  shown   in Figure~\ref{fig:lcdm_deltazs_ms} and also present significant self-calibration, especially for the lens redshift bins. Overall we find that most of the nuisance parameters are self-calibrated by the data and have tighter posteriors than the prior. This is interesting, and partly due to the fact that the DESC SRD requirements on the redshift and shear calibration parameters are rather conservative (similar to current values in Stage-III experiments). While we only show the posteriors for $\Lambda$CDM we have checked the ones for the $w_0-w_a$CDM model, which exhibit a very similar behavior.

\section{CosmoDC2 datavectors} \label{app:cosmodc2_datavectors}

\begin{figure*}
\begin{center}
\includegraphics[width=0.9\textwidth]{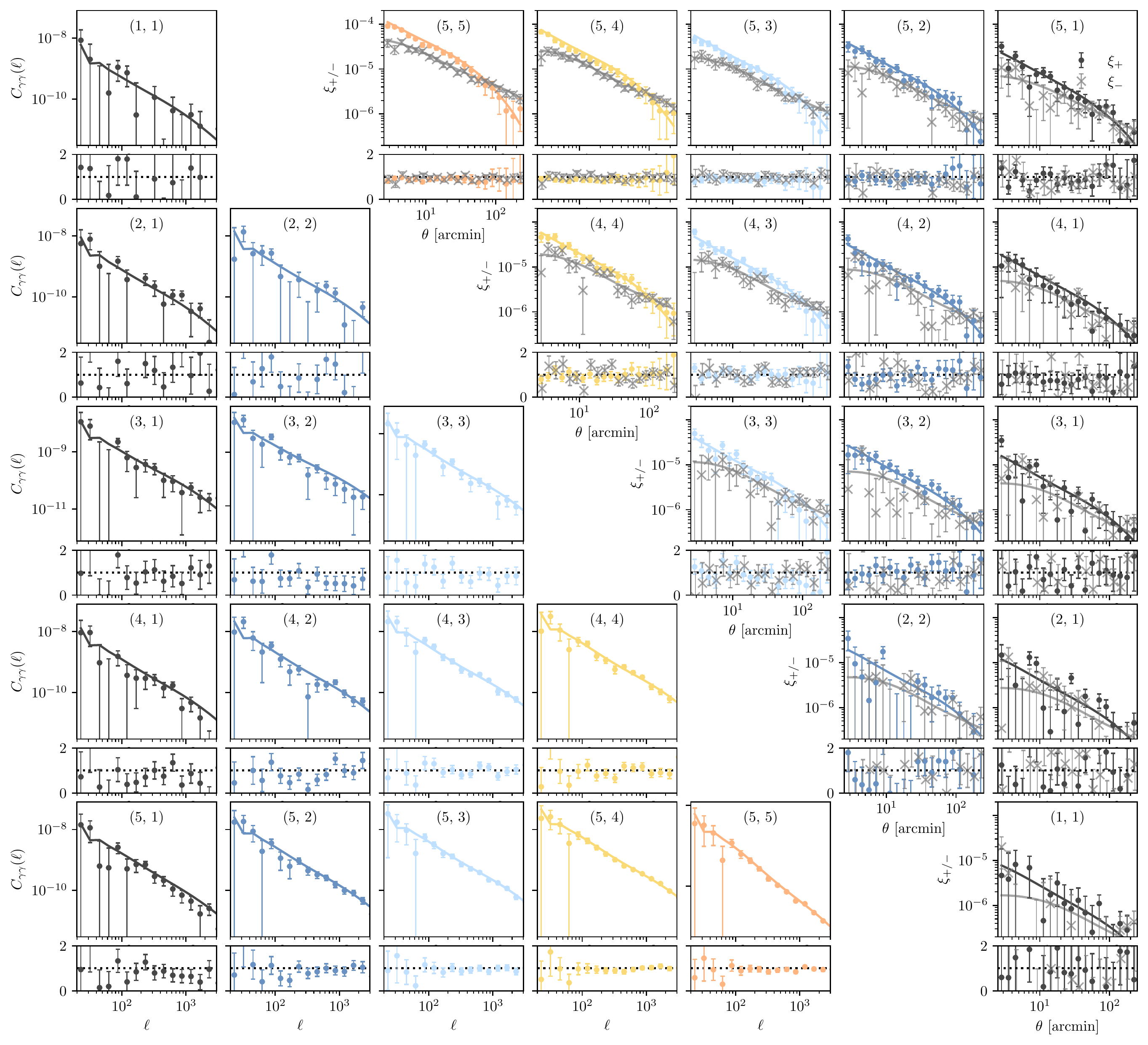}
\caption{Analogous figure to Fig.~\ref{fig:cosmicshear} for the \textsc{CosmoDC2} simulation.} 
\label{fig:cosmicshear_cosmodc2}
\end{center}
\end{figure*}

\begin{figure*}
\begin{center}
\includegraphics[width=0.68\textwidth]{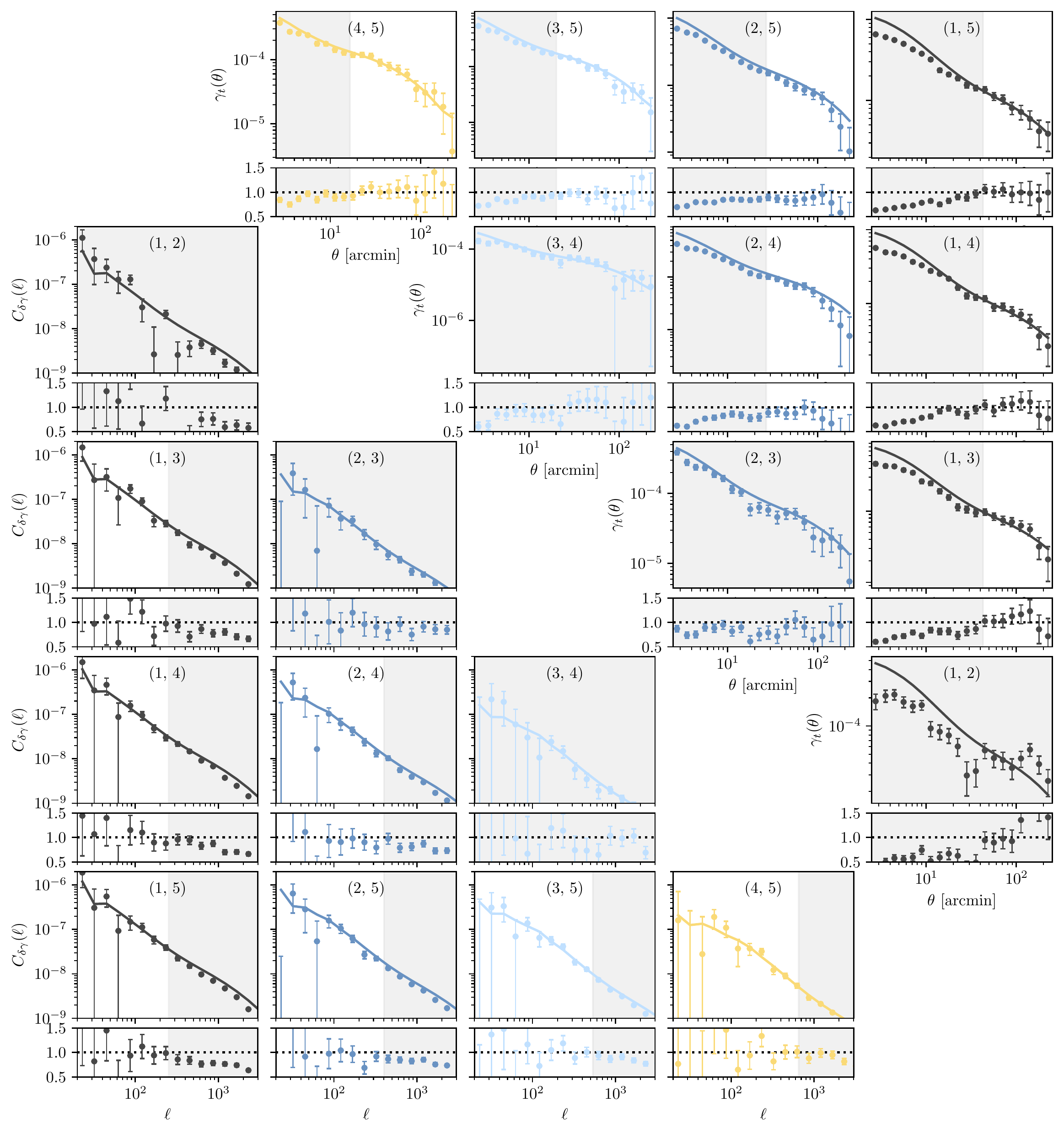}
\caption{Analogous figure to Fig.~\ref{fig:ggl} for the \textsc{CosmoDC2} simulation.}
\label{fig:ggl_cosmodc2}
\end{center}
\end{figure*}

\begin{figure*}
\begin{center}
\includegraphics[width=0.77\textwidth]{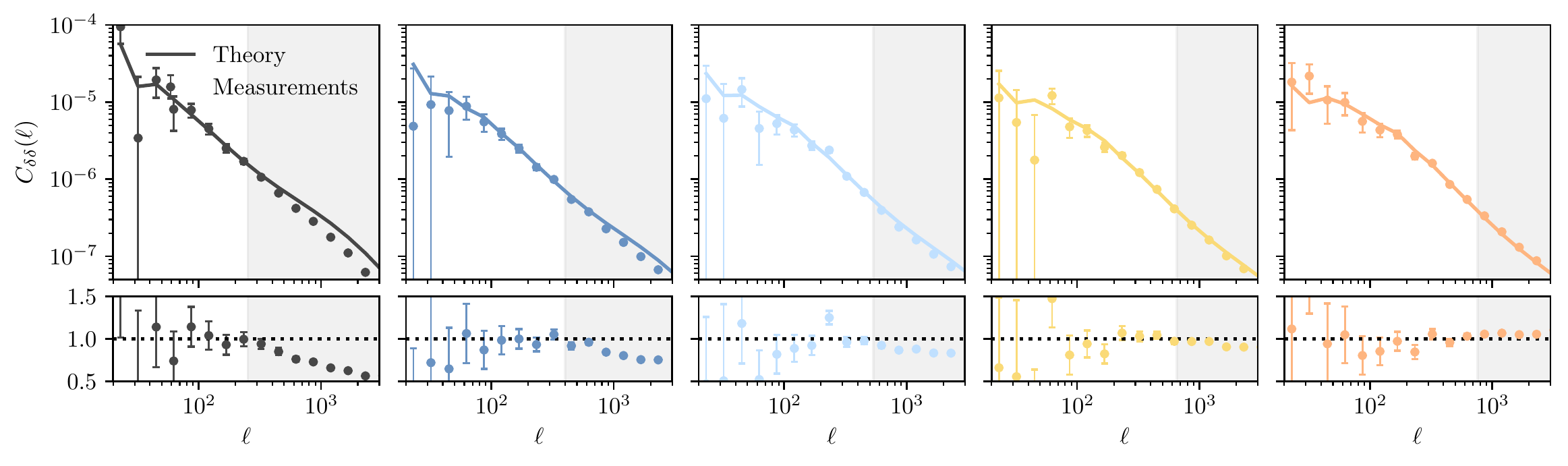} \\
\includegraphics[width=0.77\textwidth]{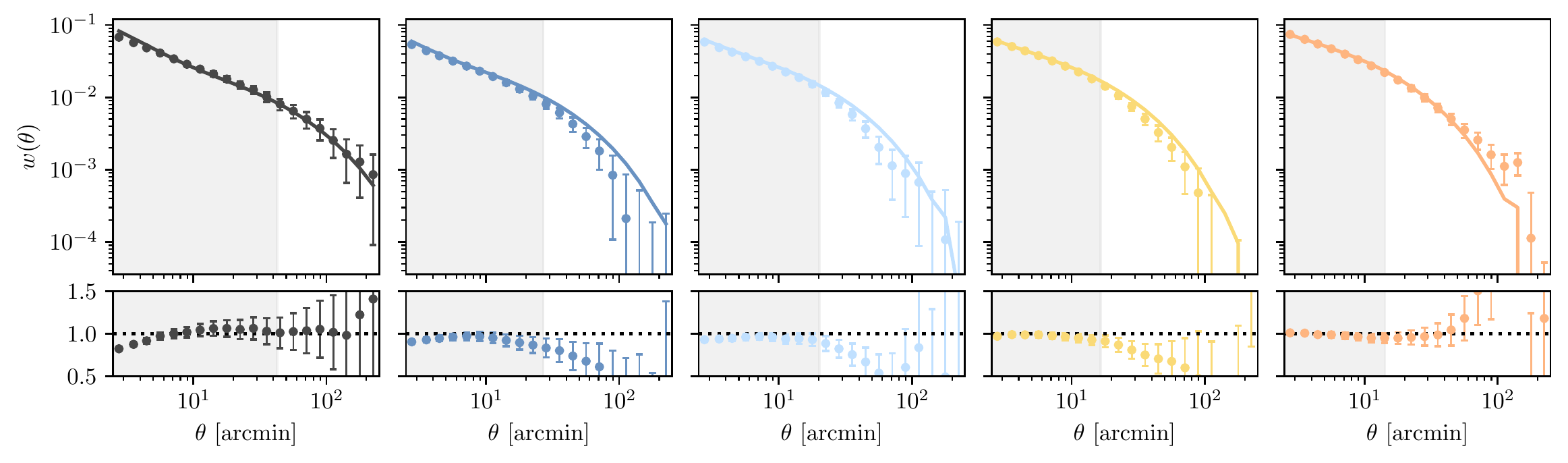}
   \caption{Analogous figure to Fig.~\ref{fig:clustering} for the \textsc{CosmoDC2} simulation. Note that these measurements are highly correlated and that we find good agreement between the theory and the measurements using the PTE metric as shown in  Table~\ref{tab:cosmodc2_sample}.} \label{fig:clustering_cosmodc2}
\end{center}
\end{figure*}

In Figures~\ref{fig:cosmicshear_cosmodc2}, \ref{fig:ggl_cosmodc2} and \ref{fig:clustering_cosmodc2} we display the measurements for cosmic shear, galaxy-galaxy lensing and galaxy clustering, respectively, for the \textsc{CosmoDC2} simulation. Visually inspecting the measurements, we observe that the large scales suffer from some sharp features below $\ell\sim50$, which probably indicates that such large scales should be used carefully given the relatively small area of \textsc{CosmoDC2}. We also observe some apparent deviations between the theory and the measurements for the real space galaxy clustering case. However, we note that these measurements are highly correlated and that we find good agreement between the theory and the measurements within the scale cuts using the $\chi^2$ and PTE metrics described in Section~\ref{sec:validation} and shown in  Table~\ref{tab:cosmodc2_sample}. 

\section{Aftermath of the code comparison with DESCQA and validation process}\label{app:descqa}
This appendix complements the discussion from Section~\ref{sec:discsussion}, where we mention that comparing different pieces of the output with external codes often exposes problems that would otherwise be hard to track. Here we illustrate this point by giving examples of the aftermath of such a comparison, in particular of the one comparing the real-space two-point correlation function measurements and theory data vectors from \textsc{TXPipe}'s output to the ones from another DESC code, DESCQA\footnote{\url{https://github.com/LSSTDESC/descqa}}, previously validated in \citet{Kovacs2021}. From this comparison we identified:
\begin{itemize}
\item A bug when constructing the redshift distributions in \textsc{TXPipe} related to an offset in the redshift binning vs. the histogram, by half a bin. This produced noticeable differences in the theory data vectors yielded by each code. 
\item A bug in the mean shear subtraction. The mean shear across the whole sample per each redshift bin was not being subtracted correctly.  
\item A bug related to a mislabeling/typographical error.
A name mismatch in the configuration file made us think we were using \texttt{bin\_slop=0} (a parameter in \texttt{TreeCorr} that when non-zero introduces some randomness in the result while speeding up the calculation), when in reality this option was not being transferred to \texttt{TreeCorr}.
\end{itemize}
 Other examples of issues that were identified in the process of validating the two-point measurements and theory data vectors for the Gaussian simulations include:
 \begin{itemize}
\item Bugs related to the ingestion of the input catalogs: Initially the catalog that was fed in to \textsc{TXPipe} was already masked, and the catalog level mask was interacting with the mask applied within \textsc{TXPipe} in unexpected ways. 
\item Bugs related to using patches in several stages, which are needed to handle such large catalogs. For instance,
clashing cache name directories  produced identical outputs from different inputs, which was identified and then fixed by making a cache directory based on creation time with an absolute path. 
\item Some redshift distributions produced by \textsc{TXPipe} were susceptible to numerical errors when being integrated to produce theory predictions, which led to incorrect theory data vectors. 
 \end{itemize}
\end{document}

%% file: authorlist.tex
\author{J.~Prat\altaffilmark{*, 1, 2}, J.~Zuntz\altaffilmark{$\dagger$, 3}, Y.~Omori\altaffilmark{1, 2}, C.~Chang\altaffilmark{1, 2}, T.~Tr\"oster\altaffilmark{4}, E.~Pedersen\altaffilmark{5}, C.~Garc\'ia-Garc\'ia\altaffilmark{6}, E.~Phillips-Longley\altaffilmark{7}, J.~Sanchez\altaffilmark{8}, D.~Alonso\altaffilmark{6}, X.~Fang\altaffilmark{9}, E.~Gawiser\altaffilmark{10}, K.~Heitmann\altaffilmark{11}, M.~Ishak\altaffilmark{12}, M.~Jarvis\altaffilmark{13}, E.~Kovacs\altaffilmark{11}, P.~Larsen\altaffilmark{11}, Y.-Y.~Mao\altaffilmark{14},
L.~Medina Varela\altaffilmark{12}, 
M.~Paterno\altaffilmark{15}, S.~D.~P.~Vitenti\altaffilmark{16},
Z.~Zhang\altaffilmark{1} and The LSST Dark Energy Science Collaboration. 
\textit{Author affiliations may be found before the references.}}

\email[$^\ast$: Corresponding author: ]{jprat@uchicago.edu}
\email[$^\dagger$: Corresponding author: ]{joe.zuntz@ed.ac.uk }
%\collaboration{The LSST Dark Energy Science Collaboration, LSST DESC}

%% file: author_contributions.tex
\subsection{Author contributions}
Judit Prat performed the main analysis with TXPipe and FireCrown, produced all the plots in the paper and led the paper writing. Joe Zuntz created, developed, and managed the TXPipe pipeline software, contributed to the text and plots in the paper, and discussed issues arising. Yuuki Omori generated Gaussian realizations of the SRD sample, and contributed to the measurement and covariance validation. Chihway Chang contributed to defining overall scope and text in the paper, and the development and testing of TXPipe and Firecrown. Tilman Tr\"oster developed the firecrown likelihood and contributed code to firecrown and CCL to enable cosmological parameter inference. Eske Pederson worked on TXPipe as an active member of the development team, implemented the Jackknife code into TXPipe together with Judit Prat.  Carlos Garcia-Garcia implemented the \textsc{NaMaster} covariance (NKA) into \textsc{TJPCov} and its interface with \textsc{TXPipe} and helped with the Fourier space measurements. Emily Phillips-Longley contributed code to TXPipe, specifically surrounding two-point calculation, shear calibration and null tests, and contributed to ongoing testing of the pipeline on catalogs. Javier Sanchez contributed to harmonic-space pipeline stages, and analytic noise estimation of the 2-point harmonic-space data-vector. David Alonso contributed code to \textsc{CCL} and \textsc{NaMaster}, aided with covariance validation and provided feedback on paper results. Xiao Fang contributed to the testing of covariance. Eric Gawiser is a DESC builder, gave feedback on paper results and their scientific implications and contributed to planned development of 3$\times$2pt pipeline. Katrin Heitmann is a DESC builder, carried out the simulation underlying \textsc{cosmoDC2} and contributed to the \textsc{cosmoDC2} effort. Mustapha Ishak contributed to Google Jax extension work for the \texttt{TXNoiseMaps} stage and corresponding text in paper; tested the code and created issues; reviewed self-calibration pull request for TXPipe; made suggestions and comments for the paper writing. As builder: worked on CCL coding; and earlier version of Firecrown. Mike Jarvis developed new features in \textsc{TreeCorr}, largely for specific use within TXPipe, including new patch-based algorithms for computing covariances and for MPI usage. Eve Kovacs contributed to the development and validation of the \textsc{cosmoDC2} measurements used in the paper. Patrica Larsen contributed to the development of the \textsc{cosmoDC2} catalogs, advised on the \textsc{cosmoDC2} application. Yao-Yuan Mao is a DESC builder, contributed to the data access and validation of \textsc{cosmoDC2} used in the paper. Medina Varela  actively attended the Friday hack sessions for TXPipe, worked on the validation of the self-calibration for Intrinsic Alignments with Eske Pedersen, as well as prototyped a Google Jax extension for the \texttt{TXNoiseMaps}  stage. Marc Paterno is a \textsc{fireCrown} developer. Sandro Dias Pinto Vitenti is a \textsc{fireCrown} developer.  Zhuoqi Zhang contributed to the covariance validation in \textsc{TXPipe} (with vs. without mask effects).

%% file: affiliations.tex
\subsection{Affiliations}
%\scriptsize
$^{1}$ University of Chicago, 5801 S Ellis Ave, Chicago, IL 60637, USA\\ 
$^{2}$ Kavli Institute for Cosmological Physics, University of Chicago,
Chicago, IL 60637, USA\\ 
$^{3}$ Institute of Astronomy, Royal Observatory Edinburgh, University of Edinburgh, Edinburgh EH9 3HJ, United Kingdom\\
$^{4}$ Eidgenoessische Technische Hochschule Zuerich, R\"{a}mistrasse 101, 8092 Z\"{u}rich, Switzerland, Switzerland\\
$^{5}$ Harvard University, Cambridge, MA 02138, USA\\
$^{6}$ Department of Physics, University of Oxford, Denys Wilkinson Building, Keble Road, Oxford OX1 3RH, United Kingdom\\
$^{7}$ Duke University, Durham, NC 27708, USA\\
$^{8}$ Space Telescope Science Institute, 3700 San Martin Dr., Baltimore, MD   21211, USA\\
$^{9}$ University of California Berkeley, 366 Physics North, MC7300, Berkeley, CA 94720, USA \\
$^{10}$ Department of Physics and Astronomy, Rutgers University, Piscataway, NJ 08854, USA\\
$^{11}$ Argonne National Laboratory, 9700 S Cass Ave, Lemont, IL 60439, USA\\
$^{12}$ University of Texas, Dallas, 800 W Campbell Rd, Richardson, TX 75080, USA\\
$^{13}$ University of Pennsylvania, Philadelphia, PA 19104, USA\\
$^{14}$ University of Utah, Department of Physics and Astronomy 115 South 1400 East, Salt Lake City, UT 84112-0830, USA\\
$^{14}$ University of Utah, Department of Physics and Astronomy 115 South 1400 East, Salt Lake City, UT 84112-0830, USA\\
$^{15}$ Fermi National Accelerator Laboratory, P.O.\ Box 500, Batavia, IL 60510-5011, USA \\
$^{16}$ Laborat\'orio Interinstitucional de e-Astronomia - LIneA, Rua Gal. Jos\'e Cristino 77, Rio de Janeiro, RJ - 20921-40, Brazil